\documentclass[12pt,oneside]{report}
\usepackage{psu}
\begin{document}

\clearpage
\doublespace
\pagestyle{empty}
\setcounter{page}{-2} 
\noindentmode

\hfill

\begin{center}DISSERTATION APPROVAL\end{center}

The abstract and dissertation of Faisal Shah Khan for the Doctor of Philosophy in Mathematical Sciences were presented April 22, 2009 and accepted by the dissertation committee and the doctoral program.

\vspace{.375in}

\begin{singlespace}
\hspace{-.07in}\begin{tabular*}{4in}{ll}COMMITTEE APPROVALS: & 
\begin{tabular*}{3.4in}[t]{l}
\hline Steven Bleiler, Chair \\[.5in] 
\hline Bin Jiang \\[.5in]
\hline Gerardo Lafferriere \\[.5in]
\hline Marek Perkowski \\ [.5in]
 \hline Bryant York \\   Representative of the Office of Graduate Studies
\end{tabular*}
\end{tabular*}\\
\end{singlespace}

\vspace{.6in}
\begin{singlespace}
\hspace{-.07in}\begin{tabular*}{4in}{rl}
DOCTORAL PROGRAM APPROVAL: &
\begin{tabular*}{2.75in}[t]{l}
\hline Steven Bleiler, Director \\
 Mathematical Sciences Ph.D. Program
\end{tabular*}
\end{tabular*}\end{singlespace}


\clearpage
\setcounter{page}{-1} 

\begin{center}
ABSTRACT
\end{center}
\hfill

An abstract of the dissertation of Faisal Shah Khan for the Doctor of Philosophy in Mathematical Sciences presented April 22, 2009.

\vspace{.25in}
Title: Quantum Multiplexers, Parrondo Games, and Proper Quantization

\indentmode
\vspace{.25in}

A quantum logic gate of particular interest to both electrical
engineers and game theorists is the quantum multiplexer. This shared
interest is due to the facts that an arbitrary quantum logic gate may be
expressed, up to arbitrary accuracy, via a circuit consisting entirely
of variations of the quantum multiplexer, and that certain one player
games, the history dependent Parrondo games, can be quantized as games
via a particular variation of the quantum multiplexer. However, to date
all such quantizations have lacked a certain fundamental game theoretic
property.

The main result in this dissertation is the development of
quantizations of history dependent quantum Parrondo games that satisfy
this fundamental game theoretic property.  Our approach also yields
fresh insight as to what should be considered as the proper quantum
analogue of a classical Markov process and gives the first game
theoretic measures of multiplexer behavior.

\newpage
\singlespace
\setcounter{page}{0} 

\noindentmode

\begin{center}

\hfill

\vspace{78pt}
\doublespace
QUANTUM MULTIPLEXERS, PARRONDO GAMES, AND PROPER QUANTIZATION

\singlespace
\vspace{56pt}

by

\vspace{12pt}

FAISAL SHAH KHAN	

\vspace{104pt}

A dissertation submitted in partial fulfillment of the \\
requirements for the degree of

\vspace{80pt}

DOCTOR OF PHILOSOPHY

in

MATHEMATICAL SCIENCES

\vspace{42pt}

Portland State University

2009

\end{center}


\clearpage
\pagenumbering{roman}
\pagestyle{plain}
\doublespace
\indentmode

\begin{center}
\vspace*{0.4\textheight}

\emph{This dissertation is dedicated to the following individuals: foremost, to my wife Seema and to my sons Arsalaan and Armaan for exhibiting a remarkable sense of humor toward my ``changed'' state of being during the days I wrote up this document. To my late father Haroon Shah Khan whose words ``study finance, you clearly cannot do math!'' inspired me to embark on this mathematical journey in the first place. To my brother Farrukh whose habitual philosophical ramblings and his ability to inspire through truly fantastical science fictional ideas and stories (not to mention financial support during my undergraduate studies) gave me the intellectual fortitude to reach this ultimate stage in my student career. And finally, to my mother Safia Bano and my sisters Farah, Fakhra, and Fadia, for putting up with me.}
\end{center}


\clearpage
\indentmode

\chapter*{\rm ACKNOWLEDGEMENTS}
I thank Professors Marek Perkowski and Steven Bleiler for their guidance and support without which this work would not have been possible. This dissertation was composed with LaTex using a modified version of the thesis template created by Mark Paskin.

\addcontentsline{toc}{chapter}{Acknowledgements}


\clearpage
\doublespace
\indentmode
\tableofcontents


\clearpage
\doublespace
\setstretch{1}
\listoffigures


\clearpage
\indentmode
\listoftables


\clearpage
\pagenumbering{arabic}
\doublespace
\indentmode
\pagestyle{plain}


\pagestyle{fancy}
\renewcommand{\sectionmark}[1]{}
\renewcommand{\chaptermark}[1]{%
 \markboth{\chaptername
 \ \thechapter.\ #1}{}} 
\fancyhead[L]{\, \rm \selectfont \nouppercase{\leftmark} \normalfont }
\fancyfoot[C]{\rm \selectfont \thepage \normalfont}
\fancyhead[R]{}


%

\chapter{Quantum Mechanics and Computation}
Advances in computation technology over the last two decades have roughly followed Moore's Law, which asserts that the number of transistors on a microprocessor doubles approximately every two years. Extrapolating this trend, somewhere between the years of 2020 and 2030 circuits on a microprocessor will measure on an atomic scale. At this scale, quantum mechanical effects will materialize, and virtually every aspect of microprocessor design and engineering will be required to account for these effects. 

To this end, \emph{quantum information theory} studies information processing under a quantum mechanical model. One goal of the theory is the development of quantum computers with the potential to harness quantum mechanical effects for superior computational capability. In addition, attention will have to be paid to quantum mechanical effects that may obstruct coherent computation. 

The study of possible development of quantum computers falls under the theory of quantum computation, an implementation of quantum information theory. 
Quantum computation model quantum information units, called {\it qudits,} as elements of a projective $d$-dimensional complex Hilbert space. Physical operations on the qudits are represented by unitary matrices and are viewed as quantum logic circuits.
Major results in quantum computing demonstrate properties of quantum information that are not endemic to classical information. Contemporary data implies that in various aspects, quantized 
information offers advantages over classical information. For example, the 
Deutsch-Jozsa quantum algorithm \cite{DJ} determines whether a function of $n$ binary variables has a specific 
property or not in only one evaluation of the function, compared to the $2^{n-1}+1$ evaluations required by the deterministic non-quantum algorithm. 
Similarly, Grover's quantum search algorithm \cite{Grover} searches a list in time that 
is quadratic rather than exponential in the number of elements in the list, and 
Shor's period finding quantum algorithm \cite{Shor} gives a polynomial time algorithm for 
factoring integers. The last two are well known results in quantum 
computation and they show that quantum algorithms have the potential to out perform
classical algorithms for practical problems. 

Quantum game theory offers an exciting and relatively new game theoretic perspective on quantum information. Typically, research in the subject looks for different than usual behavior of the payoff function of a game when the game is played in a quantum mechanical setting. In multi-player games played in a quantum mechanical setting, the different than usual behavior of the payoff function studied is typically the occurrence of Nash equilibria that are absent in the original game \cite{Eisert, Landsburg, Marinatto}. Because quantum game theory has traditionally been heuristic in nature, confusion about and controversy over the relevance of ``quantum games'' to game theory abounds. A resolution to this confusion and controversy has been recently proposed by Bleiler in \cite{Bleiler} via a mathematically formal approach to quantum game theory.  

Using Bleiler's mathematically formal approach to quantum game theory as a stepping stone, this dissertation promotes the philosophy that quantum game theory should be used to gain insights into quantum computation. To this end, the reader is provided with a basic introduction to quantum computation and quantum mechanics in the remaining sections of this chapter. In Chapter \ref{Bleilerformalism}, the Bleiler formalism is reproduced to give readers a mathematically formal game-theoretic perspective on quantum games. Chapter \ref{Proper} presents the main results, which are construction and and game theoretic analysis of quantum versions of certain one player games, known as history dependent Parrondo games, and their randomized sequences using the Bleiler formalism as a blueprint. These constructions utilize a particular version of a quantum logic circuit known as the quantum multiplexer. The connection between quantum game theory and quantum computation is made apparent in Chapter \ref{CosSin}, where the importance of the quantum multiplexer to quantum computation is established via abstract realization of an arbitrary quantum logic circuit in terms of circuits composed entirely of quantum multiplexers. Chapter \ref{Quaterncoord} may be treated as a stand alone chapter; it proposes the analysis of quantum circuits acting on exactly two quantum informational units (qubits) via quaternionic coordinates. 
\section{Introduction to Quantum Computation}\label{Quantum Computation}
Like geometry, quantum mechanics is best viewed axiomatically. For the axioms of and basic facts about quantum mechanics, the reader is referred to \cite{Nielsen, Oskin, Bone}. These axioms and some of the basic facts appear explicitly in the next section during the development of one qubit quantum computation. 
%

A  $d$-ary quantum digit, or {\it qudit} for short, is a vector in a complex projective $d$-dimensional Hilbert space $\mathcal{H}_{d}$, called the state space of the digit, equipped with the orthogonal {\it computational basis}
$$
\left\{\ket{0},\ket{1}, \dots \ket{d-1}\right\}
$$
where $\ket{i}=(0,0,\dots,1, \dots, 0)^{T}$ with a 1 in the $(i+1)$-st coordinate, for $0 \leq i \leq (d-1)$. To pass from classical to quantum computing, replace a classical $d$-ary digit (dit) with a qudit as an information unit. 
The replacement amounts to identifying all possible values of the dit with the elements of the computational basis of the state space of the corresponding qudit. This identification enlarges the set of operations on the dit to include quantum operations which, by the axioms of quantum mechanics, are represented by unitary operators on the state space.
One then typically explores whether this enlargement results in any computational advantages or enhancements. 

To be more specific, unitary operators can be used to create complex projective linear combinations of the basis qudits. In other words, a qudit $\ket{a}$ in $\mathcal{H}_{d}$ can be expressed as a complex projective linear combination of the basis qudits
$$
\ket{a}=\sum_{i=0}^{d-1}{x_{i}\ket{i}}, \quad x_{i} \in \mathbb{C} 
$$ 
where $\ket{a} \equiv \lambda\ket{a}$ for any non-zero complex number $\lambda$. Physicists call this complex number $\lambda$ a phase. Up to phase, the state $\ket{a}$ can be normalized; that is, $\ket{a}$ can be expresses with
$$
\sum_{i=0}^{d-1}{\left|x_{i}\right|^{2}}=1
$$
The measurement axioms of quantum mechanics say that the real number $\left|x_{i}\right|^{2}$ is the probability that the state vector $\ket{a}$ will be observed in $i$-th basis state upon measurement with respect to that basis. Typical considerations in quantum computing are whether evolutions of the state space offer computational enhancements. 

When considering several qudits at once, the axioms of quantum mechanics tell us to consider their joint state space. When the state spaces of $n$ qudits of different $d$-valued dimensions are combined, they do so via their tensor product as per the axioms of quantum mechanics and the result is a $n$ qudit \emph{hybrid} state space 
$$
\mathcal{H}=\mathcal{H}_{d_{1}} \otimes \mathcal{H}_{d_{2}} \otimes \dots \otimes \mathcal{H}_{d_{n}}
$$
where $\mathcal{H}_{d_{i}}$ is the state space of the $d_{i}$-valued qudit. The computational basis for $\mathcal{H}$ consists of all possible tensor products of the computational basis vectors of the component state spaces $\mathcal{H}_{d_{i}}$. If $d_i=d$ for each $i$, the resulting state space $\mathcal{H}_{d}^{\otimes{n}}$ is that of $n$ $d$-valued qudits. 

Once a basis for the state space has been chosen, a unitary operator on it is represented by a unitary matrix. For the hybrid state space $\mathcal{H}$, an evolution matrix will be of size $(d_{1}d_{2} \ldots d_{N}) \times (d_{1} d_{2} \ldots d_{N})$, while the evolution matrix for $\mathcal{H}_{d}^{\otimes{n}}$ will have size $d^{n} \times d^{n}$. 

Consider a two dimensional state space $\mathcal{H}_2$. This is the state space of a quantum system which gives two possible outcomes upon measurement. An example of such a system would be one that describes the spin states of an electron. Topologically, $\mathcal{H}_2=\mathbb{C}P^1$. The two possible states of the system form the computational basis for the state space. These orthogonal basis state are viewed as the two possible values a bit of information can take on. Call the elements of $\mathcal{H}_2$ {\it qubits}, short for binary quantum digit. The resulting 2-valued quantum computing has traditionally been the most active area of research. The basics of 2-valued quantum computing are reviewed in the following sections. Higher valued quantum computing has seen much research activity recently as well. The reader is referred to chapter 2 for a discussion of certain aspects of $d$-valued quantum computing and relevant references.  

\subsection{One Qubit Quantum Computing}\label{one qubit qcomp}

Let $\left| {b_0} \right\rangle$ and $\left| {b_1 }\right\rangle$ be an orthogonal basis for $\mathcal{H}_2$. Then the states of the qubit are projective linear combinations of these basis elements over $\mathbb{C}$: 
$$
\left| \psi \right\rangle =\alpha_0\left| {b_0 } \right\rangle 
+\alpha_1\left| {b_1 } \right\rangle
$$
with $\alpha_0, \alpha_1 \in \mathbb{C}$ satisfying, without loss of generality, $\left|\alpha_0\right|^2+\left|\alpha_1\right|^2=1$. These projective complex linear combinations are also called {\it superpositions} of the states $\ket{0}$ and $\ket{1}$. The computational basis is the set 
$$
B_{\mbox{comp}}=\left\{ {\left( {{\begin{array}{*{20}c}
 1 \hfill \\
 0 \hfill \\
\end{array} }} \right),\left( {{\begin{array}{*{20}c}
 0 \hfill \\
 1 \hfill \\
\end{array} }} \right)} \right\}
$$
which gives the convention of labeling the 
basis with Boolean names, with
$$
\left| {b_0} \right\rangle =\left| 0 
\right\rangle =\left( {{\begin{array}{*{20}c}
 1 \hfill \\
 0 \hfill \\
\end{array} }} \right)\;\mbox{and } \left| {b_1 } \right\rangle =\left| 1 
\right\rangle =\left( {{\begin{array}{*{20}c}
 0 \hfill \\
 1 \hfill \\
\end{array} }} \right).$$

But note that these are only names. For example, in the spin state model for an electron, one 
might imagine that $\left| 0 \right\rangle$  is being represented by an 
up-spin while $\left| 1\right\rangle $ by a down-spin.
The key is that there 
is an abstraction between the technology (spin state or other quantum 
phenomena) and the logical meaning. This same detachment is true in 
classical computers where we traditionally call a high positive voltage 
``1'' and a low ground potential ``0''.

Let $\left| {\psi _1 } \right\rangle =\alpha_0\left| 0 \right\rangle +\alpha_1\left| 1 
\right\rangle$
and 
\begin{equation}\label{unitary mat}
U=\left( {{\begin{array}{*{20}c}
 0 \hfill & -\overline{\eta} \hfill \\
 \eta \hfill & 0 \hfill \\
\end{array} }} \right)
\end{equation}
be a special unitary operator which, by axioms of quantum mechanics, corresponds to a physical operation. Further, suppose that $\eta$ is a complex root of unity other than $\pm 1$, the use of which will be justified shortly. The matrix $U$ acts on $\left|{\psi _1} \right\rangle$ as follows.
\begin{equation}\label{eqA}
U\left| 
{\psi _1 } \right\rangle =\left( {{\begin{array}{*{20}c}
 0 \hfill & -\overline{\eta} \hfill \\
 \eta \hfill & 0 \hfill \\
\end{array} }} \right)\left( {{\begin{array}{*{20}c}
 \alpha_0 \hfill \\
 \alpha_1 \hfill \\
\end{array} }} \right)=\left( {{\begin{array}{*{20}c}
 -\overline{\eta}\alpha_1 \hfill \\
 \eta \alpha_0\hfill \\
\end{array} }} \right)=-\overline{\eta}\alpha_1\left| 0 \right\rangle +\eta \alpha_0\left| 1 \right\rangle.
\end{equation}
Up to multiplication by unitary phase, the operator $U$ interchanges the coefficients of the basis states of $\mathbb{C}P^{1}$. In particular, $U$ sends the state $\left|0\right\rangle$ to the state $\eta\left|1\right\rangle$
$$
U\left| 0 \right\rangle =\left( {{\begin{array}{*{20}c}
 0 \hfill & -\overline{\eta} \hfill \\
 \eta \hfill & 0 \hfill \\
\end{array} }} \right)\left( {{\begin{array}{*{20}c}
 1 \hfill \\
 0 \hfill \\
\end{array} }} \right)=\left( {{\begin{array}{*{20}c}
 0 \hfill \\
 \eta \hfill \\
\end{array} }} \right)=\eta\left| 1 \right\rangle
$$
and the state $\ket{1}$ to the state $-\overline{\eta}\ket{0}$
$$
U\left| 1 \right\rangle 
=\left( {{\begin{array}{*{20}c}
 0 \hfill & -\overline{\eta} \hfill \\
 \eta \hfill & 0 \hfill \\
\end{array} }} \right)\left( {{\begin{array}{*{20}c}
 0 \hfill \\
 1 \hfill \\
\end{array} }} \right)=\left( {{\begin{array}{*{20}c}
 -\overline{\eta} \hfill \\
 0 \hfill \\
\end{array} }} \right)=-\overline{\eta}\left| 0 \right\rangle.
$$

\begin{figure}
\centerline{\includegraphics[scale=0.25]{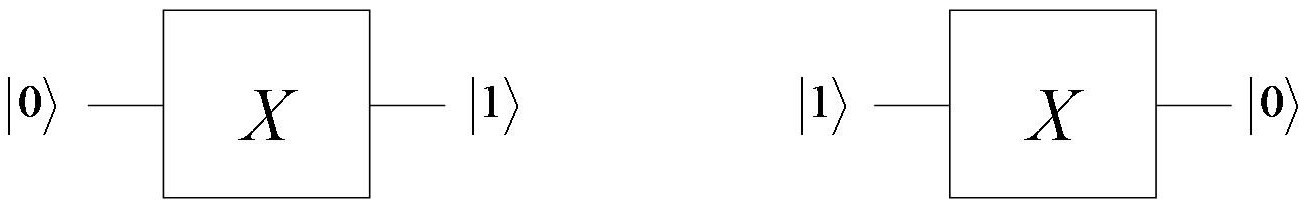}}
\caption{\small{Inverter or the NOT quantum logic gate $U$. The wires carry quantum information, namely qubits.}}
\label{inverteU}
\end{figure}

\begin{figure}
\centerline{\includegraphics[scale=0.25]{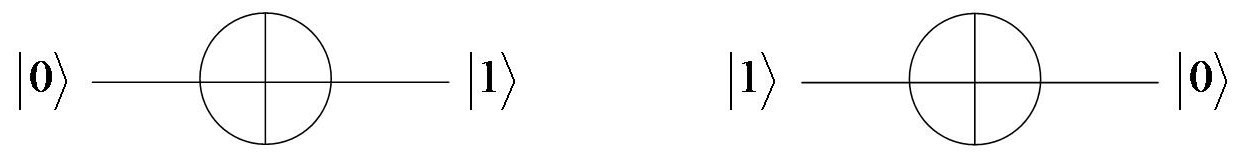}}
\caption{\small{Standard notation for the NOT gate.}}
\label{NOT}
\end{figure}
This action of $U$ is interpreted as that of a {\it quantum logic gate} that inverts, up to unitary phase, the logical values $\left| 0 \right\rangle$ and $\left| 1 \right\rangle$; that is, the gate $U$ is a quantum version of the NOT gate in classical logic. This point of view allows one to view quantum mechanics as a theory of quantum computation. Standard notation for the NOT gate is given in Figure \ref{NOT}.

In the quantum theory of games, one frequently views a qubit as a ``quantum coin'' and hence the gate $U$ can be interpreted as the quantum mechanical analog of flipping over a coin, while the $2 \times 2$ identity matrix is the analog of leaving the coin un-flipped. In certain quantum games, such as the ones found in \cite{Landsburg, Ahmed}, the flipping and un-flipping actions of players on the so-called maximally entangled state of two qubits
$$
\frac{1}{\sqrt 2}\left(\ket{00}+\ket{11}\right)
$$
are considered. For the purpose of analysis of the quantum game, these actions are required to produce an orthogonal basis of the joint state space, and this happens only when $\eta$ is an appropriate root of unity other than $\pm 1$.

\subsection{The One Qubit Hadamard Quantum Logic Gate}

Quantum computing literature gives many interesting examples of one qubit gates. The focus here will be on the one qubit Hadamard gate described by
the special unitary matrix 
$$
H=\frac{i}{\sqrt 2 }\left( {{\begin{array}{*{20}c}
 1 \hfill & 1 \hfill \\
 1 \hfill & {-1} \hfill \\
\end{array} }} \right).
$$
Application of the gate $H$ to either basis states $\left| 0 \right\rangle$ and $\left| 1\right\rangle$ creates an {\it equal superposition} of the basis state, that is, a superposition that will appear in each basis state with equal probability upon measurement with respect to the basis. 
\begin{align*}
H\left| 0 \right\rangle &=\frac{i}{\sqrt 2 }\left( {{\begin{array}{*{20}c}
 1 \hfill & 1 \hfill \\
 1 \hfill & {-1} \hfill \\
\end{array} }} \right)\left( {{\begin{array}{*{20}c}
 1 \hfill \\
 0 \hfill \\
\end{array} }} \right) \\
&=\frac{i}{\sqrt 2 }\left( {{\begin{array}{*{20}c}
 1 \hfill \\
 1 \hfill \\
\end{array} }} \right)=\frac{i}{\sqrt 2 }\left( {{\begin{array}{*{20}c}
 1 \hfill \\
 0 \hfill \\
\end{array} }} \right)+\frac{i}{\sqrt 2 }\left( {{\begin{array}{*{20}c}
 0 \hfill \\
 1 \hfill \\
\end{array} }} \right)=\frac{i}{\sqrt 2 }\left| 0 \right\rangle 
+\frac{i}{\sqrt 2 }\left| 1 \right\rangle 
\end{align*}
and
\begin{align*}
H\left| 1 \right\rangle &=\frac{i}{\sqrt 2 }\left( {{\begin{array}{*{20}c}
 1 \hfill & 1 \hfill \\
 1 \hfill & {-1} \hfill \\
\end{array} }} \right)\left( {{\begin{array}{*{20}c}
 0 \hfill \\
 1 \hfill \\
\end{array} }} \right) \\
&=\frac{i}{\sqrt 2 }\left( {{\begin{array}{*{20}c}
 1 \hfill \\
 {-1} \hfill \\
\end{array} }} \right)=\frac{i}{\sqrt 2 }\left( {{\begin{array}{*{20}c}
 1 \hfill \\
 0 \hfill \\
\end{array} }} \right)-\frac{i}{\sqrt 2 }\left( {{\begin{array}{*{20}c}
 0 \hfill \\
 1 \hfill \\
\end{array} }} \right)=\frac{i}{\sqrt 2 }\left| 0 \right\rangle 
-\frac{i}{\sqrt 2 }\left| 1 \right\rangle. 
\end{align*}

\begin{figure}
\centerline{\includegraphics[scale=0.25]{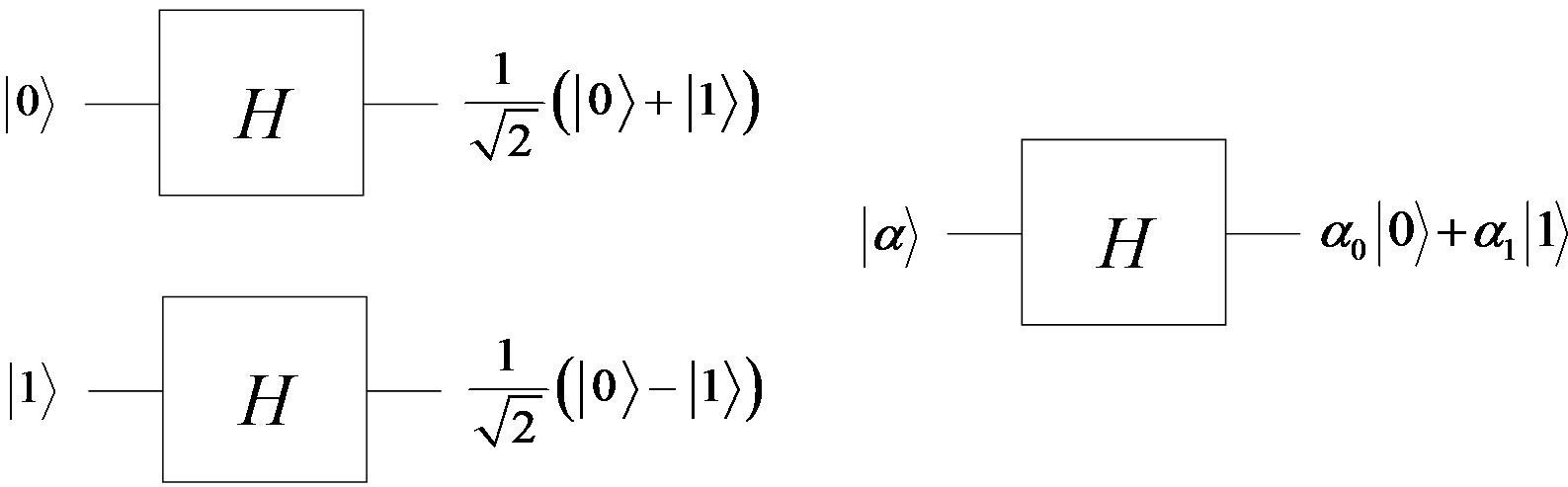}}
\caption{\small{The Hadamard gate $H$ that puts a basis state into an equal superposition of the basis states, and an arbitrary state $\left|\alpha\right\rangle$ into the superposition $\alpha_0\left|0\right\rangle+\alpha_1\left|1\right\rangle$.}}
\label{Hadamard}
\end{figure}


\subsection{Measurement}
In Equation (\ref{eqA}), if both $\alpha_0, \alpha_1 \neq 0$, then the how does one interpret the complex projective linear combination $-\overline{\eta}\alpha_1\left| 0 \right\rangle +\eta \alpha_0\left| 1 \right\rangle$ of the basis states in the context of computing? The answer comes from quantum mechanics' axiom of measurement which allows a probabilistic interpretation of such complex projective linear combinations as follows. Upon measurement with respect to the orthogonal basis $\left\{\left| 0 \right\rangle, \left| 1 \right\rangle\right\}$, the combination is observed to be in the basis state  $\left| 0 \right\rangle$ with probability $\left| \alpha_1\right|^2$ and in the basis state $\left| 1 \right\rangle$ with probability $\left| \alpha_1\right|^2$ (remember that $\eta$ is a unit complex number so $\left|\eta\right|^2=\left|\overline{\eta}\right|^2=1$).
Two important measurement operators are 
\[
M_0 =
\left( {{\begin{array}{*{20}c}
 1 \hfill & 0 \hfill \\
 0 \hfill & 0 \hfill \\
\end{array} }} \right), 
M_1 =
\left( {{\begin{array}{*{20}c}
 0 \hfill & 0 \hfill \\
 0 \hfill & 1 \hfill \\
\end{array} }} \right)
\]
The measurement operator $M_0$ projects a complex projective linear combination onto the basis state $\left| 0 \right\rangle$ while $M_1$ projects onto the basis state $\left| 1 \right\rangle$. For example, let 
$$
\left| \psi \right\rangle =\alpha_0\left| 0 \right\rangle +\alpha_1\left| 1 
\right\rangle
$$
Then the probability of measuring the complex projective linear combination $\left| \psi \right\rangle$ in the basis state $\left| 0 \right\rangle$ is
$$
p(\left| 0 \right\rangle)=
\left( {{\begin{array}{*{20}c}
 {\overline{a} } \hfill & {\overline{b} } \hfill \\
\end{array} }} \right)\left( {{\begin{array}{*{20}c}
 1 \hfill & 0 \hfill \\
 0 \hfill & 0 \hfill \\
\end{array} }} \right)\left( {{\begin{array}{*{20}c}
 a \hfill \\
 b \hfill \\
\end{array} }} \right)\\
$$
$$
=\left( {{\begin{array}{*{20}c}
 {\overline{a} } \hfill & {\overline{b} } \hfill \\
\end{array} }} \right)\left( {{\begin{array}{*{20}c}
 a \hfill \\
 0 \hfill \\
\end{array} }} \right)=\left| a \right|^2.
$$
Note that measurement operators are {\it not} quantum logic gates as they are non-unitary, but rather are projections onto the basis states. 

\section{Quantum Computing with Multiple Qubits}

Quantum computing can be extended to multiple qubits via the creation of composite state spaces from the state spaces of many individual qubits. 

For example, consider two qubits $\left| {\psi _1 } \right\rangle =a\left| 0 \right\rangle +b\left| 1 
\right\rangle \mbox{ and }\left| {\psi _2 } \right\rangle =c\left| 0 
\right\rangle +d\left| 1 \right\rangle $, both written with respect to the computational basis. Then the {\it joint state} of the total 
system is given by:
\[
\begin{array}{l}
 \left| {\psi _1 } \right\rangle \otimes \left| {\psi _2 } \right\rangle 
=\left| {\psi _1 \psi _2 } \right\rangle =ac\left| 0 \right\rangle \otimes 
\left| 0 \right\rangle +ad\left| 0 \right\rangle \otimes \left| 1 
\right\rangle +bc\left| 1 \right\rangle \otimes \left| 0 \right\rangle 
+bd\left| 1 \right\rangle \otimes \left| 1 \right\rangle \\ 
 \mbox{ } \\ 
 \mbox{ }=ac\left( {{\begin{array}{*{20}c}
 1 \hfill \\
 0 \hfill \\
\end{array} }} \right)\otimes \left( {{\begin{array}{*{20}c}
 1 \hfill \\
 0 \hfill \\
\end{array} }} \right)+ad\left( {{\begin{array}{*{20}c}
 1 \hfill \\
 0 \hfill \\
\end{array} }} \right)\otimes \left({{\begin{array}{*{20}c}
 0 \hfill \\
 1 \hfill \\
\end{array} }} \right)+bc\left( {{\begin{array}{*{20}c}
 0 \hfill \\
 1 \hfill \\
\end{array} }} \right)\otimes \left( {{\begin{array}{*{20}c}
 1 \hfill \\
 0 \hfill \\
\end{array} }} \right)\\
\quad +bd\left( {{\begin{array}{*{20}c}
 0 \hfill \\
 1 \hfill \\
\end{array} }} \right)\otimes \left( {{\begin{array}{*{20}c}
 0 \hfill \\
 1 \hfill \\
\end{array} }} \right) \\ 
 \mbox{ } \\ 
 \mbox{ }=ac\left( {{\begin{array}{*{20}c}
 1 \hfill \\
 0 \hfill \\
 0 \hfill \\
 0 \hfill \\
\end{array} }} \right)+ad\left( {{\begin{array}{*{20}c}
 0 \hfill \\
 1 \hfill \\
 0 \hfill \\
 0 \hfill \\
\end{array} }} \right)+bc\left( {{\begin{array}{*{20}c}
 0 \hfill \\
 0 \hfill \\
 1 \hfill \\
 0 \hfill \\
\end{array} }} \right)+bd\left( {{\begin{array}{*{20}c}
 0 \hfill \\
 0 \hfill \\
 0 \hfill \\
 1 \hfill \\
\end{array} }} \right) \\ 
 \\ 
 \mbox{ }=ac\left| {00} \right\rangle +ad\left| {01} \right\rangle +bc\left| 
{10} \right\rangle +bd\left| {11} \right\rangle \\ 
 \end{array}
\]

\subsection{Two Qubit Quantum Gates}\label{two qubit qcomp}

\begin{figure}
\centerline{\includegraphics[scale=0.23]{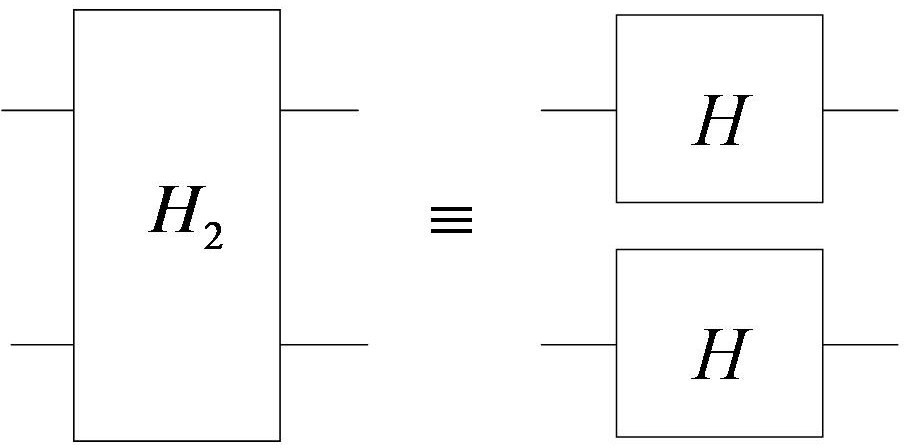}}
\caption{\small{The two qubit Hadamard gate is just the tensor product of the one qubit Hadamard gates acting on each qubit. In general, multiqubits gates can be created via the tensor product of one qubit gates. However, it is not always true that a multiqubit gate is equal to the tensor product of one qubit gates. Consider for example the CNOT gate of Figure \ref{CNOT}.}}
\label{2qubitHadamard}
\end{figure} 

An easy way to obtain two qubit quantum gates is by 
producing the tensor product of two one qubit gates. That is, if $U_1$ and $U_2$ are one qubit gates, then 
$$
U=U_1 \otimes U_2
$$
is a two qubit gate. Two qubit gates such as $U$ above that are tensor products of one qubit gates act locally on each qubit due to the bi-linearity of the tensor product. Nonetheless, such gates are crucial to quantum computing. For example, the two qubit Hadamard gate defined as
$$
H_2=H \otimes H=\frac{i}{\sqrt 2}\left( {\begin{array}{*{20}c}
 1  & 1  \\
 1  & -1 \\
\end{array} } \right) \otimes \frac{i}{\sqrt 2}\left( {\begin{array}{*{20}c}
 1  & 1  \\
 1  & -1 \\
\end{array} } \right) 
$$

$$
=-\frac{1}{2}\left( {\begin{array}{*{20}c}
 1\left( \begin{array}{*{20}c}
 1  & 1  \\
 1  & -1 \\
\end{array}  \right)  & 1 \left( {\begin{array}{*{20}c}
 1  & 1  \\
 1  & -1 \\
\end{array} } \right)  \\
 1\left( {\begin{array}{*{20}c}
 1  & 1  \\
 1  & -1 \\
\end{array} } \right)  & -1\left( {\begin{array}{cc}
 1  & 1  \\
 1  & -1 \\
\end{array} } \right) 
\end{array} } \right)=
-\frac{1}{2}\left( \begin{array}{cccc}
 1  &  1 &  1  & 1  \\
 1  & -1 &  1  & -1  \\
 1  &  1 & -1  & -1  \\
 1  & -1 & -1  & 1  \\
\end{array}  \right)
$$
is essential for the creation of a particular equal superposition of two qubits which plays a crucial role in the development of quantum algorithms that out-perform classical algorithms \cite{Grover,Shor}.

%
\subsection{Controlled NOT (CNOT) gate}\label{CNOT gate}

Perhaps the most important two qubit gate is the controlled NOT (CNOT) gate. Its importance lies in its property of forming, together with one qubit gates, sets of universal quantum logic gates. Informally, a set of quantum logic gates is {\it universal} if any quantum logic gate may be approximated by the gates in the set to arbitrary accuracy. For a detailed discussion of universality, the reader is refer

The CNOT gate acts as a NOT gate on the second qubit (target qubit) if the first 
qubit (control qubit) is in the computational basis state $\left| 1 
\right\rangle $. So when passing through the gate the states $\left| {00} \right\rangle$ and $\left| {01} \right\rangle$ are unaltered, while the state $\left| {10} 
\right\rangle $is sent to $\left| {11} \right\rangle $ and vice versa. In the joint computational basis, the 
CNOT gate is 
$$
CNOT=\left( {{\begin{array}{*{20}c}
 1 \hfill & 0 \hfill & 0 \hfill & 0 \hfill \\
 0 \hfill & 1 \hfill & 0 \hfill & 0 \hfill \\
 0 \hfill & 0 \hfill & 0 \hfill & 1 \hfill \\
 0 \hfill & 0 \hfill & 1 \hfill & 0 \hfill \\
\end{array} }} \right)
$$
Note that the CNOT gate is {\it not} the tensor product of any pair of one qubit gates. Indeed, there are plenty of other two and multiqubit gates that are not tensor products of one qubit gates. This property of quantum logic gates is one more reason that quantum logic circuit synthesis is a much studied subject. 

\begin{figure}
\centerline{\includegraphics[scale=0.22]{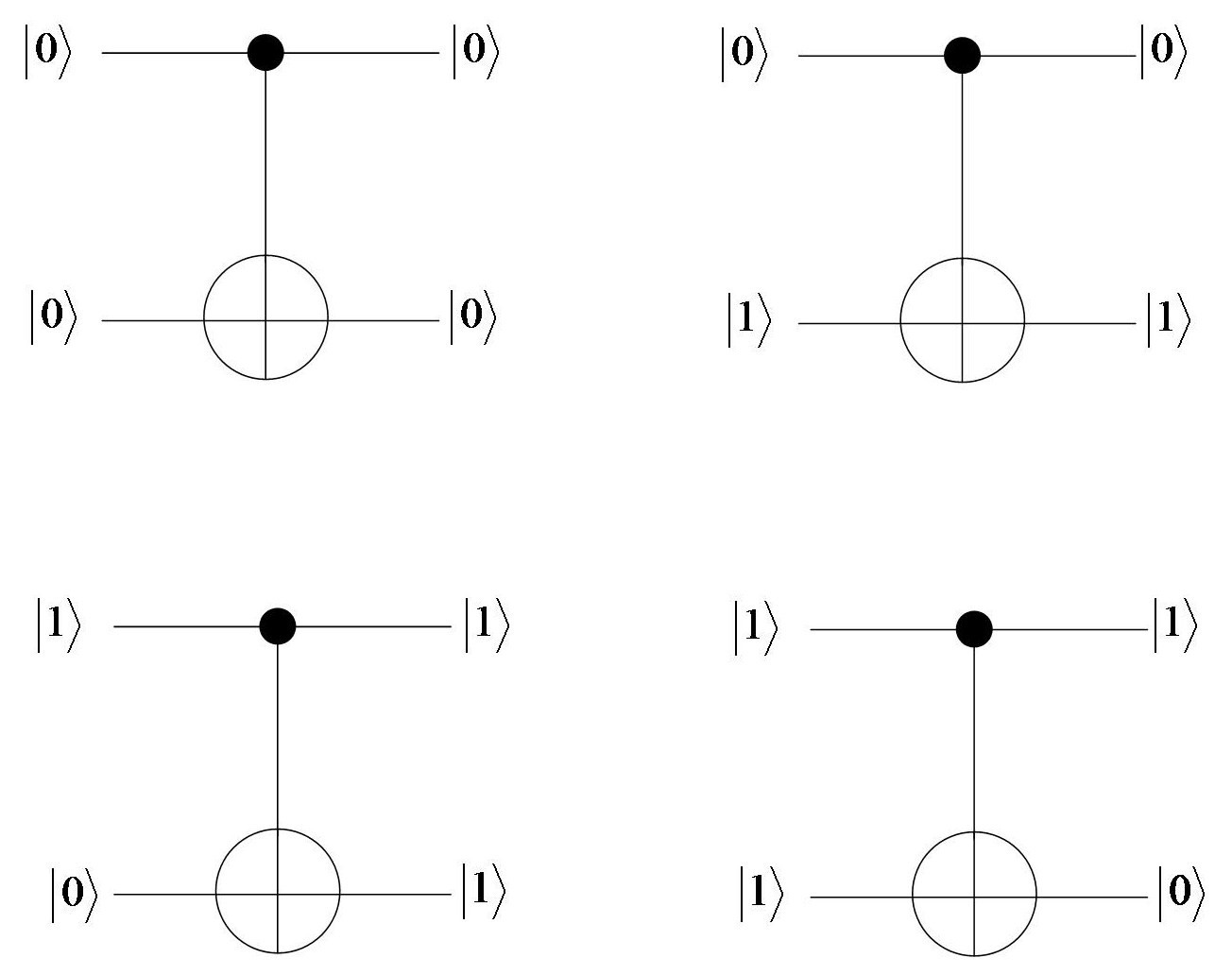}}
\caption{\small{The controlled NOT (CNOT) gate. The vectors $\left| {00} \right\rangle$ and 
$\left| {01} \right\rangle$ are unaltered, while the vector $\left| {10} 
\right\rangle$ is sent to $\left| {11} \right\rangle$ and vice versa.}}
\label{CNOT}
\end{figure} 

\subsection{Entanglement}\label{entanglemt}

Entanglement is a uniquely quantum phenomenon. Entanglement is a property of 
a multi-qubit system and can be thought of as a resource. 
To explain entanglement, let us examine a so-called {\it EPR 
pair} of qubits named after Einstein, Podolsky, and Rosen. The CNOT gate will be 
used in this example.

\begin{figure}
\centerline{\includegraphics[scale=0.25]{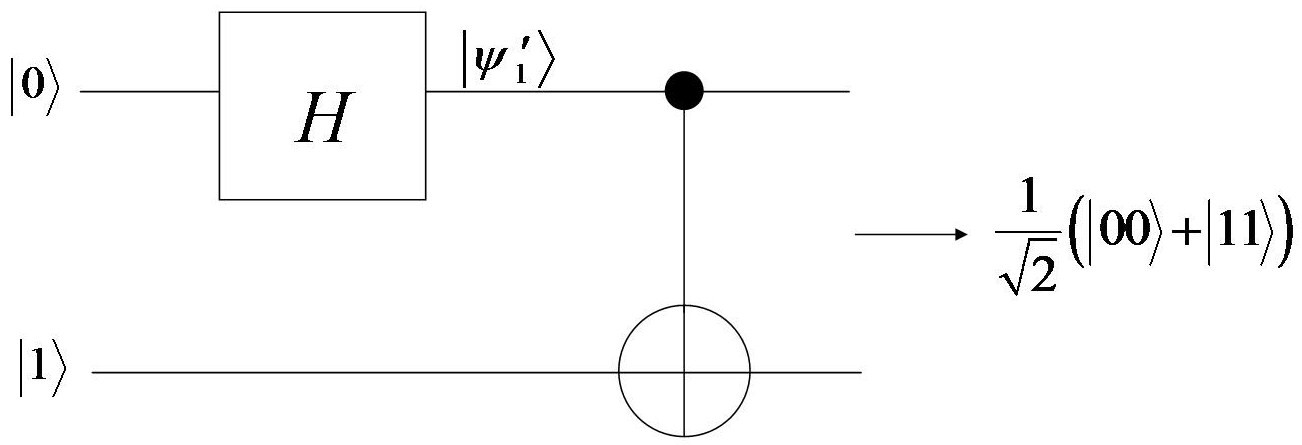}}
\caption{\small{The Hadamard gate $H$ that puts a basis state into an equal superposition of the basis states, and an arbitrary state $\left|\alpha\right\rangle$ into the superposition $\alpha_0\left|0\right\rangle+\alpha_1\left|1\right\rangle$.}}
\label{Entangling}
\end{figure}

We begin with two qubits $\left|  \psi_1 \right\rangle=\left| 0 \right\rangle$ and $\left| \psi_2 \right\rangle=\left|1 \right\rangle$. Apply the Hadamard gate to $\left|\psi_1 \right\rangle$ to get
\[
\left| {{\psi'}_1 } \right\rangle =H\left| {\psi _1 } \right\rangle 
=\frac{i}{\sqrt 2 }\left| 0 \right\rangle +\frac{i}{\sqrt 2 }\left| 1 
\right\rangle 
\]
The joint state-space vector is the tensor product
\[
\left| {{\psi }'_1 } \right\rangle \otimes \left| {\psi _2 } \right\rangle 
=\left| {{\psi }'_1 \psi _2 } \right\rangle =\frac{i}{\sqrt 2 }\left| {00} 
\right\rangle +(0)\left| {01} \right\rangle +\frac{i}{\sqrt 2 }\left| {10} 
\right\rangle +(0)\left| {11} \right\rangle 
\]
Now apply the CNOT gate to this joint state of the two qubits. This gives
\begin{align*}
\left( 
{{\begin{array}{*{20}c}
 1 \hfill & 0 \hfill & 0 \hfill & 0 \hfill \\
 0 \hfill & 1 \hfill & 0 \hfill & 0 \hfill \\
 0 \hfill & 0 \hfill & 0 \hfill & 1 \hfill \\
 0 \hfill & 0 \hfill & 1 \hfill & 0 \hfill \\
\end{array} }} \right) \left( {{\begin{array}{*{20}c}
 {\frac{i}{\sqrt 2 }} \hfill \\
 0 \hfill \\
 {\frac{i}{\sqrt 2 }} \hfill \\
 0 \hfill \\
\end{array} }} \right)=\left( {{\begin{array}{*{20}c}
 {\frac{i}{\sqrt 2 }} \hfill \\
 0 \hfill \\
 0 \hfill \\
 {\frac{i}{\sqrt 2 }} \hfill \\
\end{array} }} \right)&=\frac{i}{\sqrt 2 }\left( {{\begin{array}{*{20}c}
 1 \hfill \\
 0 \hfill \\
 0 \hfill \\
 0 \hfill \\
\end{array} }} \right)+\frac{i}{\sqrt 2 }\left( {{\begin{array}{*{20}c}
 0 \hfill \\
 0 \hfill \\
 0 \hfill \\
 1 \hfill \\
\end{array} }} \right)\\
&= \frac{i}{\sqrt 2 }\left| {00} \right\rangle 
+\frac{i}{\sqrt 2 }\left| {11} \right\rangle 
\end{align*}
The final joint state above has the property that it cannot be 
built up from the tensor product of states in the component spaces of each qubit. That is, 
\[
\frac{i}{\sqrt 2 }\left| {00} \right\rangle +\frac{i}{\sqrt 2 }\left| {11} 
\right\rangle \neq \mbox{ }\left| {\phi _1 } \right\rangle \otimes \left| {\phi 
_2 } \right\rangle .
\].
 
To illustrate why entanglement is so strange, let's consider performing a 
measurement just prior to applying the CNOT gate. The two measurement operators 
(for obtaining a $\left| {00} \right\rangle $or a$\left| {11} \right\rangle 
)$ are:
\[
M_{00 } =\left( {{\begin{array}{*{20}c}
 1 \hfill & 0 \hfill & 0 \hfill & 0 \hfill \\
 0 \hfill & 0 \hfill & 0 \hfill & 0 \hfill \\
 0 \hfill & 0 \hfill & 1 \hfill & 0 \hfill \\
 0 \hfill & 0 \hfill & 0 \hfill & 0 \hfill \\
\end{array} }} \right)\mbox{ and }M_{11 } =\left( {{\begin{array}{*{20}c}
 0 \hfill & 0 \hfill & 0 \hfill & 0 \hfill \\
 0 \hfill & 1 \hfill & 0 \hfill & 0 \hfill \\
 0 \hfill & 0 \hfill & 0 \hfill & 0 \hfill \\
 0 \hfill & 0 \hfill & 0 \hfill & 1 \hfill \\
\end{array} }} \right)
\]
Just prior to the CNOT the system is in the state 
$$
\frac{i}{\sqrt 2 }\left| {00} 
\right\rangle +0\left| {01} \right\rangle +\frac{i}{\sqrt 2 }\left| {10} 
\right\rangle +0\left| {11} \right\rangle,
$$
therefore
\[
p(0)
=\left(
{{\begin{array}{*{20}c}
 {\frac{\overline{i}}{\sqrt 2 }} \hfill & 0 \hfill & {\frac{\overline{i}}{\sqrt 2 }} \hfill & 0 
\hfill \\
\end{array} }} \right)\left({{\begin{array}{*{20}c}
 1 \hfill & 0 \hfill & 0 \hfill & 0 \hfill \\
 0 \hfill & 0 \hfill & 0 \hfill & 0 \hfill \\
 0 \hfill & 0 \hfill & 1 \hfill & 0 \hfill \\
 0 \hfill & 0 \hfill & 0 \hfill & 0 \hfill \\
\end{array} }} \right)\mbox{ }\left( {{\begin{array}{*{20}c}
 {\frac{i}{\sqrt 2 }} \hfill \\
 0 \hfill \\
 {\frac{i}{\sqrt 2 }} \hfill \\
 0 \hfill \\
\end{array} }} \right)=1
\]
Hence the result of measuring will clearly be $\left| 0 \right\rangle $. 
After the measurement, we have
\[
\left| {\psi _1 ^\prime \psi _2 } \right\rangle 
=\frac{\left( 
{{\begin{array}{*{20}c}
 {\frac{i}{\sqrt 2 }} \hfill \\
 0 \hfill \\
 {\frac{i}{\sqrt 2 }} \hfill \\
 0 \hfill \\
\end{array} }} \right)}{1}
\]
and we see that measurement had no effect on the first qubit and it remains in 
a superposition of $\left| 0 \right\rangle \mbox{ and }\left| 1 \right\rangle 
$. Now consider the same measurement but just after the CNOT gate is 
applied, with the joint state $\left| \psi _3
\right\rangle =\frac{i}{\sqrt 2 }\left| {00} \right\rangle +\frac{i}{\sqrt 2 
}\left| {11} \right\rangle $.
\[
p(0)=
=\left({{\begin{array}{*{20}c}
 {\frac{\overline{i}}{\sqrt 2 }} \hfill & 0 \hfill & 0 \hfill & {\frac{\overline{i}}{\sqrt 2 }} 
\hfill \\
\end{array} }} \right)\left( {{\begin{array}{*{20}c}
 1 \hfill & 0 \hfill & 0 \hfill & 0 \hfill \\
 0 \hfill & 0 \hfill & 0 \hfill & 0 \hfill \\
 0 \hfill & 0 \hfill & 1 \hfill & 0 \hfill \\
 0 \hfill & 0 \hfill & 0 \hfill & 0 \hfill \\
\end{array} }} \right)\mbox{ }\left( {{\begin{array}{*{20}c}
 {\frac{i}{\sqrt 2 }} \hfill \\
 0 \hfill \\
 0 \hfill \\
 {\frac{i}{\sqrt 2 }} \hfill \\
\end{array} }} \right)=\frac{1}{2}
\]
Hence, after the \textit{CNOT }gate is applied we have only a 50{\%} chance of 
obtaining $\left| 0 \right\rangle $. Of particular interest to our 
discussion, however, is what happens to the state vector of the system after 
measurement.
\[
\frac{\left( 
{{\begin{array}{*{20}c}
 {\frac{i}{\sqrt 2 }} \hfill \\
 0 \hfill \\
 0 \hfill \\
 0 \hfill \\
\end{array} }} \right)}{\sqrt {\frac{1}{2}} }=i\left( {{\begin{array}{*{20}c}
 1 \hfill \\
 0 \hfill \\
 0 \hfill \\
 0 \hfill \\
\end{array} }} \right) =i\left(
{{\begin{array}{*{20}c}
 1 \hfill \\
 0 \hfill \\
\end{array} }} \right)\otimes \left( {{\begin{array}{*{20}c}
 1 \hfill \\
 0 \hfill \\
\end{array} }} \right)=i\left| {00} \right\rangle
\]
This is the remarkable thing about entanglement. By measuring one qubit we 
can affect the probability of the state observations of the other qubits in a system! The state of the other
qubit $\left| {{\psi }'_1 } \right\rangle =\frac{1}{\sqrt 2 }\left| 0 
\right\rangle +\frac{1}{\sqrt 2 }\left| 1 \right\rangle $ is changed to 
$\left| 0 \right\rangle$ after 
the measurement. 

Quoting Oskin \cite{Oskin} regarding entanglement: 
\begin{quote}
``How to think about this process (entanglement) in an abstract way is an 
open challenge in quantum computing. The difficulty is the lack of any 
classical analog. One useful, but imprecise way to think about entanglement, 
superposition and measurement is that superposition ``is'' quantum 
information. Entanglement links that information across quantum bits, but 
does not create any more of it. Measurement ``destroys'' quantum information 
turning it into classical. Thus think of an EPR pair as having as much 
``superposition'' as an un-entangled set of qubits, one in a superposition 
between zero and one, and another in a pure state. The superposition in the 
EPR pair is simply linked across qubits instead of being isolated in one.''
\end{quote}

\chapter{A Formal Approach to Quantum Games}\label{Bleilerformalism}

One way to view a game is as a function. We view here quantum games as extensions of such functions. For a detailed and formal introduction to game theory the reader is referred to \cite{binmore} and \cite{Myerson}. The following discussion on quantum games that follows is motivated by a mathematical formalism for ``quantum mixtures'' developed by S. Bleiler in \cite{Bleiler} and reproduced in section \ref{Bleiler formalism} below.  

Recall that a key goal in the study of multi-player, non-cooperative  games is the identification of potential Nash equilibria. Informally, a Nash equilibrium occurs when each player chooses to play a strategy that is a best reply to the choice of strategies of all the other players. In other words, unilateral deviation from the choice of strategy at a Nash equilibrium by any player is detrimental to that player's payoff in the game. However, in finite classical games, Nash equilibria may not exist. In such situations, classical game theory calls upon the players to randomize between their strategic choices, also known as mixing strategies. For finite games, Nash proved \cite{Nash} that this gives rise to Nash equilibria in the ``mixed game'' that simply do not exist in the original game. Formally, the mixed game is the result of an extension of the payoff function of the original game to a larger set of strategies for each player. 

The Bleiler formalism for quantum mixtures views quantum game theory in this light. That is, this formalism views quantum game theory as an exercise in the extension of the payoff function of a game with the goal of finding Nash equilibria with higher payoffs that were un-attainable in the original game or its ``classical extensions''. The extensions dealt with in quantum game theory are referred to as a {\it quantization protocols}. This mathematically formal perspective provides a game theoretic context in which many issues in quantum game theory can be discussed and potentially resolved. For example, critics of quantum game theory wonder whether instances of Nash equilibira with higher payoffs in certain quantum games are just Nash equilibria of some other classical game theoretic construction realized quantum mechanically. This point of view implies that quantum game theory is essentially a study in expensive ways to generate classical game theoretic results and offers nothing ``new'' to game theory. 

Such criticism is addressed in the Bleiler formalism which points out that any quantum game that contains the original or the classical game as an embedded subgame has the potential to offer something new to the game's analysis. When a quantum game has this property, it is referred to in the Bleiler formalism as a {\it proper quantization} of the original game. When a quantum game carries an embedded copy of the mixed version of the original game, the formalism refers to it as a {\it complete quantization} of the original game. Much of the current work in quantum game theory can be characterized as calling upon the players to use the higher orders of randomization given by quantum superpositions and randomized quantum superpositions. Call these {\it quantum strategies} and {\it mixed quantum strategies}, respectively. If the quantization of the game is proper or complete, then any new Nash equilibria with higher payoffs that result from the use of quantum or mixed quantum strategies can be meaningfully compared with the Nash equilibria of the original game. 

A detailed review of the Bleiler formalism follows.

\section{The Bleiler Formalism for Quantum Mixtures}\label{Bleiler formalism}

\begin{definition} {\rm
Given a set $\{ 1, 2, \cdots, n \}$ of players, for each player a set $S_i$ $(i=1, \cdots, n)$ of so-called \emph{pure strategies}, and a set $\Omega_i$ $(i=1, \cdots, n)$ of \emph{possible outcomes}, a \emph{game} $G$ is a vector-valued function whose domain is the Cartesian product of the $S_i$'s and whose range is the Cartesian product of the $\Omega_i$'s. In symbols 
$$
G: \prod_{i=1}^n S_i \longrightarrow \prod_{i=1}^n \Omega_i                      
$$
The function $G$ is sometimes referred to as the \emph{payoff function}.}
\end{definition}

Here a \emph{play} of the game is a choice by each player of a particular strategy $s_i$ the collection of which forms a \emph{strategy profile} $(s_1, \cdots, s_n)$ whose corresponding \emph{outcome profile} is $G(s_1, \cdots, s_n)=(\omega_1, \cdots, \omega_n)$, where the $\omega_i$'s represent each player's individual outcome. Note that by assigning a real valued \emph{utility} to each player which quantifies that player's preferences over the various outcomes, we can without loss of generality, assume that the $\Omega_i$'s are all copies of $\mathbb{R}$, the field of real numbers.

In game theory, players' concern is the identification of a strategy that guarantees a maximal utility. 
For a fixed $(n-1)$-tuple of opponents' strategies, rational players seek a \emph{best reply}, that is a strategy $s^*$ that delivers a utility at least as great, if not greater, than any other strategy $s$. When every player can identify such a strategy, the resulting strategy profile is called a {\it Nash equilibrium}. Formally,

\begin{definition}{\rm 
Let $s_{-i}$ be a strategy profile of all players except player $i$. A \emph{Nash equilibrium} (NE) for the game $G$ is a strategy profile $(s_i^*,s_{-i})$ such that 
$$
G(s_i^*,s_{-i}) \geq G(s_i,s_{-i})
$$
where for all $i$, $s_i, s^*_i \in S_i$ and $s_i^* \neq s_i$.}
\end{definition}

Other ways of expressing this concept include the observation that no player can increase his or her payoffs by unilaterally deviating from his or her equilibrium strategy, or that at equilibrium all of a player's opponents are indifferent to that player's strategic choice. As an example, consider the Prisoner's Dilemma, a two player game where each player has exactly two strategies (a so-called $2 \times 2$ or \emph{bimatrix} game) and whose payoff function is indicated in Table \ref{tab:PrisonerSDilemma}. The rows of Table \ref{tab:PrisonerSDilemma} contain the strategies of player 1 while the columns contain the strategies of player 2. 

Note that for player 1 the pure strategy $s_2$ always delivers a higher outcome than the strategy $s_1$ (say $s_2$ \emph{strongly dominates} $s_1$) and for player 2 the strategy $t_2$ strongly dominates $t_1$. Hence the pair $(s_2, t_2)$ is a (unique) Nash Equilibrium. 

However, games need not have equilibria amongst the pure strategy profiles as exemplified by the $2\times2$ game of Simplified Poker whose payoff function is given in Table \ref{tab:SP}.

\begin{table}[h]
	\centering
	  \begin{tabular}{r|r|r|}
	    &$t_1$&$t_2$\\
		  \hline
      $s_1$&$(3,3)$&$(0,5)$\\
      \hline
			$s_2$&$(5,0)$&$(1,1)$\\
			\hline
	  \end{tabular}
	\caption{Prisoner's Dilemma}
	\label{tab:PrisonerSDilemma}
\end{table}

\begin{table}
	\centering
	  \begin{tabular}{|r|r|r|}
	   	 \hline
	    &$t_1$&$t_2$\\
		  \hline
      $s_1$&$(5/4,-5/4)$&$(0,0)$\\
      \hline
			$s_2$&$(0,0)$&$(5/2,-5/2)$\\
			\hline
	  \end{tabular}
	\caption{Simplified Poker.}
	\label{tab:SP}
\end{table}

As remarked above, the game theoretic formalism now calls upon the theorist to extend the game $G$ by enlarging the domain and extending the payoff function. Of course, the question of if and how a given function extends is a time honored problem in mathematics and the careful application of the mathematics of extension is what will drive the formalism for quantization. Returning to classical game theory, a standard extension at this point is to consider for each player the set of mixed strategies.

\begin{definition}{\rm 
A \emph{mixed strategy} for player $i$ is an element of the set of probability distributions over the set of pure strategies $S_i$.}
\end{definition}

For a given set $X$, denote the probability distributions over $X$ by $\Delta(X)$ and note that when $X$ is finite, with $k$ elements say, the set $\Delta(X)$ is just the $k-1$ dimensional simplex $\Delta^{(k-1)}$ over $X$, i.e., the set of real convex linear combinations of elements of $X$. Of course, we can embed $X$ into $\Delta(X)$ by considering the element $x$ as mapped to the probability distribution which assigns 1 to $x$ and 0 to everything else. For a given game $G$, denote this embedding of $S_i$ into $\Delta(S_i)$ by $e_i$.

Let $p=\left(p_1, \dots, p_n\right)$ be a mixed strategy profile. Then $p$ induces the \emph{product distribution} over the product $\prod S_i$. Taking the push out by $G$ of the product distribution (i.e., given a probability distribution over strategy profiles, replace the profiles with their images under $G$) then gives a probability distribution over the image of $G$, ${\rm Im}G$. Following this by the expectation operator $E$, we obtain the {\it expected outcome of $p$}. Now our game $G$ can be extended to a new, larger game $G^{mix}$.

\begin{definition}{\rm 
 Assigning the expected outcome to each mixed strategy profile we obtain the extended game 
$$
G^{mix}: \prod \Delta(S_i) \rightarrow \prod \Omega_i
$$}
\end{definition}

\begin{figure}
\centerline{\includegraphics[scale=0.22]{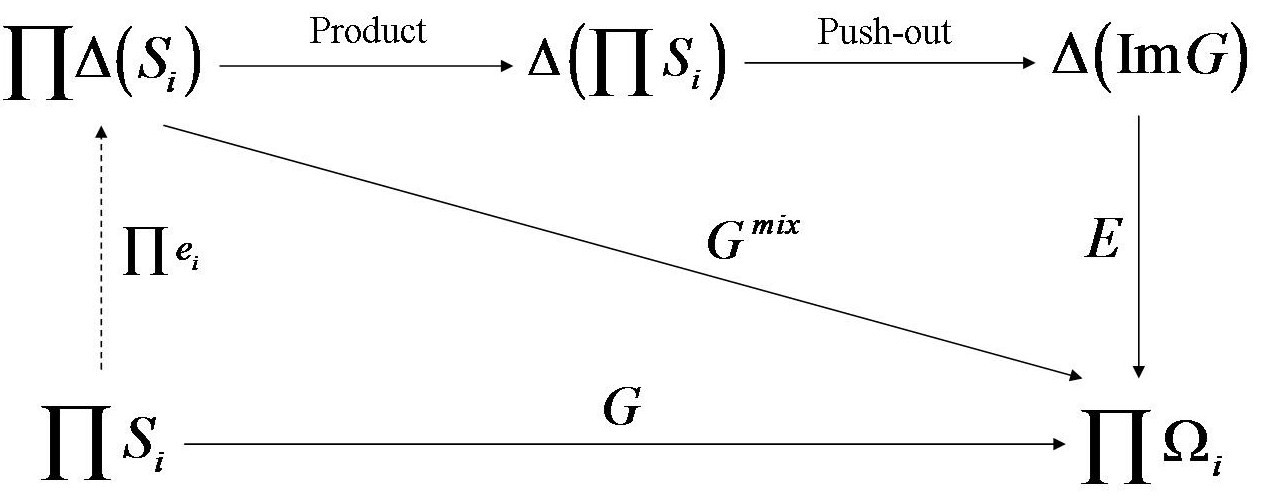}}
\caption{\small{Extension of the game $G$ to $G^{mix}$.}}
\label{Gmix}
\end{figure}

Note $G^{mix}$ is a true extension of $G$ as $G^{mix} \circ \Pi e_i = G$; that is, the diagram in Figure \ref{Gmix} is commutative.

As remarked above, Nash's famous theorem  \cite{Nash} says that if the $S_i$ are all finite, then there always exists an equilibrium in $G^{mix}$. Unfortunately, this equilibrium is called a \emph{mixed strategy equilibrium for $G$}, when it is not an equilibrium of $G$ at all, the abusive terminology confusing $G$ with its image, Im$G$. 
%

\subsection{Quantization}

The Bleiler formalism asserts that some of the controversies surrounding quantum game theory may be resolved if one focuses on the quantization of the {\it payoffs} of the original game $G$, and expresses the quantized version of $G$ as a (proper) extension of the original payout function in the set-theoretic sense, just as in the classical case. 

Classically, probability distributions over the outcomes of a game $G$ were constructed. Now the goal is to pass to a more general notion of randomization, that of quantum superposition. Begin then with a Hilbert space $\mathcal{H}$ that is a complex vector space equipped with an inner product. For the purpose here assume that $\mathcal{H}$ is finite dimensional, and that there exists a finite set $X$ which is in one-to-one correspondence with an orthogonal basis $\mathcal{B}$ of $\mathcal{H}$. 
%
When the context is clear as to the basis to which the set $X$ is identified, denote the set of quantum superpositions for $X$ as $\mathcal{Q}S(X)$. Of course, it is also possible to define quantum superpositions for infinite sets, but for the purpose here, one need not be so general. What follows can be easily generalized to the infinite case. 

As mentioned above, the underlying space of complex linear combinations is a Hilbert space; therefore, we can assign a length to each quantum superposition and, up to phase, always represent a given quantum superposition by another that has length~1. 

For each quantum superposition of $X$ we can obtain a probability distribution over $X$ by assigning to each component the ratio of the square of the length of its coefficient to the square of the length of the combination. This assignment is in fact functional, and is abusively referred to as measurement. Formally:

\begin{definition}{\rm 
{\it Quantum measurement with respect to $X$} is the function 
$$
q^{meas}_X: \mathcal{Q}S(X) \longrightarrow  \Delta(X)
$$
given by 
$$
\alpha x + \beta y \longmapsto \left(\frac{\left|\alpha\right|^2}{\left|\alpha\right|^2+\left|\beta\right|^2}, \frac{\left|\beta\right|^2}{\left|\alpha\right|^2+\left|\beta\right|^2}\right)
$$
}
\end{definition}
Note that geometrically, quantum measurement is defined by projecting a normalized quantum superposition onto the various elements of the normalized basis $\mathcal{B}$. Denote quantum measurement by $q^{meas}$ if the set $X$ is clear from the context.

Now given a finite $n$-player game $G$, suppose we have a collection $\mathcal{Q}_1, \dots, \mathcal{Q}_n$ of non-empty sets and a \emph{protocol}, that is, a function $\Theta:\prod \mathcal{Q}_i \rightarrow \mathcal{Q}S(\rm{Im}G)$. Quantum measurement $q_{\rm{Im}G}^{meas}$ then gives a probability distribution over $\rm{Im}G$. Just as in the mixed strategy case we can then form a new game $G^{\Theta}$ by applying the expectation operator $E$. 

\begin{definition}{\rm 
Assigning the expected outcome to each probability distribution over Im$G$ that results from quantum measurement, we obtain the quantized game
$$
G^{\Theta}: \prod \mathcal{Q}_i \rightarrow \prod \Omega_i
$$
}
\end{definition}

\begin{figure}
\centerline{\includegraphics[scale=0.22]{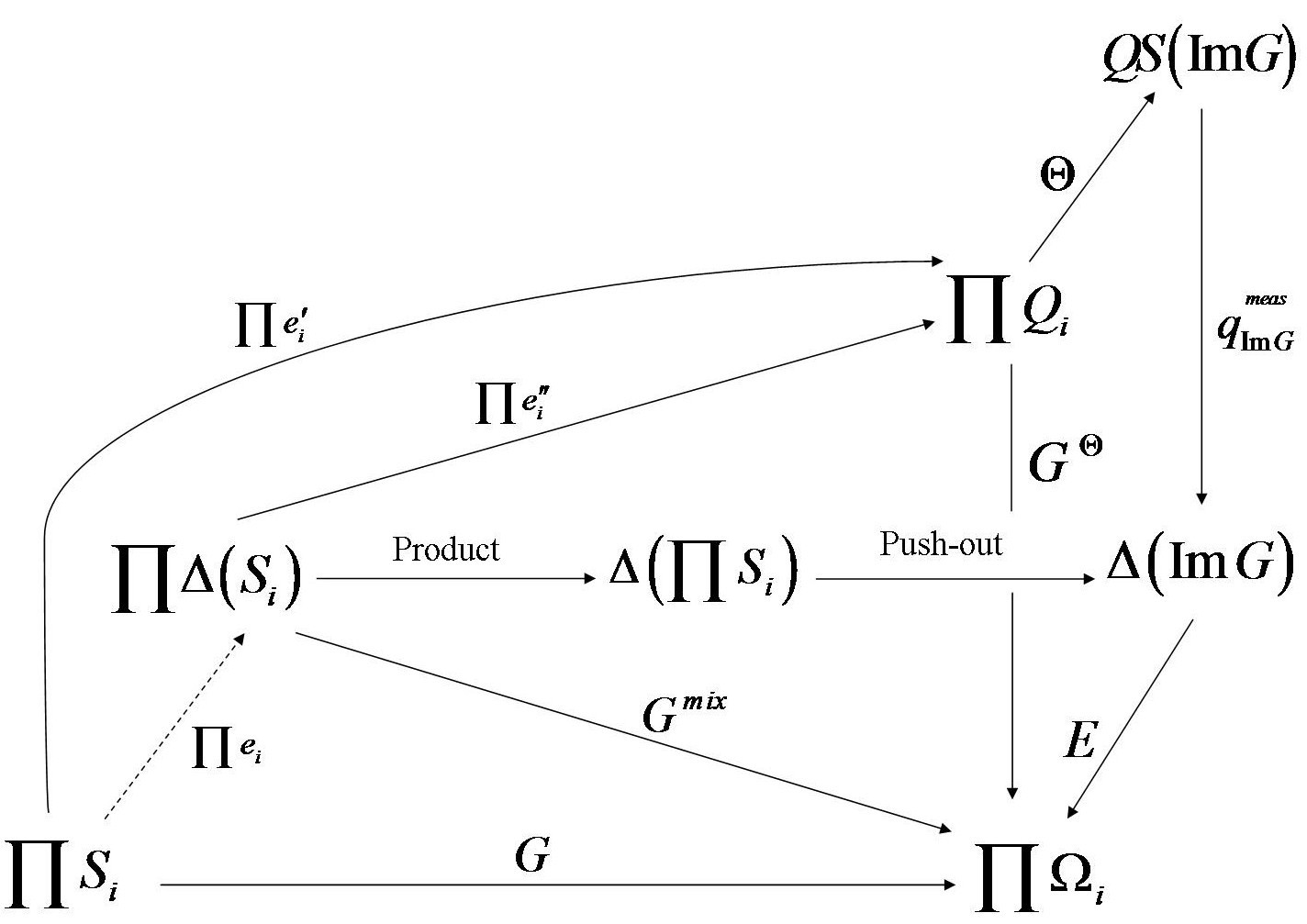}}
\caption{\small{Extension of the game $G$ to $G^{\Theta}$.}}
\label{Gquant}
\end{figure}

Call the game $G^{\Theta}$ thus defined to be the {\it quantization of $G$ by the protocol $\Theta$}. Call the $\mathcal{Q}_i$'s sets of {\it pure quantum strategies} for $G^{\Theta}$. Moreover, if there exist embeddings $e'_i:S_i \rightarrow \mathcal{Q}_i$ such that $G^{\Theta} \circ \prod e'_i=G$, call $G^{\Theta}$ a {\it proper} quantization of $G$. If there exist embeddings $e''_i:\Delta(S_i) \rightarrow \mathcal{Q}_i$ such that $G^{\Theta} \circ \prod e''_i=G^{mix}$, call $G^{\Theta}$ a {\it complete} quantization of $G$. These definitions are summed up in the commutative diagram of Figure \ref{Gquant}. Note that for proper quantizations, the original game is obtained by restricting the quantization to the image of $\prod e'_i$. For general extensions, the Game Theory literature refers to this as ``recovering'' the game $G$. 

\begin{figure}
\centerline{\includegraphics[scale=0.22]{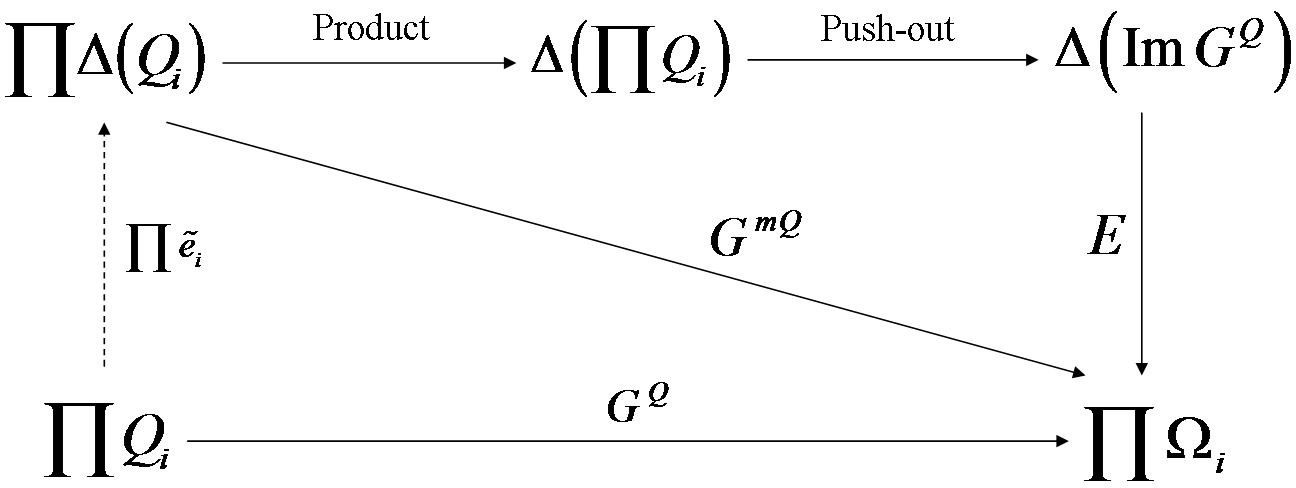}}
\caption{\small{Extension of the game $G^{\Theta}$ to $G^{m\Theta}$.}}
\label{GmQ}
\end{figure}

It follows from the definitions of $G^{mix}$ and $G^{\Theta}$ that a complete quantization is proper. Furthermore, note that finding a mathematically proper quantization of a game $G$ is now just a typical problem of extending a function. It is also worth noting here that nothing prohibits us from having a quantized game $G^{\Theta}$ play the role of $G$ in the classical situation and by considering the probability distributions over the $Q_i$, creating a yet larger game $G^{m\Theta}$, the {\it mixed quantization of G with respect to the protocol $\Theta$}. For a proper quantization of $G$, $G^{m\Theta}$ is an even larger extension of $G$. The game $G^{m\Theta}$ is described in the commutative diagram of Figure \ref{GmQ}.

In many cases, the $Q_i$ of the quantization protocols are expressed as quantum operations. These operations require a state to ``operate'' on. In this situation the definition of protocol additionally requires the definition of an ``initial state'' together with the family of quantum operations which act upon this state, along with a specific definition of how these quantum operations are to act. As exemplified in the next chapter, different choices for the initial state can give rise to very different protocols sharing a common selection and action of quantum operations. When a protocol $\Theta$ depends on a specific initial state $I$, the protocol is then denoted by $\Theta_I$. 

\begin{figure}
\centerline{\includegraphics[scale=0.22]{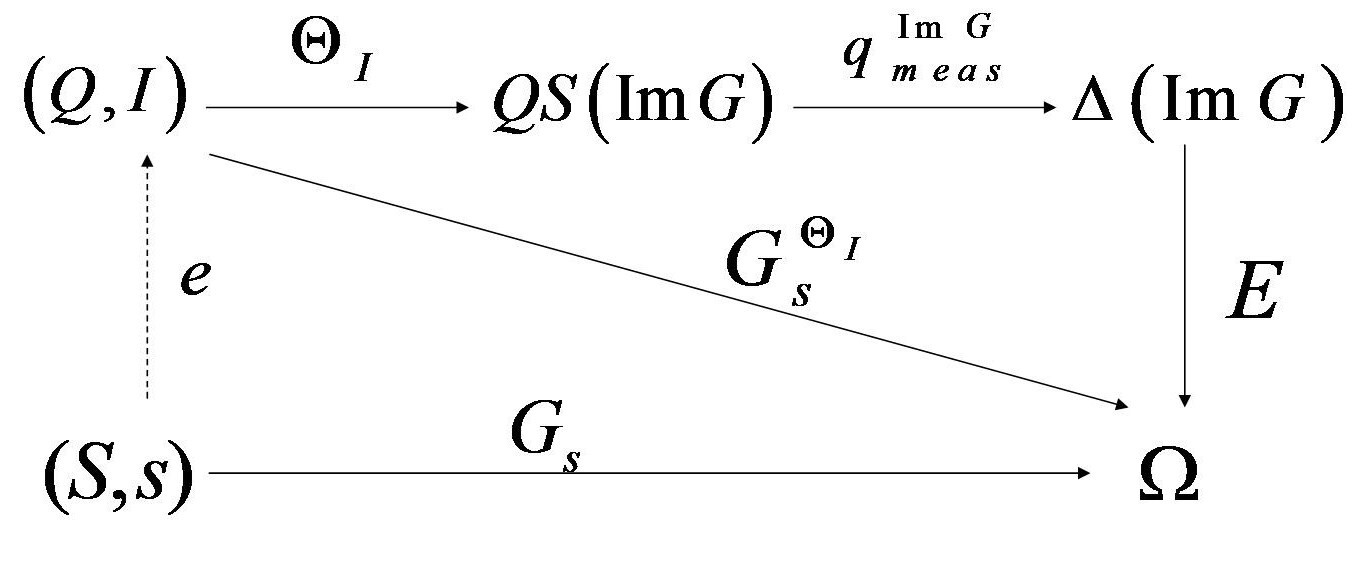}}
\caption{\small{Proper quantization of a one player game with strategy space $S$ via the protocol $\Theta$ and quantum strategy space $Q$.}}
\label{FormalParrondo}
\end{figure}

In subsequent sections, a version of this formalism adapted to one player games will be utilized. The underlying quantization paradigm being the replacement of probability distributions by the more general notion of quantum superposition followed by measurement. The functional diagram for proper quantization that will be utilized is given in Figure \ref{FormalParrondo} where the commutativity of the diagram requires that $E \circ (q^{{\rm Im}G}_{meas}) \circ \Theta \circ e =G^{\Theta} \circ e = G$. Incorporating the discussion above, when games $G_s$ and protocols $\Theta_I$ depend on a given initial states $s$ and $I$, respectively, the initial states $s$ and $I$ are regarded as part of the single player's strategic choice. In these cases, the embedding $e$ of $S$ into $Q$ additionally requires the mapping of the initial state $s$ of $G_s$ to the initial state $I$ of the protocol $\Theta_I$. The resulting quantum game is denoted by $G^{\Theta_I}_{s}$.

\chapter{Properly Quantizing History Dependent Parrondo Games}\label{Proper}

A major insight about quantized games that results from the Bleiler formalism discussed in Chapter \ref{Bleilerformalism} is that for the quantization of a game to be game-theoretically significant, it must be proper. Previous work on the quantization of the history dependent Parrondo game by Flitney, Ng, and Abbott (FNA) \cite{Flitney:02} produced quantizations that are not proper. In this chapter, after recalling the basic facts regarding Parrondo games and the FNA quantization protocols, proper quantizations for the history dependent Parrondo game and their randomized sequences are constructed.

\section{Parrondo Games}

Parrondo et. al first formulated such games in \cite{Parrondo:99}. The subject of Parrondo games has seen much research activity since then. Parrondo games typically involve the flipping of biased coins and yield only expected payoffs. A Parrondo game whose expected payoff is positive is said to be {\it winning}. If the expected payoff is negative, the game is said to be {\it losing}, and if the expected payoff is $0$, the game is said to be {\it fair}. 

Parrondo games are of interest because sequences of such games occasionally exhibit the \emph{Parrondo effect}; that is, when two or more losing games are appropriately sequenced, the resulting combined game is winning. Frequently, this sequence is {\it randomized} which means that the game played at each stage of the sequence is chosen at random with respect to a particular probability distribution over the games being sequenced. A comprehensive survey of Parrondo games and the Parrondo effect by Abbott and Harmer can be found in \cite{Harmer}.  

Earlier work on the quantization of Parrondo games can be found in \cite{Meyer1} where Meyer offers an analysis of a quantization of a particular type of Parrondo game, and in \cite{Flitney:02} where Abbott, Flitney, and Ng (AFN) propose quantizations of a different type of Parrondo game. The authors of both papers quantize their original game via their own particular quantization protocols, and further, model the game sequences as iterations of their protocols. In each of these protocols, quantum actions are performed on a collection of initial states of a quantum system. At the end, a measurement of certain specific states is made and, from the resulting probability distributions, an expected payoff computed. 

\subsection{Capital Dependent Parrondo Games}\label{CDPG}

In \cite{Parrondo:99}, Parrondo et al describe two types of coin flipping games which have the property that if individually repeated, the games result in a decreasing expected payoff to the player, yet when the two games are played in a deterministic or probabilistic sequence repeatedly, the expected payoff to the player increases over time.

Suppose that $X(t)=0, 1, 2, \dots $ is the capital available to the player. If the player wins a game, then the capital increases by one, and if the player loses, then the capital decreases by one. The simplest type of this game, referred to in the literature as game $A$, is determined by a biased coin with probability of gain $p$. That is, the capital increases by one with probability $p$ and decreases by one with probability $1-p$. Another game, called game $B$, is defined by two biased coins. The choice of which coin is to be played in an instance of the game $B$ is determined by the congruence modulo 3 of the capital, $X(t)$, available to the player in that instance. Hence, game $B$ is defined by the rules given in table \ref{table:gameB}.

\begin{table}
\caption{Game B} 
\centering 
\begin{tabular}{c c c} 
\hline\hline 
& Prob. of gain & Prob. of loss \\ [0.5ex] 
\hline 
$X(t)\equiv 0 \hspace{.05in} \mbox{mod} \hspace{.05in} 3$  &$p_{1}$ & $1-p_1$ \\ 
$X(t)\equiv 1 \hspace{.05in} {\rm or} \hspace{.05in} 2 \hspace{.05in} \mbox{mod} \hspace{.05in} 3$ & $p_{2}$ & $1-p_2$ \\[1ex] 
\hline 
\end{tabular}
\label{table:gameB} 
\end{table}

Parrondo et al set $p=\frac{1}{2}-\epsilon$, $p_1=\frac{1}{10}-\epsilon$, $p_2=\frac{3}{4}-\epsilon$, for $\epsilon >0$ as an example of games $A$ and $B$ which are losing if played individually or in a fixed sequence, but which, when combined in a randomized sequence with the uniform distribution over the two games, is winning. Both games $A$ and $B$ are losing, winning, or fair as $\epsilon >0$, $\epsilon < 0$ and $\epsilon = 0$, respectively. Parrondo et al consider in detail the case when both games $A$ and $B$ are fair. The game $B$ is analyzed as a Markov process $Y(t) \equiv X(t)$ (mod 3), that is, $Y(t)$ is equal to the remainder upon dividing the capital $X(t)$ by 3. A transition matrix for game $B$ is thus given by 
\begin{equation}\label{tran gameB}
T=\left(\begin{array}{ccc}
0 & \frac{1}{4} & \frac{3}{4} \\ \\
\frac{1}{10} & 0 & \frac{1}{4}\\ \\
\frac{9}{10} & \frac{3}{4} & 0
\end{array}\right).
\end{equation}
The stationary state for this Markov process can be computed from the matrix equation
\begin{equation}\label{matrix eq}
\left(\begin{array}{ccc}
0 & \frac{1}{4} & \frac{3}{4} \\ \\
\frac{1}{10} & 0 & \frac{1}{4}\\ \\
\frac{9}{10} & \frac{3}{4} & 0
\end{array}\right)
\left(\begin{array}{c}
\pi_0 \\ \\ \pi_1 \\ \\ \pi_2
\end{array}\right)
=
\left(\begin{array}{c}
\pi_0 \\ \\ \pi_1\\ \\ \pi_2
\end{array}\right)
\end{equation}
where $\pi_i$ is the probability of the capital $X(t)$ taking on a value congruent to $i$ (mod 3), $i=0,1,2$. The matrix Equation (\ref{matrix eq}) gives rise to the following system of equations
\begin{align}
\frac{1}{4}\pi_1	+ \frac{3}{4}\pi_2 = \pi_0\\
\frac{1}{10}\pi_0	+ \frac{1}{4}\pi_2 = \pi_1\\
\frac{9}{10}\pi_0	+ \frac{3}{4}\pi_1 = \pi_2
\end{align}
which has the following solution.
$$
\pi_0=\pi_0, \quad \pi_1=\frac{2}{5}\pi_0, \quad \pi_2=\frac{6}{5}\pi_0.
$$
Since the game is assumed to be fair, $p_1\pi_0+p_2\pi_1+p_2\pi_2=\frac{1}{2}$, and one computes $\pi_0=\frac{5}{13}$, $\pi_1=\frac{2}{13}$, $\pi_2=\frac{6}{13}$. 

Now if the fair games $A$ and $B$ are played in a randomized sequence, the resulting capital can be increasing. To see this, let $q$ be the probability with which the game $A$ is played. Then game $B$ is played with probability $(1-q)$. Again, analyze the Markov sequence $Y(t) \equiv X(t)$ (mod 3), but this time the transition matrix is 
\begin{equation}\label{matrix eq alternating generic}
T'=\left(\begin{array}{ccc}
0 & \frac{1}{2}q+\frac{1}{4}(1-q) & \frac{1}{2}q+\frac{3}{4}(1-q) \\ \\
\frac{1}{2}q+\frac{1}{10}(1-q) & 0 & \frac{1}{2}q+\frac{1}{4}(1-q)\\ \\
\frac{1}{2}q+\frac{9}{10}(1-q) & \frac{1}{2}q+\frac{3}{4}(1-q) & 0
\end{array}\right)
\end{equation}
To sequence these games via the uniform distribution, set $q=\frac{1}{2}$ and get 
\begin{equation}\label{matrix eq alternating}
T'=\left(\begin{array}{ccc}
0 & \frac{1}{4}+\frac{1}{8} & \frac{1}{4}+\frac{3}{8} \\ \\
\frac{1}{4}+\frac{1}{20} & 0 & \frac{1}{4}+\frac{1}{8}\\ \\
\frac{1}{4}+\frac{9}{20} & \frac{1}{4}+\frac{3}{8} & 0
\end{array}\right)
=
\left(\begin{array}{ccc}
0 & \frac{3}{8} & \frac{5}{8} \\ \\
\frac{3}{10} & 0 & \frac{3}{8}\\ \\
\frac{7}{10} & \frac{5}{8} & 0
\end{array}\right)
\end{equation}

Computing the stationary state $(\pi'_0, \pi'_1, \pi'_2)^T$ for the case in which each game $A$ and $B$ is fair, gives $\pi'_0=\frac{245}{709}$, $\pi'_1=\frac{180}{709}$, and $\pi'_2=\frac{284}{709}$ up to a normalization constant. Note that $\pi'_0=\frac{245}{709}$ is larger than $\frac{5}{13}$, and thus the capital increases.

\subsection{A History Dependent Parrondo Game}\label{HD game}
The history dependent Parrondo game, introduced in \cite{Parrondo:99} by Parrondo et al, is again a biased coin flipping game, where now the choice of the biased coin depends on the history of the game thus far, as opposed to the modular value of the capital. A history dependent Parrondo game $B'$ with a two stage history is reproduced in Table \ref{table:gameB HD}.

As above, let $X(t)$ be the capital available to the player at time $t$. At stage $t$, this capital goes up or down by one unit, the probability of gain determined by the biased coin used at that stage. Obtain a Markov process by setting  
\begin{equation}\label{Markov process}
Y(t)=\left(\begin{array}{c}
X(t)-X(t-1) \\ X(t-1)-X(t-2)
\end{array}\right).
\end{equation}
\begin{table}
\begin{center}
\begin{tabular}{c|c|c|c|c}
 Before last & Last& Coin & Prob. of gain & Prob. of loss \\
$t-2$ & $t-1$& ~ & at $t$ & at $t$ \\\hline
gain & gain & $B_1'$ & $p_1$ & $1-p_1$\\ gain & loss & $B_2'$ & $p_2$
& $1-p_2$ \\ loss & gain & $B_3'$ & $p_3$ & $1-p_3$ \\ loss & loss &
$B_4'$ & $p_4$ & $1-p_4$ \\
\end{tabular}
\end{center}
\caption{History dependent game $B'$.} 
\label{table:gameB HD}
\end{table}
This allows one to analyze the long term behavior of the capital in game $B'$ via the stationary state of the process $Y(t)$. The transition matrix for this process is 
\begin{equation}\label{transition matrix}
X=\left(\begin{array}{cccc}
p_1 & 0 & p_3 & 0 \\  1-p_1 & 0 & 1-p_3 & 0 \\ 0 & p_2 & 0 & p_4 \\ 0 & 1-p_2 & 0 & 1-p_4 
\end{array}\right)
\end{equation}
The stationary state can be computed from the following equations
$$
p_1\pi_1+p_3\pi_3=\pi_1
$$
$$
(1-p_1)\pi_1+(1-p_3)\pi_3=\pi_2
$$
$$
p_2\pi_2+p_4\pi_4=\pi_3
$$
$$
(1-p_2)\pi_2+(1-p_4)\pi_4=\pi_4
$$
and is given by 
\begin{equation}\label{stationary state}
s=\left(\begin{array}{c}
\pi_1 \\ \pi_2 \\\pi_3 \\\pi_4 
\end{array}\right)=\frac{1}{N}\left(\begin{array}{c}
p_3p_4  \\ p_4(1-p_1) \\ p_4(1-p_1)  \\ (1-p_1)(1-p_2)
\end{array}\right)
\end{equation}
after setting the free variable $v_4=(1-p_1)(1-p_2)$ and normalization constant
$$
N=\sqrt{\sum^4_{j=1}(\pi_j)^2}=\sqrt{(p_3p_4)^2+2\left[(1-p_1)p_4\right]^2+\left[(1-p_1)(1-p_2)\right]^2}
$$
which simplifies to 
$$ 
N=(1-p_1)(2p_4+1-p_2)+p_3p_4.
$$ 

Consequently, the probability of gain in a generic run of the game $B'$ is 
\begin{equation}\label{p of win B'}
p^{B'}_{\rm gain}=\frac{1}{N}\sum_{j=1}^{4}\pi_{j}p_{j}=\frac{p_4\left(p_3+1-p_1\right)}{\left(1-p_1\right)\left(2p_4+1-p_2\right)+p_3p_4}
\end{equation}
\noindent where $\pi_{j}$ is the probability that a certain history $j$, represented in binary format, will occur, while $p_{j}$ is the probability of gain upon the flip of the last coin corresponding to history $j$. The expression for $p^{B'}_{\rm gain}$ simplifies to  
\begin{equation}\label{p win class}
p^{B'}_{\rm gain} =1/(2+x/y)
\end{equation}
with
\begin{equation}\label{s condition}
y=p_4(p_3+1-p_1)>0
\end{equation}
for any choice of the probabilities $p_1, \dots p_4$, and
\begin{equation}\label{c condition}
x=(1-p_1)(1-p_2)-p_3p_4.
\end{equation}
Therefore, game $B'$ obeys the following
rule: if $x < 0$, $B'$ is winning, that is, has positive expected payoff; if $x= 0$, $B'$ is
fair; and if $x> 0$, $B'$ is losing, that is, has negative expected payoff. 

Before proceeding further, it is useful to view the preceding ideas in a more formal game theoretic context. For this, consider the Parrondo games as one player games in normal form, that is, as a function, where the one player's strategic choices in part correspond to the biases of the coins. For a history dependent Parrondo game with two historical stages, Parrondo et al refer to these choices as a ``choice of rules.'' However, the mere choice of biases for the coins is not enough to determine a unique normal form for these history dependent Parrondo games. In particular, an initial probability distribution over the allowable histories is also required. Although any specific distribution suffices to uniquely determine such a normal form, as the structure of the game is given by a Markov process, there is a natural choice for this initial distribution. Though this issue is not discussed by Parrondo et al, these authors immediately focus on this natural choice, namely, the distribution corresponding to the stationary state of the Markov process representing the game. 

Now, the normal form of these history dependent Parrondo games maps the tuple $\left(P, s\right)$ into the element 
$$
\left(\pi_1p_1, \pi_1(1-p_1), \pi_2p_2, \pi_2(1-p_2), \pi_3p_3, \pi_3(1-p_3), \pi_4p_4,\pi_4(1-p_4) \right)
$$ 
of the probability payoff space $[0,1]^{\times 8}$, where $s=(\pi_1,\pi_2,\pi_3,\pi_4)  \in \Delta({\rm hist}G)$ is the stationary state of the Markov process with transition matrix defined by $P=(p_1, p_2, p_3, p_4)$ $\in [0,1]^{\times 4}$, as in Equation (\ref{transition matrix}). Formally, 
\begin{equation}\label{formal payoff funct}
G_s: [0,1]^{\times 4} \times \Delta({\rm hist}G) \rightarrow [0,1]^{\times 8} 
\end{equation}
\begin{equation}\label{formal payoff funct detail}
G_s: (P,s) \mapsto \left(\pi_1p_1, \pi_1(1-p_1), \pi_2p_2, \pi_2(1-p_2), \pi_3p_3, \pi_3(1-p_3), \pi_4p_4,\pi_4(1-p_4) \right)
\end{equation}
The outcomes {\it winning}, {\it breaking even}, or {\it losing} to the player occur when $p_{\rm gain}^{B'} > \frac{1}{2}$, $p_{\rm gain}^{B'} = \frac{1}{2}$, and $p_{\rm gain}^{B'} < \frac{1}{2}$, respectively. 

Note that in this more formal game theoretic context for history dependent Parrondo games, the dependence of these games on the initial probability distribution $s$ is made clear. This initial probability distribution plays the role of the initial state $s$ for the classical game $G_s$ appearing in the proper quantization discussion at the end of chapter 2. 

\subsection{Randomized Combinations of History Dependent Parrondo Games}\label{rand A and B'}
Consider now the two stage history dependent game obtained by randomly sequencing the games $B'$ and $B''$ where each of $B'$ and $B''$ are history dependent Parrondo games with two stage histories. This can be formally considered as a real convex linear combination of the games $B'$ and $B''$, where the coefficients on $B'$ and $B''$ are given by $r$, the probability that the game $B'$ is played at a given stage, and $(1-r)$, the probability that the game $B''$ is played at a given stage. This is because the transition matrix of the Markov process associated to the randomized sequence is obtained from the transition matrices $T'$ and $T''$ for the games $B'$ and $B''$, respectively, by taking the real convex combination $rT'+(1-r)T''$. Explicitly, let 
\begin{equation}\label{T' matrix}
T'=\left(\begin{array}{cccc}
\alpha_1 & 0 & \alpha_3 & 0 \\  1-\alpha_1 & 0 & 1-\alpha_3 & 0 \\ 0 & \alpha_2 & 0 & \alpha_4 \\ 0 & 1-\alpha_2 & 0 & 1-\alpha_4 
\end{array}\right)
\end{equation} 
and
\begin{equation} \label{T'' matrix}
T''=\left(\begin{array}{cccc}
\beta_1 & 0 & \beta_3 & 0 \\  1-\beta_1 & 0 & 1-\beta_3 & 0 \\ 0 & \beta_2 & 0 & \beta_4 \\ 0 & 1-\beta_2 & 0 & 1-\beta_4 
\end{array}\right).
\end{equation} 
with $\alpha_j,\beta_j \in [0,1]$ representing the probability of gain for the $j$ coin in games $B'$ and $B''$ respectively. Then the transition matrix $rT'+(1-r)T''$ of the Markov process for the randomized sequence of $B'$ and $B''$ consists of entries $t_j=r\alpha_j+(1-r)(\beta_j)$ and $1-t_j=r(1-\alpha_j)+(1-r)(1-\beta_j)$ in the appropriate locations. Call this randomized sequence of games $B'$ and $B''$ the history dependent game $B'B''$ with probability of gain $t_j$. The stable state, computed in exactly the same fashion as the stable state for the game $B'$ in section \ref{HD game} above, has form
\begin{equation}\label{stat state mix general}
\tau=\left(\begin{array}{c}
\tau_1 \\ \tau_2 \\ \tau_3 \\ \tau_4 
\end{array}\right)=\frac{1}{R}\left(\begin{array}{c}
t_3t_4 \\ t_4(1-t_1) \\ t_4(1-t_1) \\ (1-t_1)(1-t_2)
\end{array}\right)
\end{equation}
with $R=\sum^{4}_{j=1}\tau_j$ a normalization constant. Using the stable state, the probability of gain in the game $B'B''$ is computed to be 
\begin{equation}\label{p of gain in B'B''}
p^{B'B''}_{\rm gain}=\frac{1}{R}\sum_{j=1}^{4}\tau_{j}t_{j}=\frac{t_4\left(t_3+1-t_1\right)}{\left(1-t_1\right)\left(2t_4+1-t_2\right)+t_3t_4}.
\end{equation}
Just as in case of the game $B'$, the expression for $p^{B'B''}_{\rm gain}$ reduces to  
\begin{equation}\label{p win class B' and B''}
p^{B'B''}_{\rm gain} =1/(2+x'/y')
\end{equation}
with 
\begin{equation}\label{s condition B'B''}
y'=t_4(t_3+1-t_1)>0
\end{equation}
for any choice of the probabilities $t_1, \dots t_4$, and
\begin{equation}\label{c condition B'B''}
x'=(1-t_1)(1-t_2)-t_3t_4.
\end{equation}
The game $B'B''$ therefore behaves entirely like the game $B'$, following the 
rule: if $x' < 0$, $B'B''$ is winning, that is, has positive expected payoff; if $x'= 0$, $B'B''$ is
fair; and $x'> 0$, $B'B''$ is losing, that is, has negative expected payoff.

It is therefore possible to adjust the values of the $\alpha_j$ and $\beta_j$ in games $B'$ and $B''$ so that they are individually losing, but the combined game $B'B''$ is now winning. This is the Parrondo effect. In the present example, the Parrondo effect occurs when 
\begin{equation}\label{D}
(1-\alpha_3)(1-\alpha_4) > \alpha_1\alpha_2
\end{equation}
\begin{equation}\label{E}
(1-\beta_3)(1-\beta_4)>\beta_1\beta_2
\end{equation}
and
\begin{equation}\label{F}
(1-t_3)(1-t_4)<t_1t_2.
\end{equation}
The reader is referred to \cite{Kay:03} for a detailed analysis of the values of the parameters which lead to the Parrondo effect in such games.

Restricting to the original work of Parrondo et al, a special case occurs when we consider one of the games in the randomized sequence to be of type $A$. That is, flipping a single biased coin which on the surface appears to have no history dependence. However, note that such a game may be interpreted as a history dependent Parrondo game with a two stage history where the coin used in $A$ is employed for every history. Call such a history dependent game $A'$. The transition matrix for $A'$ takes the form
\begin{equation}\label{tran mat A'}
\Delta=\left(\begin{array}{cccc}
p & 0 & p & 0 \\  1-p & 0 & 1-p & 0 \\ 0 & p& 0 & p\\ 0 & 1-p& 0 & 1-p
\end{array}\right).
\end{equation} 
Now, forming randomized sequences of games $A'$ and $B'$ is seen to agree with the forming of convex linear combinations mentioned above. In particular, as analyzed in \cite{Parrondo} if games $A'$ and $B'$ are now sequenced randomly with equal probability, the Markov process for the randomized sequence is given with transition matrix containing the entries $q_j=\frac{1}{2}(\alpha_j+p)$ and $1-q_j=\frac{1}{2}[(1-\alpha_j)+(1-p)]$ in the appropriate locations (recall that the probability of win for game $A$ is $p$), and has stationary state
\begin{equation}\label{stat state mix}
\rho=\left(\begin{array}{c}
\rho_1 \\ \rho_2 \\ \rho_3 \\ \rho_4 
\end{array}\right)=\frac{1}{M}\left(\begin{array}{c}
q_3q_4 \\ q_4(1-q_1) \\ q_4(1-q_1) \\ (1-q_1)(1-q_2)
\end{array}\right)
\end{equation}
Denote this randomized sequence of games $A'$ and $B'$ by $A'B'$.
The probability of gain in the game $A'B'$ is
\begin{equation}\label{p of win in AB'}
p^{A'B'}_{\rm gain}=\frac{1}{M}\sum_{j=1}^{4}\rho_{j}q_{j}=\frac{q_4\left(q_3+1-q_1\right)}{\left(1-q_1\right)\left(2q_4+1-q_2\right)+q_3q_4}
\end{equation}
As in the more general case of the game $B'B''$, it is now possible to adjust the values of the parameters $p$ and $p_j$'s in games $A'$ and $B'$ so that they are individually losing, but the combined game $A'B'$ is now winning. This happens when 
\begin{equation}\label{A}
1-p > p
\end{equation}
\begin{equation}\label{B}
(1-\alpha_3)(1-\alpha_4)>\alpha_1\alpha_2
\end{equation}
and
\begin{equation}\label{C}
(1-q_3)(1-q_4)<q_1q_2.
\end{equation}
Parrondo et al show in \cite{Parrondo} that when $p=\frac{1}{2}-\epsilon$, $\alpha_1=\frac{9}{10}-\epsilon$, $\alpha_2=\alpha_3=\frac{1}{4}-\epsilon$, $\alpha_4=\frac{7}{10}-\epsilon$, and $\epsilon <\frac{1}{168}$, the inequalities (\ref{A})-(\ref{C}) are satisfied. This is Parrondo et al's original example of the Parrondo effect for history dependent Parrondo games. 

%

\section{The FNA Quantization of Parrondo Games}\label{AFN}

\begin{figure}
\centerline{\includegraphics[scale=0.25]{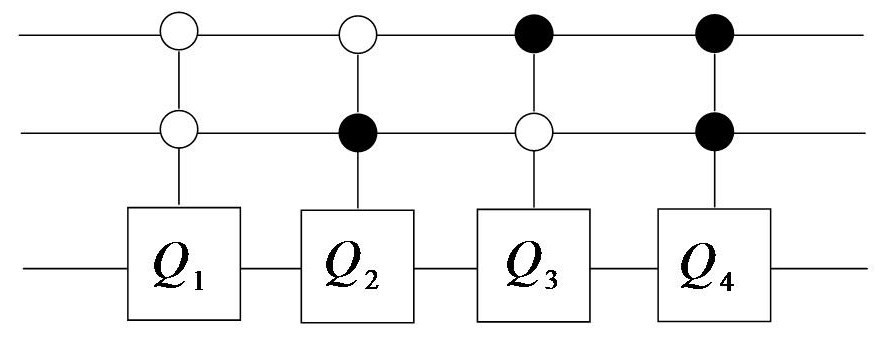}}
\caption{\small{Part of the quantization protocol for the history dependent Parrondo game. The first two wires represent the history qubits.}}
\label{HD quant}
\end{figure}

In \cite{Flitney:02}, Flitney, Ng, and Abbott quantize the type $A'$ Parrondo game by considering the action of an element of $SU(2)$ on a qubit and interpret this as ``flipping'' a biased quantum coin. They consider history dependent games with $(n-1)$ stage histories, and in the language of the Bleiler formalism, quantize these games via a family of protocols. In every protocol, $n$ qubits are required and the unitary operator representing the entire game is a $2^n \times 2^n$ block diagonal matrix with the $2 \times 2$ blocks composed of arbitrary elements of $SU(2)$. In the language of quantum logic circuits, this is a quantum multiplexer \cite{FSK:06}. The first $(n-1)$ qubits represent the history of the game via controls, as illustrated in Figure \ref{HD quant} for a two stage history dependent game similar to the game $B'$ given in Table \ref{table:gameB HD}. Each protocol is defined as the action of the quantum multiplexer on the $n$ qubits.

The quantum multiplexer illustrated in Figure \ref{HD quant}, where the elements $Q_1 \dots Q_4$ are elements of $SU(2)$, operates as follows. When the basis of the state space $(\mathbb{C}P^1)^{\otimes 3}$ of three qubits is the computational basis 
$$
\mathcal{B}=\left\{ \ket{000}, \ket{001}, \ket{010}, \ket{011}, \ket{100}, \ket{101}, \ket{110}, \ket{111} \right\}.
$$
the quantum multiplexer takes on the form of an $8 \times 8$ block diagonal matrix of the form 
\begin{equation}\label{qmuxq}
Q=\left( {{\begin{array}{*{20}c}
 {Q_1 } \hfill & 0 \hfill & 0 \hfill & 0 \hfill \\
 0 \hfill & {Q_2 } \hfill & 0 \hfill & 0 \hfill \\
 0 \hfill & 0 \hfill & {Q_3} \hfill & 0 \hfill \\
 0 \hfill & 0 \hfill & 0 \hfill & {Q_4 } \hfill \\
\end{array} }} \right),
\end{equation}
where each $Q_j \in SU(2)$. That is
\begin{equation}\label{eqn:eq16}
Q_{j}=\left( {{\begin{array}{*{20}c}
a_j \hfil & -\overline{b}_j \hfill \\
b_j \hfill & \overline{a}_j \hfill \\
\end{array}} } \right)
\end{equation}
with $a_j, b_j \in \mathbb{C}$ satisfying $\left|a_j\right|^2+\left|b_j\right|^2=1$. 

For further description of the workings of the quantum multiplexer, the following convention, found in D. Meyer's original work \cite{Meyer}, will be used. Let a ``win'' or ``gain'' for a player be represented by the action ``No Flip'' which is the identity element of $SU(2)$. For example, in Meyer's quantum penny flip game, the ``quantum coin'' is in the initial state of ``Head'' represented by $\ket{0}$ and a gain for the player using the quantum strategies occurs when the final orientation state of the coin is observed to be $\ket{0}$. This is contrast to the convention in FNA \cite{Flitney:02} where $\ket{1}$ represents a gain. 

Now the first two qubits of an element of $\mathcal{B}$ represent a history of the classical game, with $\ket{0}$ representing gain ($G$) and the $\ket{1}$ representing loss ($L$). The blocks $Q_j$ act on the third qubit in the circuit under the control of the history represented by the binary configuration of the first two qubits. For example, if the first two qubits are in the joint state $\ket{00}$, the $SU(2)$ action $Q_{1}$ is applied to the third qubit. Similarly, for the other three basic initial joint states of the first two qubits. This models the historical dependence of the game by having the history $(G,G)$ correspond to the initial joint state $\ket{00}$ of the first two qubits, the history $(G,L)$ correspond to the initial joint state $\ket{01}$, the history $(L,G)$ correspond to the initial joint state $\ket{10}$, and the history $(L,L)$ correspond to the initial joint state $\ket{11}$. Thus, an appropriate action is taken for each history. 

Recall from section \ref{HD game} that the evaluation of the behavior of the classical history dependent Parrondo game requires more than just the Markov process. The evaluation also requires the stable state and a payoff rule. Note that the results of applying the quantum multiplexer depends entirely on the initial state on which it acts. That is, different initial states result in differing final states. The payoff rule used by Abbott, Flitney, and Ng resembles that for the classical game in that the quantized versions are winning when the expectation greater than $0$ (gain capital), fair if the expectation is equal to $0$ (break even), and losing if the expectation is less than $0$ (lose capital). Further, as in the classical game this question is decided by examining the probability of gain versus the probability of loss. In particular, if the probability of gain is greater than $\frac{1}{2}$, the quantum game is winning. 

\subsection{Problems with the FNA Protocol}

The FNA quantization protocols for the history dependent game attempt to replace the classical biases of the coins in the game with arbitrary elements of $SU(2)$ and the stable state of Markov process describing the dynamics of the game with certain initial states of the qubits on which a quantum multiplexer, composed of the arbitrary elements of $SU(2)$, acts. The problems with the FNA quantization protocols are two-fold. First, the attempted embedding of the classical history dependent game into the quantized game by replacing the biases of the classical coins with arbitrary $SU(2)$ elements, turns out to be {\it relational} rather than functional. That is, Equations (\ref{qmuxq}) and (\ref{eqn:eq16}) together give a family of quantum multiplexers that the classical game maps into via the embedding. This relational mapping makes it impossible to recover the classical game by restricting the quantized game to the image of the embedding. Therefore, the FNA quantization of the history dependent Parrondo game is not proper. 

The second problem arises from the choice of initial state. No attempt is made to produce an analog of the stable state of a Markov process. Instead, the authors mention the obvious fact that different initial states will produce different results, and in particular consider two arbitrary initial states, one the maximally entangled state $\frac{1}{\sqrt{2}}\left(\left|000\right\rangle+\left|111\right\rangle\right)$, the other the basic state $\ket{000}$. In the latter, the authors assert that the quantum game behaves like a classical game with fixed initial history $(L, L)$, according to their convention in which $\ket{0}$ represents loss. Note that even if the  this is not a proper quantization of any classical history dependent game as it fails to incorporate the other histories represented in the stable state. For 
$$
\left|000\right\rangle=
\left( {{\begin{array}{c}
1\\
0\\
0\\
0\\
0\\
0\\
0\\ 
0
\end{array}} } \right)
$$
and when acted upon by the quantum multiplexer in Equation (\ref{qmuxq}) produces the output
$$
\left( {{\begin{array}{c}
a_1\\
b_1\\
0\\
0\\
0\\
0\\
0\\ 
0
\end{array}} } \right)
$$
which makes the failure of the protocol to incorporate the other histories apparent. 

In the former, 
a similar situation occurs where only the histories $\ket{000}$ and $\ket{111}$ are incorporated. This protocol is also not proper as only the histories $(L,L)$ and $(G,G)$ are non-trivially represented in the initial state. For
$$
\frac{1}{\sqrt{2}}\left(\left|000\right\rangle+\left|111\right\rangle\right)=
\frac{1}{\sqrt{2}}\left( {{\begin{array}{c}
1\\
0\\
0\\
0\\
0\\
0\\
0\\ 
1
\end{array}} } \right)
$$
and when acted upon by the quantum multiplexer in Equation (\ref{qmuxq}) produces the output
$$
\frac{1}{\sqrt{2}}\left( {{\begin{array}{c}
a_1\\
b_1\\
0\\
0\\
0\\
0\\
-\overline{b_4}\\ 
\overline{a_4} 
\end{array}} } \right)
$$
from which, again, the failure of the protocol to incorporate the other histories is apparent. 

Moreover, both quantization protocols fail to reproduce the Markovian dynamics and the payoff function of the original game.

Flitney et al also consider various ``sequences'' of the quantum games $A'$ and $B'$, where $B'$ is played with three qubits and quantized using the maximally entangled initial state. These sequences are defined by compositions of the unitary operators defining the games. Indeed, these sequences now produce the results presented in \cite{Flitney:02}. These results are certainly novel and perhaps carry scientific significance; however, they fail to carry game-theoretic significance as, with respect to the classical Parrondo games, each arises from a quantization that is {\it not} proper. 

In light of the Bleiler formalism discussed in chapter \ref{Bleilerformalism}, constructing proper quantizations of games is a fundamental problem for quantum theory of games. In the following section, a proper quantization paradigm is developed for both history dependent Parrondo games and randomized sequences of such. 

\section{Properly Quantizing History Dependent Parrondo Games}\label{prop quantum A and B'}

Consider the history dependent game $B'$ with only 2 histories. As in the FNA protocol, the quantization protocol for this game uses a three qubit quantum multiplexer with matrix representation
$$
Q=\left( {{\begin{array}{*{20}c}
 {Q_1 } \hfill & 0 \hfill & 0 \hfill & 0 \hfill \\
 0 \hfill & {Q_2 } \hfill & 0 \hfill & 0 \hfill \\
 0 \hfill & 0 \hfill & {Q_3} \hfill & 0 \hfill \\
 0 \hfill & 0 \hfill & 0 \hfill & {Q_4 } \hfill \\
\end{array} }} \right)
$$
with each $Q_j \in SU(2)$, together with an initial state.

To reproduce the classical game, first embed the four classical coins that define the game $B'$ into blocks of the matrix $Q$ corresponding to the appropriate history. The embedding is via superpositions of the embeddings of the classical actions of ``No Flip'' and ``Flip'' on the coins into $SU(2)$ given either by 
\begin{equation}\label{basic embed 1}
N=\left( {{\begin{array}{*{20}c}
1 & 0 \\
0 & 1 \\
\end{array}} } \right), \quad F=\left( {{\begin{array}{*{20}c}
0 & -\overline{\eta} \\
\eta & 0 \\
\end{array}} } \right)
\end{equation}
or by 
\begin{equation}\label{basic embed 2}
N^{*}=\left( {{\begin{array}{*{20}c}
i & 0 \\
0 & \overline{i} \\
\end{array}} } \right), \quad F^{*}=\left( {{\begin{array}{*{20}c}
0 & -\overline{i\eta} \\
i\eta & 0 \\
\end{array}} } \right)
\end{equation}
with $\eta^6=1$. Call the embeddings in equations (\ref{basic embed 1}) {\it basic embeddings of type 1} and the embedding in equations (\ref{basic embed 2}) {\it basis embeddings of type 2}. Choosing the basic embedding of type 1 embeds the $j^{\rm {th}}$ coin into $SU(2)$ as 
\begin{equation}\label{classical embed B}
Q_{j}=\sqrt{p_j}N+\sqrt{(1-p_j)}F=\left( {{\begin{array}{*{20}c}
{\sqrt p_j} & -\sqrt{1-p_j}\overline{\eta} \\
\sqrt{1-p_j}\eta & {\sqrt p_j} \\
\end{array}} } \right)
\end{equation}
where $p_j$ is the probability of gain when the $j^{\rm {th}}$ coin is played in the classical game $B'$ given in Table \ref{table:gameB HD}. Note that the probabilities $p_j$ of gaining are associated with the classical action $N$ in line with Meyer's original convention from \cite{Meyer} where $\ket{0}$ represents a gain. Hence, the elements of the subset 
$$
\mathcal{W}=\left(\ket{000}, \ket{010}, \ket{100}, \ket{110}\right)
$$ 
of $\mathcal{B}$ all represent possible gaining outcomes in the game. The probability of gain in the quantized game is therefore the sum of the coefficients of the elements of $\mathcal{W}$ that result from measurement. 

Next, set the initial state $I$ equal to 
\begin{equation}\label{unentangled state}
\frac{1}{\sqrt{\sum_{j=1}^{n}\pi_j}}\left( {{\begin{array}{c}
{\sqrt \pi_1}\\
0\\
{\sqrt \pi_2}\\
0\\
{\sqrt \pi_3}\\
0\\
{\sqrt \pi_4}\\ 
0
\end{array}} } \right),
\end{equation}
where the $\pi_j$ are the probabilities with which the histories occur in the classical game, as computed from the stationary state of the Markovian process of section \ref{HD game}. The quantum multiplexer $Q$ acts on $I$ to produce the final state
\begin{equation}\label{output}
F_I=\frac{1}{\sqrt{\sum_{j=1}^{4}\pi_j}}\left( {{\begin{array}{c}
\sqrt{p_1\pi_1} \\
\eta\sqrt{(1-p_1)\pi_1}\\
\sqrt{p_2\pi_2} \\
\eta\sqrt{(1-p_2)\pi_2}\\
\sqrt{p_3\pi_3} \\
\eta\sqrt{(1-p_3)\pi_3}\\
\sqrt{p_4\pi_4} \\ 
\eta\sqrt{(1-p_4)\pi_4}
\end{array}} } \right). 
\end{equation}
Measuring the state $F_I$ in the observational basis and adding together the resulting coefficients of the elements of the set $\mathcal{W}'$ gives the probability of gain in the quantized game to be 
\begin{equation}\label{quantum prob gain}
p^{QB'}_{\rm gain}=\frac{1}{\sum_{j=1}^{4}\pi_j}\left(\sum_{j=1}^4p_j\pi_j\right)=\frac{1}{N}\left(\sum_{j=1}^4p_j\pi_j\right)
\end{equation}
which is equal to the probability of gain in the classical game. 

\begin{figure}
\centerline{\includegraphics[scale=0.23]{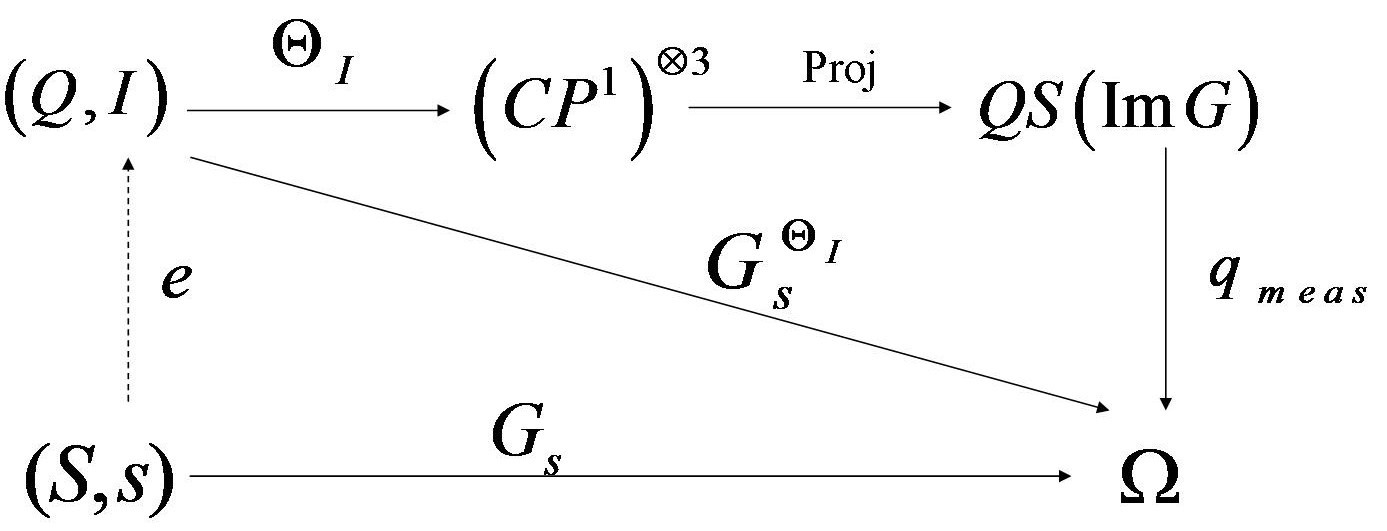}}
\caption{\small{Proper Quantization, using the embedding $e$, of the History Dependent Game via the quantization protocol $\Theta_I$}.}
\label{FormalParrondoquant}
\end{figure}

This proper quantization paradigm is based on the philosophy first discussed at the end of chapter \ref{Bleilerformalism}. That is, a proper quantization of a classical game $G_s$ that depends on an initial state $s$ requires that $s$ be embedded into an initial state $I$ on which the quantum multiplexer acts. Here, the initial state $s=(\pi_1, \pi_2, \pi_3, \pi_4) \in [0,1]^{\times 4}$ embeds as the initial state $I \in (\mathbb{C}P^1)^{\otimes 3}$ given in expression (\ref{unentangled state}). The resulting game $G^{\Theta_I}_s$ is the quantization of the classical game $G_s$ by the protocol $\Theta_I$ which maps the tuple $(Q,I)$, with $Q=(Q_1, Q_2, Q_3, Q_4) \in [SU(2)]^{\times 4}$ to $F_I \in (\mathbb{C}P^1)^{\otimes 3}$ given in Equation (\ref{output}). Formally,
\begin{equation}\label{quantum payoff}
\Theta_I: [SU(2)]^{\times 4} \times (\mathbb{C}P^1)^{\otimes 3} \rightarrow (\mathbb{C}P^1)^{\otimes 3}
\end{equation}
\begin{equation}\label{quantum payoff detail}
\Theta_I: (Q,I) \mapsto F_I
\end{equation}
By projecting on to the gaining basis $\mathcal{W}$, one now gets a quantum superposition over the image Im$G$ of the game $G$. Finally, quantum measurement produces Im$G$. Call $Proj$ the function that projects $F_I$ on to $\mathcal{W}$, and denote quantum measurement by $q_{meas}$. Then 
\begin{equation}\label{quantum payoff comp}
G_s^{\Theta_I}= q_{meas} \circ Proj \circ \Theta_I: (Q,I) \mapsto {\rm Im}G
\end{equation}
is a proper quantization of the payoff function of the normal form of classical history dependent game $G_s$ given in Equations (\ref{formal payoff funct}) and (\ref{formal payoff funct detail}). Equation (\ref{quantum payoff comp}) can be expressed by the commutative diagram of Figure \ref{FormalParrondoquant}, which the reader is urged to compare and contrast with Figure \ref{FormalParrondo} in chapter \ref{Bleilerformalism}.

Note that by embedding $s$ into $I$, the notion of randomization via probability distributions is generalized in the quantum game to the higher order  notion of randomization via quantum superpositions plus measurement. In particular, the probability distribution $P=(p_1, p_2, p_3, p_4) \in [0,1]^{\times 4}$ that defines the Markov process associated with the game is replaced with the quantum multiplexer $Q=(Q_1, Q_2, Q_3, Q_4) \in [SU(2)]^{\times 4}$ associated with the quantized game, and the stable state $s$ of the Markov process is replaced with an initial evaluative state $I$ of the quantum multiplexer.

\section{Properly Quantizing Randomized Sequences of History Dependent Parrondo Games}\label{sequences}

Recall from section \ref{rand A and B'} that randomized sequences of games $B'$ and $B''$ are analyzed via a Markov process with transition matrix equal to a real convex combination of the transition matrices of each game in which $B'$ is played with probability $r$ and $B''$ with probability $(1-r)$. Moreover, such a sequence is considered to by an instance of a history dependent game denoted as $B'B''$. 

Motivated by the discussion on proper quantization of the game Parrondo games $B'$ and $B''$ in section \ref{prop quantum A and B'} above, let us now consider a higher order randomization in the form of a quantum superposition of the quantum multiplexers used in the proper quantization of the the games $B'$ and $B''$ with the goal of producing a proper quantization of the game $B'B''$. 

As in section \ref{prop quantum A and B'}, associate the quantum multiplexer $Q'=(Q'_1,Q'_2,Q'_3,Q'_4)$ with the game $B'$, where
$$
Q'_{j}=\sqrt{\alpha_j}N+\sqrt{(1-\alpha_j)}F=\left( {{\begin{array}{*{20}c}
{\sqrt \alpha_j} & -\sqrt{1-\alpha_j}\overline{\eta} \\
\sqrt{1-\alpha_j}\eta & {\sqrt \alpha_j} \\
\end{array}} } \right),
$$
Next, associate the quantum multiplexer $Q''=(Q''_1, Q''_2, Q''_3, Q''_4)$ with the game $B''$, where 
$$
Q''_j = \sqrt{\beta_j}N^{*}+\sqrt{(1-\beta_j)}F^{*}=\left( {{\begin{array}{*{20}c}
{\sqrt \beta_j}i & -\sqrt{1-\beta_j}(\overline{i\eta}) \\
\sqrt{1-\beta_j}i\eta & {\sqrt \beta_j}\overline{i} \\
\end{array}} } \right).
$$
Now consider the quantum superposition
\begin{align}\label{mult superposition}
\Sigma&=\gamma'Q'+\gamma''Q''\\
&=\left( {{\begin{array}{*{20}c}
 {\gamma'Q'_1+\gamma''Q''_1} \hfill & 0 \hfill & 0 \hfill & 0 \hfill \\
 0 \hfill & {\gamma'Q'_2+\gamma''Q''_2} \hfill & 0 \hfill & 0 \hfill \\
 0 \hfill & 0 \hfill & {\gamma'Q'_3+\gamma''Q''_3} \hfill & 0 \hfill \\
 0 \hfill & 0 \hfill & 0 \hfill & {\gamma'Q'_4+\gamma''Q''_4} \hfill \\
\end{array} }} \right)
\end{align}
of the quantum multiplexers $Q'$ and $Q''$ with 
\begin{equation}\label{general quant game cond}
(\gamma')^2+(\gamma'')^2=1, \quad \left|\gamma'\right|^2=r, \quad \left|\gamma''\right|^2=(1-r), \quad \overline{\gamma'}\gamma''-\overline{\gamma''}\gamma'=0
\end{equation}
and
 \begin{equation}
\gamma'Q'_j+\gamma''Q''_j=\left( {{\begin{array}{*{20}c}
\gamma'{\sqrt \alpha_j}+\gamma''\sqrt{\beta_j}i & -\left(\gamma'\sqrt{1-\alpha_j}-\gamma''\sqrt{1-\beta_j}i\right)\overline{\eta} \\
\left(\gamma'\sqrt{1-\alpha_j}+\gamma''\sqrt{1-\beta_j}i\right)\eta & \gamma'{\sqrt \alpha_j}-\gamma''\sqrt{\beta_j}i \\
\end{array}} } \right)\\ \\
\end{equation}
Set the evaluative initial state in this case equal to 
\begin{equation}\label{AB' initial state}
I=\frac{1}{\sqrt{\sum_{j=1}^{n}\tau_j}}\left( {{\begin{array}{c}
{\sqrt \tau_1}\\
0\\
{\sqrt \tau_2}\\
0\\
{\sqrt \tau_3}\\
0\\
{\sqrt \tau_4}\\ 
0
\end{array}} } \right)
\end{equation}
where the $\tau_j$ are the probabilities that form the stationary state of the classical game $B'B''$ given in Equation (\ref{stat state mix general}). The claim is that the quantum multiplexer $\Sigma$ in Equation (\ref{mult superposition}) together with the evaluative initial state $I$ in Equation (\ref{AB' initial state}) define a proper quantization of the classical game $B'B''$ in which $B'$ is played with probability $r$ and and $B''$ is played with probability $(1-r)$. 

To check the validity of this claim, compute the output of $\Sigma$ for the evaluative initial state $I$ in Equation (\ref{AB' initial state}):
$$
\frac{1}{\sqrt{\sum_{j=1}^{n}\tau_j}}\left( {{\begin{array}{c}
{\sqrt \tau_1}(\gamma'{\sqrt \alpha_1}+\gamma''\sqrt{\beta_1}i) \\
{\sqrt \tau_1}\left(\gamma'\sqrt{1-\alpha_1}+\gamma''\sqrt{1-\beta_1}i\right)\eta \\
{\sqrt \tau_2}(\gamma'{\sqrt \alpha_2}+\gamma''\sqrt{\beta_2}i)\\
{\sqrt \tau_2}\left(\gamma'\sqrt{1-\alpha_2}+\gamma''\sqrt{1-\beta_2}i\right)\eta\\
{\sqrt \tau_3}(\gamma'{\sqrt \alpha_3}+\gamma''\sqrt{\beta_3}i)\\
{\sqrt \tau_3}\left(\gamma'\sqrt{1-\alpha_3}+\gamma''\sqrt{1-\beta_3}i\right)\eta \\
{\sqrt \tau_4}(\gamma'{\sqrt \alpha_4}+\gamma''\sqrt{\beta_4}i) \\ 
{\sqrt \tau_4}\left(\gamma'\sqrt{1-\alpha_4}+\gamma''\sqrt{1-\beta_4}i\right)\eta
\end{array}} } \right).
$$
The probability of gain produced upon measurement of this output is 
\begin{equation}\label{norm form of quant sequence}
p^{QB'B''}_{\rm gain}=\frac{1}{\sum_{j=1}^{n}\tau_j}\sum_{j=1}^4\left|{\sqrt \tau_j}(\gamma'{\sqrt \alpha_j}+\gamma''\sqrt{\beta_j}i)\right|^2
\end{equation}
which simplifies to 
\begin{equation}
\frac{1}{R}\sum_{j=1}^4\tau_j\left[\left|\gamma'\right|^2\alpha_j+\left|\gamma''\right|^2\beta_j+{\sqrt \alpha_j\beta_j}i\left(\overline{\gamma'}\gamma''-\overline{\gamma''}\gamma'\right)\right].
\end{equation}
Using the conditions set up in Equation (\ref{general quant game cond}), the previous expression further simplifies to give
$$
p^{QB'B''}_{\rm gain}=\frac{1}{R}\sum_{j=1}^4\tau_j\left[r\alpha_j+(1-r)\beta_j\right]=\frac{1}{R}\sum_{j=1}^4\tau_jt_j.
$$
which is exactly that given in Equation (\ref{p of gain in B'B''}) in section \ref{sequences} for the classical game $B'B''$.

Again, note that this proper quantization paradigm requires mapping of the initial state of the classical game $B'B''$, which is a probability distribution, into an initial state which the quantization protocol acts on, which is a higher order randomization in the form of a quantum superposition which measures appropriately with respect to the observational basis. The image of the normal form of the quantum game in $[0,1]$ agrees precisely with $p^{QB'B''}_{\rm gain}$. Note that in this proper quantization of $B'B''$, not only is the initial state of the classical game replaced by a quantum superposition, but also a probabilistic combination of the transition matrices of the classical games is replaced with a quantum superposition of the quantum multiplexers associated with each classical game. 

\subsection{A Special Case}

Recall from section \ref{rand A and B'} the classical analysis of the special case of the randomized sequence of history dependent Parrondo games, with $r=(1-r)=\frac{1}{2}$, in which one of the games is $A'$. The game $A'$ has the property that regardless of history, game $A$ is always played. Such a sequence was considered to by an instance of a history dependent game denoted by $A'B'$. In this section, a proper quantization of the randomized sequence is shown to follow as a special case of the proper quantization of the classical game $B'B''$ developed in section \ref{sequences} above.

As before, associate the quantum multiplexer $Q'=(Q'_1,Q'_2,Q'_3,Q'_4)$, where
$$
Q'_{j}=\sqrt{p_j}N+\sqrt{(1-p_j)}F=\left( {{\begin{array}{*{20}c}
{\sqrt p_j} & -\sqrt{1-p_j}\overline{\eta} \\
\sqrt{1-p_j}\eta & {\sqrt p_j} \\
\end{array}} } \right),
$$
with the game $B'$. Now, first embed the game $A$ into $SU(2)$ using basic embeddings of type 2. That is,
$$
A = \sqrt{p}N^{*}+\sqrt{(1-p)}F^{*}=\left( {{\begin{array}{*{20}c}
{\sqrt p}i & -\sqrt{1-p}(\overline{i\eta}) \\
\sqrt{1-p}i\eta & {\sqrt p}\overline{i} \\
\end{array}} } \right).
$$
The transition matrix for the game $A'$ was given in Equation (\ref{tran mat A'}) and is reproduced here:
$$
\Delta=\left(\begin{array}{cccc}
p & 0 & p & 0 \\  1-p & 0 & 1-p & 0 \\ 0 & p& 0 & p\\ 0 & 1-p& 0 & 1-p
\end{array}\right).
$$ 
The form of $\Delta$ suggests that the quantum multiplexer $Q''=(A,A,A,A)$ should be associated with the game $A'$. Now let $\gamma'=\gamma''=\frac{1}{\sqrt 2}$in Equation (\ref{mult superposition}) so that
\begin{equation}\label{mult superposition special case}
\Sigma=\frac{1}{\sqrt 2}(\Delta'+Q')=\frac{1}{\sqrt 2}\left( {{\begin{array}{*{20}c}
 {A+Q'_1} \hfill & 0 \hfill & 0 \hfill & 0 \hfill \\
 0 \hfill & {A+Q'_2} \hfill & 0 \hfill & 0 \hfill \\
 0 \hfill & 0 \hfill & {A+Q'_3} \hfill & 0 \hfill \\
 0 \hfill & 0 \hfill & 0 \hfill & {A+Q'_4} \hfill \\
\end{array} }} \right)
\end{equation}
with 
\begin{align*}
A+Q'_j&=\left( {{\begin{array}{*{20}c}
{\sqrt p}i+\sqrt{p_j} & -\left(\sqrt{1-p}(\overline{i\eta})+\sqrt{1-p_j}\overline{\eta}\right) \\
\sqrt{1-p}i\eta+\sqrt{1-p_j}\eta & {\sqrt p}\overline{i}+\sqrt{p_j} \\
\end{array}} } \right)\\ \\
&= \left( {{\begin{array}{*{20}c}
\sqrt{p_j}+{\sqrt p}i & -\left(\sqrt{1-p_j}-\sqrt{1-p}i\right)\overline{\eta} \\
\left(\sqrt{1-p_j}+\sqrt{1-p}i\right)\eta & \sqrt{p_j}-{\sqrt p}i \\
\end{array}} } \right).
\end{align*}
With the evaluative initial state 
\begin{equation}\label{AB' initial state}
I=\frac{1}{\sqrt{\sum_{j=1}^{n}\rho_j}}\left( {{\begin{array}{c}
{\sqrt \rho_1}\\
0\\
{\sqrt \rho_2}\\
0\\
{\sqrt \rho_3}\\
0\\
{\sqrt \rho_4}\\ 
0
\end{array}} } \right)
\end{equation}
where the $\rho_j$ are the probabilities that form the stationary state of the classical game $A'B'$ given in Equation (\ref{stat state mix}), the quantum multiplexer $\Sigma$ in Equation (\ref{mult superposition}) defines a proper quantization of the classical game $AB'$ when both $A$ and $B'$ are played with equal probability. 

To see this, compute the output of $\Sigma$ for the evaluative initial state $I$ in Equation (\ref{AB' initial state}):
$$
\frac{1}{\sqrt{2\sum_{j=1}^{n}\rho_j}}\left( {{\begin{array}{c}
{\sqrt \rho_1}({\sqrt p}i+\sqrt{p_1}) \\
{\sqrt \rho_1}\left(\sqrt{1-p_1}+\sqrt{1-p}i\right)\eta \\
{\sqrt \rho_2}({\sqrt p}i+\sqrt{p_2}) \\
{\sqrt \rho_2}\left(\sqrt{1-p_2}+\sqrt{1-p}i\right)\eta \\
{\sqrt \rho_3}({\sqrt p}i+\sqrt{p_3}) \\
{\sqrt \rho_3}\left(\sqrt{1-p_3}+\sqrt{1-p}i\right)\eta \\
{\sqrt \rho_4}({\sqrt p}i+\sqrt{p_4}) \\ 
{\sqrt \rho_4}\left(\sqrt{1-p_4}+\sqrt{1-p}i\right)\eta 
\end{array}} } \right).
$$
The probability of gain produced upon measurement is 
\begin{equation}\label{norm form of quant sequence}
p^Q_{\rm gain}=\frac{1}{2\sum_{j=1}^{n}\rho_j}\sum_{j=1}^4\left|{\sqrt \rho_j}({\sqrt p}i+\sqrt{p_j})\right|^2=\frac{1}{M}\sum_{j=1}^4\rho_j\left(\frac{p+p_j}{2}\right)=\frac{1}{M}\sum_{j=1}^4\rho_jq_j
\end{equation}
which is exactly that given in Equation (\ref{p of win in AB'}) in section \ref{HD game} for the classical game $A'B'$.

\section{A Second Proper Quantization of the Randomized Sequence of History Dependent Parrondo Games}
A second proper quantization of the sequence $B'B''$ can be constructed in a manner similar to that used to construct the proper quantization for $B'$ in section \ref{prop quantum A and B'}. Instead of forming a quantum superposition of the quantum multiplexers associated with each game, first embed the classical coins used in the game $B'B''$ into $SU(2)$ as  
\begin{align*}
Y_j &= \sqrt{t_j}N+\sqrt{1-t_j}F \\
&=\left( {{\begin{array}{*{20}c}
\sqrt{t_j} & -\sqrt{1-t_j}\overline{\eta} \\
\sqrt{1-t_j}\eta & \sqrt{t_j} \\
\end{array}} } \right)
\end{align*}
with
$$
t_j=r\alpha_j+(1-r)\beta_j \quad {\rm and} \quad 1-t_j=r(1-\alpha_j)+(1-r)(1-\beta_j)
$$ 
and associate the quantum multiplexer $Y=(Y_1, Y_2, Y_3, Y_4)$ with the classical game $B'B''$. Set the initial state, as in section \ref{sequences}, equal to
$$
I=\frac{1}{\sqrt{\sum_{j=1}^{n}\tau_j}}\left( {{\begin{array}{c}
{\sqrt \tau_1}\\
0\\
{\sqrt \tau_2}\\
0\\
{\sqrt \tau_3}\\
0\\
{\sqrt \tau_4}\\ 
0
\end{array}} } \right)
$$
where the $\tau_j$ are the probabilities that form the stationary state of the classical game $B'B''$ given in Equation (\ref{stat state mix general}). The output state of this protocol is 
\begin{equation}\label{AB' output state}
F_I=\frac{1}{\sqrt{\sum_{j=1}^{n}\tau_j}}\left( {{\begin{array}{c}
\sqrt{\tau_1t_1}\\
\sqrt{\tau_1(1-t_1)}\eta\\
\sqrt{\tau_2t_2}\\
\sqrt{\tau_2(1-t_2)}\eta\\
\sqrt{\tau_3t_3}\\
\sqrt{\tau_3(1-t_3)}\eta\\
\sqrt{\tau_4t_4}\\ 
\sqrt{\tau_4(1-t_4)}\eta
\end{array}} } \right)
\end{equation}
which, upon measurement produces the probability of gain 
$$
p^{QB'B''}_{\rm gain}=\frac{1}{\sum_{j=1}^{n}\tau_j}\sum_{j=1}^4\tau_jt_j
$$
which is exactly the probability of gain computed in Equation (\ref{p of win in AB'}) of section \ref{HD game} for the classical game $AB'$.

Hence, there are two approaches, both motivated by different facets of the Bleiler formalism, used here to properly quantize random sequences of Parrondo games $A$ and $B'$ in which each game occurs with equal probability. One approach, discussed in section \ref{prop quantum A and B'}, generalizes the notion of randomization between the two games via probability distributions to randomization between games via quantum superpositions. The other approach, discussed above, embeds a probabilistic combination of the games into a quantum multiplexer directly rather than via quantum superpositions of the protocols for each game. 

In the former approach, note that it was crucial that game $A$ was embedded into $SU(2)$ using basic embedding of type 2 as this allowed for the use of the broader arithmetical properties, namely factorization, of complex numbers to reproduce the classical result. In the latter on the other hand, basic embedding of type 1 sufficed. 

These two different approaches to quantizing history dependent Parrondo games raise interesting questions regarding the relationship between {\it general} quantum history dependent Parrondo games, which are quantum multiplexers with arbitrary $SU(2)$ elements forming the diagonal blocks, and the proper quantizations of the classical history dependent Parrondo games. For example, can a general quantum history dependent Parrondo game always be factored into a sum of games which correspond to embedding of some classical history dependent Parrondo games? The reader is referred to the future directions section of chapter 6 where this subject is discussed in detail.

\chapter{Quantum Logic Synthesis by Decomposition}\label{CosSin}

%
In Chapter \ref{Proper}, quantum multiplexers were used to properly quantize certain history dependent Parrondo games. In the following, quantum multiplexers will play a central role in synthesis of quantum logic circuits. 

Recent research in generalizing quantum computation from 2-valued qubits to $d$-valued qudits has shown practical advantages for scaling up a quantum computer. A further generalization leads to quantum computing with \emph{hybrid} qudits where two or more qudits have different finite dimensions. Advantages of hybrid and $d$-valued gates (circuits) and their physical realizations have been studied in detail by Muthukrishnan and Stroud \cite{MuthuStroud:04}, Daboul et. al \cite{Daboul:02}, and Bartlett et. al \cite{Bartlett}. 

Recall from section \ref{Quantum Computation} that the evolution of state space changes the state of the qudits under the action of a unitary matrix. Because evolution matrices are viewed as quantum logic gates in quantum computing, an essential idea from the theory of classical logic circuits carries over, namely, logic synthesis. One of the goals of logic synthesis is to express a given logic gate in terms of a universal set of quantum logic gates. Recall from section \ref{CNOT gate} that sets of one and two qubit (even qudit) gates are universal. Hence, the synthesis of a quantum logic gate requires that the corresponding matrix be decomposed to the level of unitary matrices acting on one or two qudits. Technological considerations for the implementation of one qudit gates might still require synthesis of these gates in terms of simpler one qudit rotation gates and two qudit controlled rotation gates. For $2$-valued quantum computing, this is easily accomplished by the well known Euler angle parameterization of a $2 \times 2$ special unitary matrix (since a unitary matrix is equivalent to a special unitary matrix up to a complex multiple). For higher valued quantum computing, Tilma et al's work in \cite{Tilma:02} shows that a one qudit gate can be synthesized in terms of an Euler angle parametrization similar to the one available for $2 \times 2$ special unitary matrices.

If the quantum system consists of multiple qudits, then a gate may be synthesized by matrix decomposition techniques such as QR factorization and the cosine-sine Decomposition (CSD). Both the acronym CSD and the term CS decomposition will be used to refer to the cosine-sine decomposition from now on. 
The CSD is used by M\"ott\"onen et. al \cite{mottonen:04} and Shende et. al \cite{shende:05} to iteratively synthesize multi-qubit quantum circuits. Khan and Perkowski \cite{FSK:05} use the CSD to develop an iterative synthesis method for 3-valued quantum logic circuits acting on $n$ qudits. Bullock et. al present a synthesis method for $n$ qudit quantum logic gates using a variation of the QR matrix factorization in \cite{bullock:04} In \cite{FSK:06}, Khan and Perkowski give a CSD based method for synthesis of $n$ qudit hybrid and $d$-valued quantum logic gates. This chapter reviews the work of these authors on quantum logic synthesis techniques based on the CS decomposition. 
%
%
\section{The Cosine-Sine Decomposition (CSD)}\label{sect:CSD}
Let the unitary matrix $\textit{W}\in \textbf{C}^{m\times m}$ be partitioned in $2 \times 2$ block form as
\begin{equation}\label{eqn:CSD matrix}
W=\bordermatrix {  &r      & m-r    \cr
                 r &W_{11} & W_{12} \cr
               m-r &W_{21} & W_{22} \cr}
\end{equation}
with $2r\leq m$. Then there exist $r \times r$ unitary
matrices $U$ and $X$, $r \times r$ real diagonal matrices $C$ and
$S$, and $(m-r) \times (m-r)$ unitary matrices $V$ and $Y$ such that
\begin{equation}\label{eqn:CSD1}
W  = \left(\begin{array}{cc}
U & 0 \\ 0 & V
\end{array}\right)
\left(\begin{array}{ccc}
  C & -S & 0 \\ S & C & 0 \\ 0 & 0 & I_{m-2r}
\end{array}\right)
\left(\begin{array}{cc}
 X & 0 \\ 0 & Y
\end{array}\right)
\end{equation}
The matrices $C$ and $S$ are the so-called cosine-sine matrices and are of the form $C$ = diag$(\cos \theta_{1}, \cos\theta_{2}, \dots,\cos\theta_{r})$, $S$ = diag$(\sin \theta_{1}$, $\sin \theta_{2},\dots,\sin \theta_{r})$ such that $\sin^{2}\theta_{i}+\cos^{2}\theta_{i}=1$ for some $\theta_{i}$, $1 \leq i \leq r$~\cite{Stewart:90}. Algorithms for computing the CSD and the angles $\theta_{i}$ are given in~\cite{Bjorck:73, Stewart:82}.
The CSD is essentially the well known singular value decomposition of a unitary matrix implemented at the block matrix level \cite{Paige:92}. Appendix B gives a review of the CS decomposition. 

The reader is advised that in the narrative that follows quantum logic gates, circuits and the corresponding unitary matrices will not be distinguished. 

\section{Synthesis of 2-valued (binary) Quantum Logic Circuits}\label{sect:2-valued CSD}

As the authors of \cite{FSK:05, mottonen:04,shende:05,Tucci:98} show, CSD gives a recursive method for synthesizing 2-valued and 3-valued $n$ qudit quantum logic gates. In the 2-valued case the CSD of a $2^{n} \times 2^{n}$ unitary matrix $W$ reduces to the form
\begin{equation}\label{eqn:CSD2}
W  = \left(\begin{array}{cc}
  U & 0 \\ 0 & V
\end{array}\right)
\left(\begin{array}{cc}
  C & -S \\ S & C \\
\end{array}\right)
\left(\begin{array}{cc}
  X & 0 \\  0 & Y
\end{array}\right)
\end{equation}
with each block matrix in the decomposition of size $2^{n-1} \times 2^{n-1}$. 

\begin{figure}
\centerline{\includegraphics[scale=0.25]{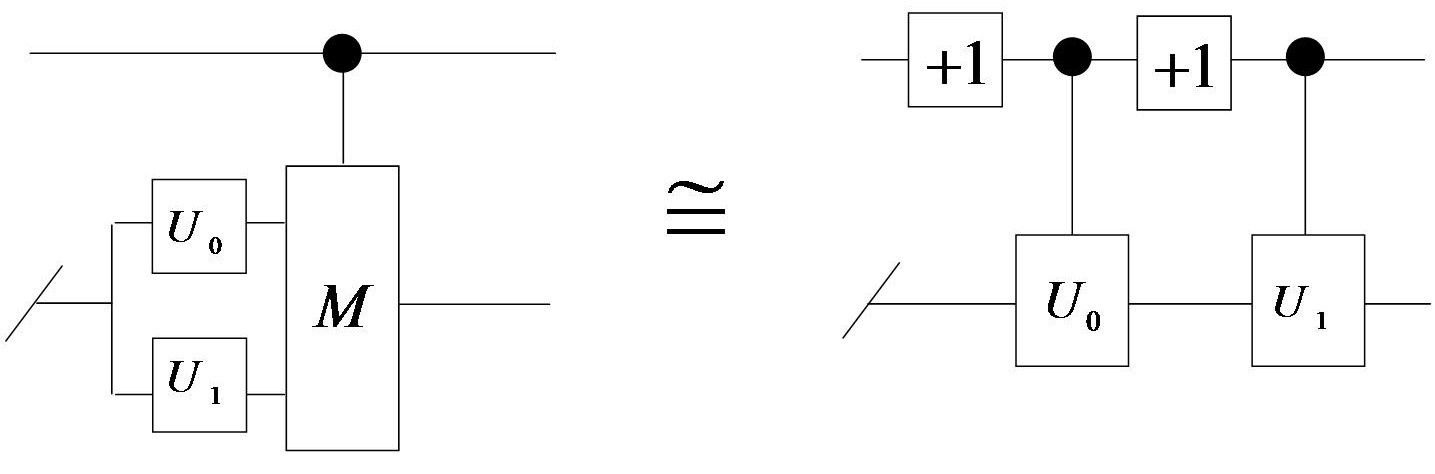}}
\caption{\small{2-valued Quantum Multiplexer $M$ controlling the lower $(n-1)$ qubits by the top qubit. The slash symbol (/) represents $(n-1)$ qubits on the second wire. The gates labeled +1 are shifters (inverters in 2-valued logic), increasing the value of the qubit by 1 mod 2 thereby allowing for control by the highest qubit value. Depending on the value of the top qubit, one of $U_{t}$ is applied to the lower qubits for $t\in\left\{0,1\right\}$.}}
\label{fig:2-valued QMUX}
\end{figure}

A {\it quantum multiplexer} is a quantum logic gate acting on $n$ qubits of which one is designated as the control qubit. If the control qubit of a quantum multiplexer is the lowest order qubit, that is, the first qubit in the joint state of $n$ qubits, the multiplexer matrix is block diagonal. Note that the lowest order qubit is represented as the top most qubit in circuit diagrams. Thus, in terms of synthesis, the block diagonal matrices in Equation (\ref{eqn:CSD2}) are quantum multiplexers \cite{shende:05}. Now, depending on whether the control qubit carries $\ket{0}$ or $\ket{1}$, the gate then performs either the top left block or the bottom right block of the $n \times n$ block diagonal matrix on the remaining $(n-1)$ qubits, respectively. A circuit diagram for a $n$ qubit quantum multiplexer with the lowest order control qubit is given in Figure \ref{fig:2-valued QMUX} where the black circle represents control via the basis state $\ket{1}$. 

Such a quantum multiplexer is expressed as
\begin{equation}\label{eqn:n qubit qmux}
\ket{a_{1}} \otimes \left(\begin{array}{cc} U_{0} & 0 \\ 0 & U_{1}
\end{array}\right)
\left(\ket{a_{2}} \otimes \dots \otimes \ket{a_{n}}\right)
\end{equation}
where $\ket{a_{i}}$ is the $i$-th qubit in the circuit, and both block matrices $U_{0}$ and $U_{1}$ are of size $2^{n-1} \times 2^{n-1}$. Depending on whether $\ket{a_{1}}=\ket{0}$ or $\ket{a_{1}}=\ket{1}$, the expression (\ref{eqn:n qubit qmux}) reduces to 
\begin{equation}\label{eqn:n qubit qmux1}
\ket{0} \otimes 
U_{0}
\left(\ket{a_{2}} \otimes \ket{a_{3}} \otimes \dots \otimes \ket{a_{n}} \right)
\end{equation}
or 
\begin{equation}\label{eqn:n qubit qmux2}
\ket{1} \otimes 
U_{1}
\left(\ket{a_{2}} \otimes \ket{a_{3}} \otimes \dots \otimes \ket{a_{n}}\right)
\end{equation}
respectively.

\begin{figure}
\centerline{\includegraphics[scale=0.27]{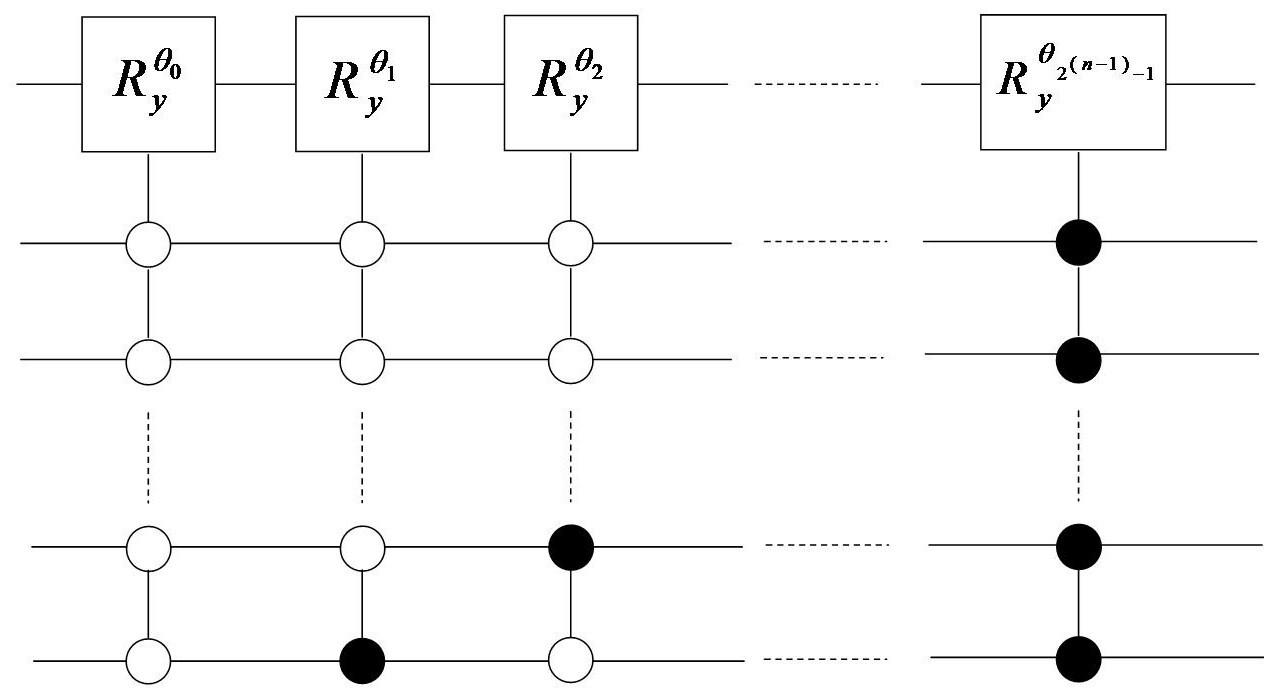}}
\caption{\small{A uniformly $(n-1)$-controlled $R_{y}$ rotation for 2-valued quantum logic. The $\circ$ control turns on for control value $\ket{0}$ and the $\bullet$ control turns on for control value $\ket{1}$. It requires $2^{n-1}$ one qubit controlled gates $R_{y}^{\theta_{i}}$ to implement a uniformly $(n-1)$-controlled $R_{y}$ rotation.}}
\label{fig:(n-1)-unifo control Ry}
\end{figure}

A {\it uniformly $(n-1)$-controlled $R_{y}$ rotation gate $R_{y}$} is composed of a sequence of $(n-1)$-fold controlled gates $R_{y}^{\theta_{i}}$, all acting on the lowest order qubit, where
\begin{equation}\label{eqn:R_{y}2}
R_{y}^{\theta_{i}}  = \left(\begin{array}{cc}
  \cos\theta_{i} & -\sin\theta_{i} \\  \sin\theta_{i} &  \cos\theta_{i}
\end{array}\right).
\end{equation}
The cosine-sine matrix in Equation (\ref{eqn:CSD2}) is realized as a uniformly $(n-1)$-controlled $R_{y}$ rotation gate, a variation of a quantum multiplexer, as shown in Figure \ref{fig:(n-1)-unifo control Ry}.
The control selecting the angle $\theta_{i}$ in the gate $R_{y}^{\theta_{i}}$ depends on which of the $({n-1})$ basis state configurations the control qubits are in at that particular stage in the circuit. In Figure \ref{fig:(n-1)-unifo control Ry}, the white circle represents control via the basis state $\ket{0}$. The $i$-th $(n-1)$-controlled gate $R_{y}^{\theta_{i}}$ may be expressed as
\begin{equation}\label{eqn:n qubit uniformly controlled R_y}
\left(\begin{array}{cc} \cos\theta_{i} & -\sin\theta_{i} \\ \sin\theta_{i} &\cos\theta_{i}
\end{array}\right)
\ket{a_{1}} \otimes \left(\ket{a_{1}} \otimes \dots \otimes \ket{a_{n}}\right)
\end{equation}
with $\theta_{i}$ taking on values from the set $\left\{\theta_{0}, \theta_{1}, \dots, \theta_{2^{n-1}-1}\right\}$ depending on the specific configuration of $\left(\ket{a_{2}} \otimes \dots \otimes \ket{a_{n}}\right)$, resulting in a specific $R_{y}^{\theta_{i}}$ for each $i$.

\begin{figure}
\centerline{\includegraphics[scale=0.23]{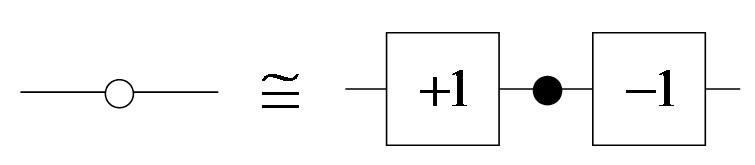}}
\caption{\small{A control by input value 0 (mod 2) realized in terms of control by the highest value 1 (mod 2).}}
\label{fig:0-control}
\end{figure}

\begin{figure}
\centerline{\includegraphics[scale=0.27]{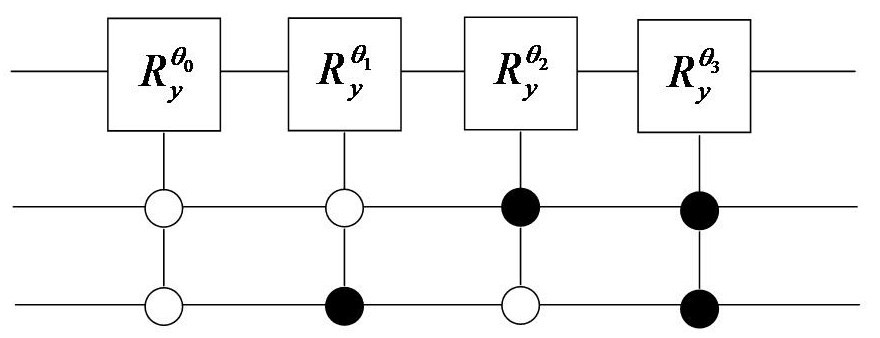}}
\caption{\small{A uniformly 2-controlled $R_{y}$ rotation in 2-valued logic: the lower two qubits are the control qubits and the top bit is the target bit.}}
\label{fig:2-unifo control Ry}
\end{figure}

As an example, consider the 3 qubit uniformly 2-controlled $R_{y}$ gate controlling the top qubit from Figure~\ref{fig:2-unifo control Ry}. Then the action of $R_{y}^{\theta_{i}}$ on the circuit is 
\begin{equation}\label{eqn:n qubit uniformly controlled R_y1}
\left(\begin{array}{cc} \cos\theta_{i} & -\sin\theta_{i} \\ \sin\theta_{i} &\cos\theta_{i}
\end{array}\right)
\ket{a_{1}} \otimes \left(\ket{a_{2}} \otimes \ket{a_{3}}\right)
\end{equation}
with $\theta_{i} \in \left\{\theta_{0}, \theta_{1}, \theta_{2}, \theta_{3}\right\}$. As $\ket{a_{2}} \otimes \ket{a_{3}}$ takes on the values from the set \\ $\left\{\ket{0}\otimes \ket{0}, \ket{0}\otimes \ket{1}, \ket{1}\otimes \ket{0}, \ket{1}\otimes \ket{1}\right\}$ in order, the expression in (\ref{eqn:n qubit uniformly controlled R_y1}) reduces to the following 4 expressions respectively.
\begin{equation}\label{eqn:n qubit uniformly controlled R_y2}
\left(\begin{array}{cc} \cos\theta_{0} & -\sin\theta_{0} \\ \sin\theta_{0} &\cos\theta_{0}
\end{array}\right)
\ket{a_{1}} \otimes
\left(\ket{0} \otimes \ket{0}\right)
\end{equation}
\begin{equation}\label{eqn:n qubit uniformly controlled R_y3}
\left(\begin{array}{cc} \cos\theta_{1} & -\sin\theta_{1} \\ \sin\theta_{1} &\cos\theta_{1}
\end{array}\right)
\ket{a_{1}} \otimes
\left(\ket{0} \otimes \ket{1}\right)
\end{equation}
\begin{equation}\label{eqn:n qubit uniformly controlled R_y4}
\left(\begin{array}{cc} \cos\theta_{2} & -\sin\theta_{2} \\ \sin\theta_{2} &\cos\theta_{2}
\end{array}\right)
\ket{a_{1}} \otimes
\left(\ket{1} \otimes \ket{0}\right)
\end{equation}
\begin{equation}\label{eqn:n qubit uniformly controlled R_y5}
\left(\begin{array}{cc} \cos\theta_{3} & -\sin\theta_{3} \\ \sin\theta_{3} &\cos\theta_{3}
\end{array}\right)
\ket{a_{1}} \otimes
\left(\ket{1} \otimes \ket{1}\right)
\end{equation}

Observe that by iterating the CSD and factoring the result each time results in a quantum circuit consisting of variations of the quantum multiplexer.

\section{CSD Synthesis of 3-valued (ternary) Quantum Logic Circuits}\label{sect:3-valued CSD}
In the 3-valued case, two applications of the CSD are needed to decompose a $3^{n} \times 3^{n}$ unitary matrix $W$ to the point where every block in the decomposition has size $3^{n-1} \times 3^{n-1}$~\cite{FSK:05}. Choose the parameters $m$ and $r$ given in Equation (\ref{eqn:CSD matrix}) as $m=3^{n}$ and $r=3^{n-1}$, so that $m-r = 3^{n}-3^{n-1}=3^{n-1}(3-1)=3^{n-1}\cdot2$. The CS decomposition of $W$ will now take the form in Equation  (\ref{eqn:CSD1}), with the matrix blocks $U$ and $X$ of size $3^{n-1} \times 3^{n-1}$ and blocks $V$ and $Y$ of size $3^{n-1}\cdot2 \times 3^{n-1}\cdot2 $. Repeating the partitioning process for the blocks $V$ and $Y$ with $m=3^{n-1}\cdot 2$ and $r=3^{n-1}$, and decomposing them with CSD followed by some matrix factoring will give rise to a decomposition of $W$ involving unitary blocks each of size $3^{n-1}$ as follows.

\begin{figure}
\centerline{\includegraphics[scale=0.23]{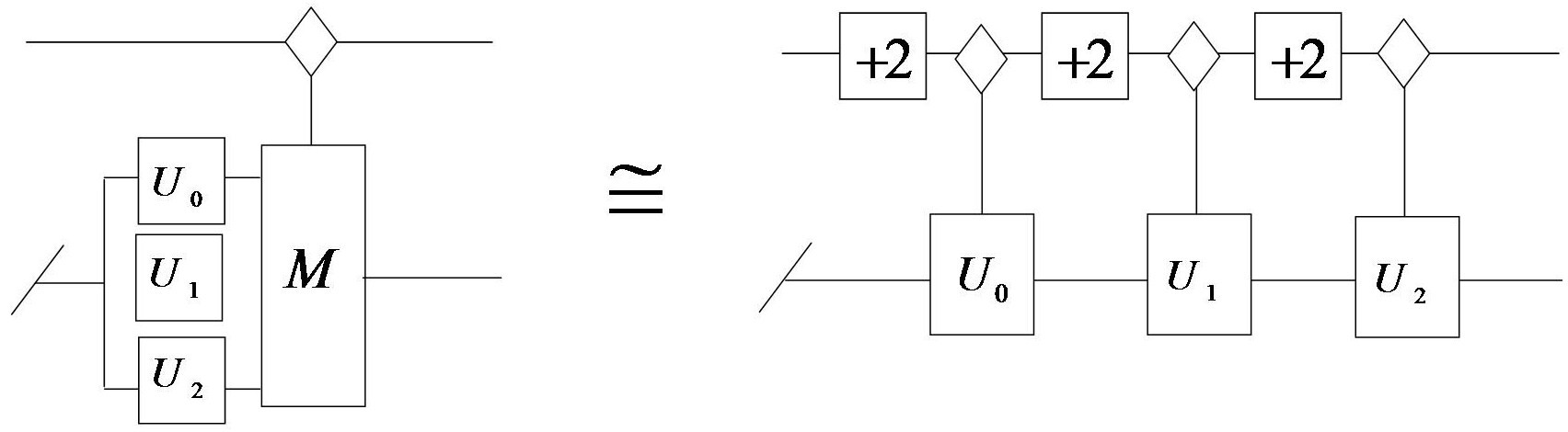}}
\caption{\small{3-valued Quantum Multiplexer $M$ controlling the lower $(n-1)$ qutrits via the top qutrit. The slash symbol (/) represents $(n-1)$ qutrits on the second wire. The gates labeled +2 are shift gates, increasing the value of the qutrit by 2 mod 3, and the control $\diamondsuit$ turns on for input $\ket{2}$. Depending on the value of the top qutrit, one of $U_{t}$ is applied to the lower qutrits for $t\in\left\{0,1,2\right\}$.}}\label{fig:3-valued QMUX}
\end{figure}
\begin{equation}\label{eqn:Ternary Decomposed CSD}
W=
ABC
\left(\begin{array}{ccc}
  C & -S & 0 \\ S & C & 0 \\  0 & 0 & I
\end{array}\right)
DEF
\end{equation}
with
\begin{equation}\label{eqn:ABC}
A =
\left(\begin{array}{ccc}
  X_{1} & 0 & 0 \\ 0 & X_{2} & 0 \\  0 & 0 & X_{3}
\end{array}\right),\hspace{0.25in}
B= \left(\begin{array}{ccc}
  I & 0 & 0 \\ 0 & C_{1} & -S_{1} \\  0 & S_{1} & C_{1}
\end{array}\right), \hspace{0.25in}
C= \left(\begin{array}{ccc}
  I & 0 & 0 \\ 0 & Z_{1} & 0 \\  0 & 0 & Z_{2}
\end{array}\right)
\end{equation}
\begin{equation}\label{eqn:DEF}
D= \left(\begin{array}{ccc}
Y_{1} & 0 & 0 \\ 0 & Y_{2} & 0 \\ 0 & 0 & Y_{3}
\end{array}\right), \hspace{0.25in}
E= \left(\begin{array}{ccc}
   I & 0 & 0 \\ 0 & C_{2} & -S_{2} \\  0 & S_{2} & C_{2}
\end{array}\right), \hspace{0.25in}
F= \left(\begin{array}{ccc}
 I & 0 & 0 \\ 0 & W_{1} & 0 \\  0 & 0 & W_{2}
\end{array}\right)
\end{equation}

\begin{figure}
\centerline{\includegraphics[scale=0.23]{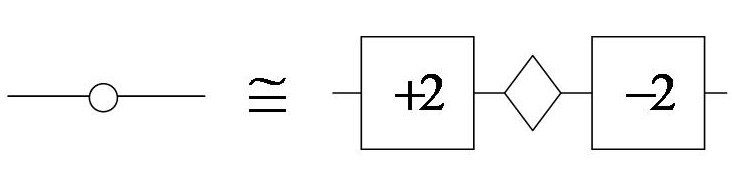}}
\caption{\small{A control by the value 0 (mod 3) realized in terms of control by the highest value 2 (mod 3).}}\label{fig:0-control mod 3}
\end{figure}

\begin{figure}
\centerline{\includegraphics[scale=0.23]{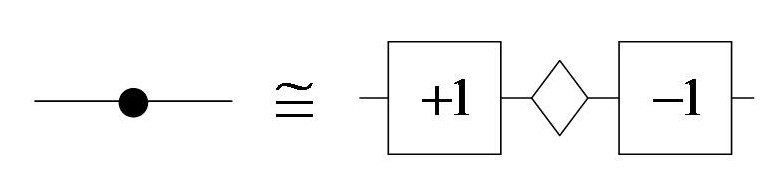}}
\caption{\small{A control by the value 1 (mod 3) realized in terms of control by the highest value 2 (mod 3).}}\label{fig:1-control mod 3}
\end{figure}

We realize the block diagonal matrices $A,C,D$ and $F$ in (\ref{eqn:ABC}) and (\ref{eqn:DEF}) as 3-valued quantum multiplexers acting on $n$ qutrits of which the lowest order qutrit (top most in a circuit diagram) is designated as the control qutrit. Depending on which of the values $\ket{0}$, $\ket{1}$, or $\ket{2}$ the control qutrit carries, the gate then performs either the top left block, the middle block, or the bottom right block respectively on the remaining $n-1$ qutrits. Figure \ref{fig:3-valued QMUX} gives the layout for a $n$ qutrit quantum multiplexer realized in terms of \emph{Muthukrishnan-Stroud} (MS) gates. The MS gate is a $d$-valued generalization of a controlled gate from 2-valued quantum logic, and allows for control of one qudit by the other via the highest value of a $d$-valued quantum system, which in the 3-valued case is 2 \cite{MuthuStroud:04}.

\begin{figure}
\centerline{\includegraphics[scale=0.23]{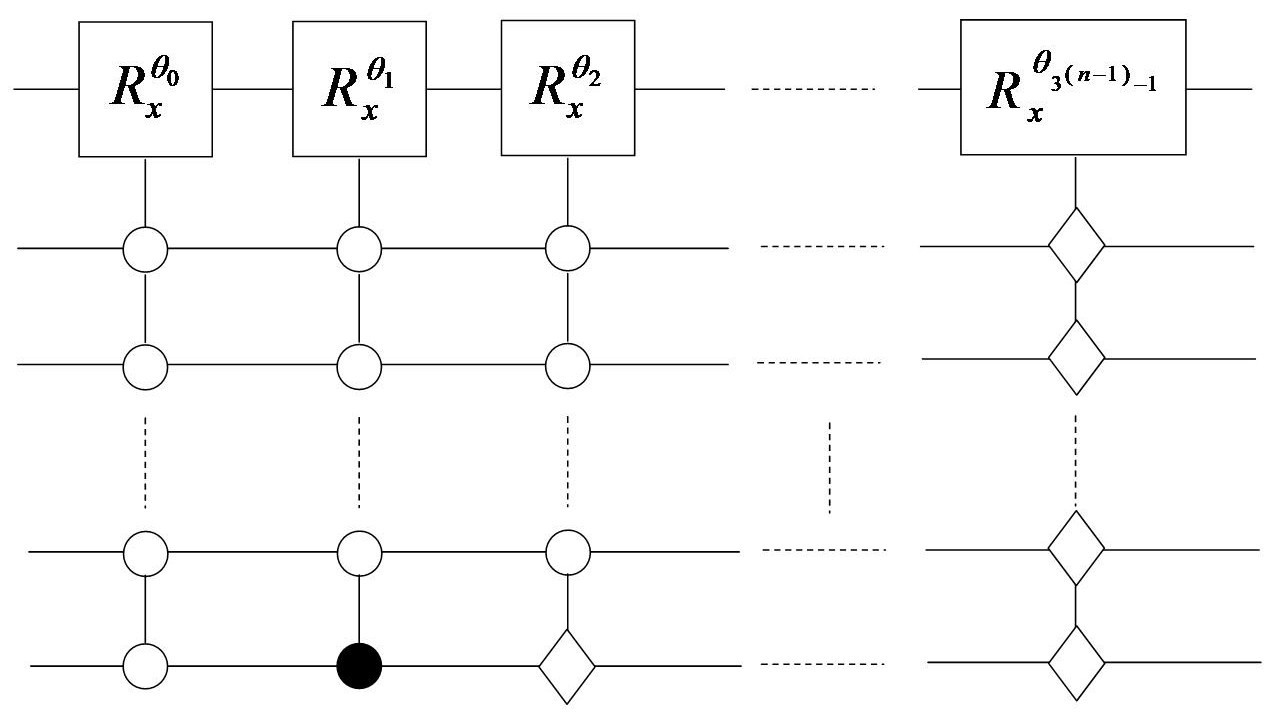}}
\caption{\small{A uniformly $(n-1)$-controlled $R_{x}$ rotation. The lower $(n-1)$ qutrits are the control qutrits. The controls $\circ$, $\bullet$, and $\diamond$ turn on for inputs $\ket{0}$, $\ket{1}$, and $\ket{2}$ respectively. It requires $3^{n-1}$ one qutrit controlled gates to implement a uniformly $(n-1)$-controlled $R_{x}$ or $R_{z}$ rotation. }}\label{fig:(n-1)-unifo control Rx}
\end{figure}

The cosine-sine matrices are realized as the uniformly $(n-1)$-controlled $R_{x}$ and $R_{z}$ rotations in $\mathbb{R}^{3}$. Similar to the 2-valued case, each $R_{x}$ and $R_{z}$ rotation is composed of a sequence of $(n-1)$-fold controlled gates $R^{\theta_{i}}_{x}$ or $R^{\phi_{i}}_{z}$, where 
\begin{equation}\label{eqn:R_{x}}
R^{\theta_{i}}_{x} = \left(\begin{array}{ccc}
 1 & 0 & 0 \\ 0 & \cos\theta_{i} & -\sin\theta_{i}  \\ 0 &  \sin\theta_{i}&  \cos\theta_{i} \\
\end{array}\right),\hspace{0.25in}
R^{\phi_{i}}_{z} = \left(\begin{array}{ccc}
\cos\phi_{i} & -\sin\phi_{i}  & 0 \\  \sin\phi_{i} &  \cos\phi_{i} & 0 \\ 0 & 0 & 1
\end{array}\right).
\end{equation}
Each $R^{\theta_{i}}_{x}$ or $R^{\phi_{i}}_{z}$ operator is applied to the top most qutrit, with the value of the angles $\theta_{i}$ and $\phi_{i}$ determined by the $(n-1)$ basis state configurations of the control qutrits. A uniformly controlled $R_{x}$ gate is shown in Figure \ref{fig:(n-1)-unifo control Rx}. Figures \ref{fig:0-control mod 3} and~\ref{fig:1-control mod 3} explain the method to create controls of maximum value. Notet that the value of the control qubit is always restored in Figures \ref{fig:0-control mod 3} and~\ref{fig:1-control mod 3}.

\section{Synthesis of Hybrid and $d$-valued Quantum Logic Circuits}\label{sect:hybrid}
It is evident from the 2 and 3-valued cases above that the CSD method of synthesis is of a general nature and can be extended to synthesis of $d$-valued gates acting on $n$ qudits. In fact, it can be generalized for synthesis of hybrid $n$ qudit gates. We propose that a $(d_{1}d_{2}\dots d_{n}) \times (d_{1}d_{2}\dots d_{n})$ block diagonal unitary matrix be regarded as a quantum multiplexer for an $n$ qudit hybrid quantum state space $\mathcal{H}=\mathcal{H}_{d_{1}} \otimes \mathcal{H}_{d_{2}} \otimes \dots \otimes \mathcal{H}_{d_{n}}$, where $\mathcal{H}_{d_{i}}$ is the state space of the $i$ qudit.

Moreover, consider a cosine-sine matrix of size $(d_{1}d_{2}\dots d_{n}) \times (d_{1}d_{2}\dots d_{n})$ of the form
\begin{equation}
\left(\begin{array}{cccc} I_{p} & 0 & 0 & 0\\ 0 & C & -S
& 0
\\ 0 & S & C  & 0 \\ 0 & 0 & 0 & I_{q}
\end{array}\right)
\end{equation}
with $I_{p}$ and $I_{q}$ both some appropriate sized identity matrices, $C$ = diag$(\cos \theta_{1}, \\ \cos\theta_{2}, \dots, \cos\theta_{t})$ and $S$ = diag$(\sin \theta_{1}$, $\sin \theta_{2},\dots,\sin \theta_{t})$ such that $\sin^{2}\theta_{i}+\cos^{2}\theta_{i}=1$ for some $\theta_{i}$ with $1 \leq i \leq t$, and $p+q+2t=(d_{1}d_{2}\dots d_{n})$. We regard this matrix as a \emph{uniformly controlled Givens rotation} matrix, a generalization of the $R_{y}$, $R_{x}$, and $R_{z}$ rotations of the 2 and 3-valued cases. A Givens rotation matrix has the general form
\begin{equation}\label{eqn:givens matrix}
G_{(i,j)}^{\theta} = \left(\begin{array}{ccccccc}
1 & \dots & 0 & \dots & 0 & \dots & 0 \\
\vdots & \ddots & \vdots &  & \vdots & & \vdots \\
0 & \dots & \cos\theta & \dots & -\sin\theta & \dots & 0 \\
\vdots &  & \vdots  & \ddots &  \vdots &  & \vdots \\
0 & \dots & \sin\theta & \dots & \cos\theta & \dots & 0 \\
\vdots & & \vdots & & \vdots & \ddots & \vdots \\
0 & \dots & 0 & \dots & 0 & \dots & 1\\
\end{array}\right)
\end{equation}
where the cosine and sine values reside in the intersection of the $i$-th and $j$-th rows and columns, and all other diagonal entries are 1~\cite{Golub:89}. Hence, a Givens rotation matrix corresponds to a rotation by some angle $\theta$ in the $ij$-th hyperplane.

Based on the preceding discussion, we give in Theorem \ref{thm1} below an iterative CSD method for synthesizing a $n$ qudit hybrid quantum circuit by decomposing the corresponding unitary matrix of size $(d_{1}d_{2}\dots d_{n}) \times (d_{1}d_{2}\dots d_{n})$ in terms of quantum multiplexers and uniformly controlled Givens rotations. As a consequence of Theorem \ref{thm1}, we give in corollary \ref{coro1} a CSD synthesis of a quantum quantum logic circuit with corresponding unitary matrix of size $d^{n} \times d^{n}$. The synthesis methods given above for 2-valued and 3-valued circuits may then be treated as special cases of the former.

\subsection{Hybrid Quantum Logic Circuits}
Consider a hybrid quantum state space of a $n$ qudits, $\mathcal{H}=\mathcal{H}_{d_{1}} \otimes \mathcal{H}_{d_{2}} \otimes \dots
\otimes \mathcal{H}_{d_{n}}$, where each qudit may be of distinct $d$-valued dimension $d_{i}$, $1 \leq i \leq n$. Since a qudit in $\mathcal{H}$ is a column vector of length $d_{1}d_{2} \ldots d_{n}$, a quantum logic gate acting on such a vector is a $(d_{1}d_{2} \ldots d_{N}) \times (d_{1}d_{2} \ldots d_{n})$ unitary matrix $W$. We will decompose $W$, using CSD iteratively, from the level of $n$ qudits to $(n-1)$ qudits in terms of quantum multiplexers and uniformly controlled Givens rotations. However, since the $d$-valued dimension may be different for each qudit, the block matrices resulting from the CS decomposition may not be of the form $d^{n-1} \times d^{n-1}$ for some $d$. Therefore, we proceed by choosing one of the qudits, $c_{d_{i}}$ of dimension $d_{i}$, to be the control qudit and order of the basis of $\mathcal{H}$ in such a way that $c_{d_{i}}$ is the highest order qudit. We will decompose $W$ with respect to $c_{d_{i}}$ so that the resulting quantum multiplexers are controlled by $c_{d_{i}}$ and the uniformly controlled Givens rotations control $c_{d_{i}}$ via the remaining $(n-1)$ qudits. We give the synthesis method in Theorem \ref{thm1} below.

\begin{theorem}\label{thm1}{\rm
Let $W$ be an $M \times M$ unitary matrix, with $M=d_{1}d_{2}\ldots d_{n}$, acting as a quantum logic gate on a quantum hybrid state space $\mathcal{H}=\mathcal{H}_{d_{1}} \otimes \mathcal{H}_{d_{2}} \otimes \dots \otimes \mathcal{H}_{d_{n}}$ of $n$ qudits. Then $W$ can be synthesized with respect to a control qudit $c_{d_{i}}$ of dimension $d_{i}$, having the highest order in $\mathcal{H}$, iteratively from level $n$ to level $(n-1)$ in terms of quantum multiplexers and uniformly controlled Givens rotations.}
\end{theorem} 

\begin{proof}
\textbf{Step 1.} At level $n$, identify a control qudit $c_{d_{i}}$ of dimension $d_{i}$. Reorder the basis of $\mathcal{H}$ so that $c_{d_{i}}$ is the highest order qudit and the new state space isomorphic to $\mathcal{H}$ is $\mathcal{\bar{H}}=\mathcal{H}_{d_{i}} \otimes \mathcal{H}_{d_{2}} \otimes \dots
\otimes \mathcal{H}_{d_{1}} \otimes \dots \otimes \mathcal{H}_{d_{n}}$.

If we choose values for the CSD parameters $m$ and $r$ as $m=\left(d_{1}d_{2}\dots d_{n}\right)$ and $r=\left(d_{1}d_{2}\ldots d_{i-1}d_{i+1}\dots d_{n}\right)$, then $m-r=d_{1}\ldots d_{i-1}d_{i+1}\dots d_{n}(d_{i}-1)$. Decomposing $W$ by CSD, we get the form in (\ref{eqn:CSD1}) with the matrix blocks  $U$ and $X$ of size $r \times r$ and blocks $V$ and $Y$ of size $(m-r) \times (m-r)$. Should $m-r$ not have the factor $(d_i-1)$, we would achieve the desired decomposition of $W$ from level of $n$ qudits to the level of $(n-1)$ qudits in terms of block matrices of size $r \times r$. The task therefore is to divide out the factor $(d_{i}-1)$ from $m-r$ by an iterative \emph{lateral decomposition} described below, that uses the CSD to cancel $(d_{i}-1)$ from $m-r$ at each iteration level leaving only blocks of size $r \times r$. 

For step 2 of the proof below, we will say that a matrix with $k$ rows and $k$ columns has size $k$ instead of $k \times k$. 

\textbf{Step 2.} \emph{Iterative Lateral Decomposition}: For the unitary matrix $W$ of size $M$, we define the $j$-th lateral decomposition of $W$ as the CS decomposition of all block matrices of size other than $r$ that result from the $(j-1)$-st lateral decomposition of $W$:

\emph{
For $0\leq j \leq (d_{i}-2)$, set \\
$m_{0}=\left(d_{1}d_{2} \ldots d_{n}\right)$ \\
$r_{0}=\left(d_{1}d_{2}\ldots d_{i-1}d_{i+1}\dots d_{n}\right)$ \\
If $j=0$ \\
Apply CSD to W\\
Else set \\
$m_{j}=m_{0}-j\cdot r_{0}$ \\
$r_{j}=r_{0}$ \\
$m_{j}-r_{j}=m_{0}-(j+1)r_{0}$\\
=$\left(d_{1}d_{2}\ldots d_{i-1}d_{i+1}\dots d_{n}\right)[d_{i}-(j+1)]$\\
$m_{j}-2r_{j}=m_{0}-(j+2)r_{0}$\\
=$\left(d_{1}d_{2}\ldots d_{i-1}d_{i+1}\dots d_{n}\right)[d_{i}-(j+2)]$\\
Apply CSD to matrix blocks of size other than $r_{0}$ from step $j-1$\\
End If\\
End For}.

When $j=0$, we call the resulting 0-th lateral decomposition the \emph{global decomposition}. Note that if $d_{i}=2$, then the algorithm for the lateral decomposition stops after the global decomposition. This suggests that whenever feasible, the control system in the quantum circuit should be 2-valued so as to reduce the number of iterations . Below we give a matrix description of the algorithm.

For $j=0$, the $0$-th lateral decomposition of $W$ will just be the CS decomposition of $W$.
\begin{equation}\label{eqn:lateral1}
 W=A_{0}^{(0)}B_{0}^{(0)}D_{0}^{(0)}            
\end{equation}
where
$$
A_{0}^{(0)}=\left(\begin{array}{cc}
 U^{(0)}_{0} & 0 \\ 0 & V^{(0)}_{0}
\end{array}\right),
B_{0}^{(0)}=\left(\begin{array}{ccc}
  C^{(0)}_{0} & -S^{(0)}_{0} & 0 \\ S^{(0)}_{0} & C^{(0)}_{0} & 0 \\ 0 & 0 & I_{m_{0}-2r_{0}}
\end{array}\right)
D_{0}^{(0)}=\left(\begin{array}{cc}
  X^{(0)}_{0} & 0 \\  0 & Y^{(0)}_{0}
\end{array}\right)
$$
with $U^{(0)}_{0}$, $X^{(0)}_{0}$, $C^{(0)}_{0}$, and $S^{(0)}_{0}$ all of the
desired size $r_{0}$, while $V^{(0)}_{0}$ and $Y^{(0)}_{0}$ are of size
$m_{0}-r_{0}$. The superscripts label the iteration step, in this case $j=0$. The
subscript is used to distinguish between the various matrix blocks $U, V, X, Y, C,
S$, that occur at the various levels of iteration. The 0-th lateral decomposition in the form from Equation 
(\ref{eqn:lateral1}) is called the \emph{global decomposition} of $W$. 

For $j=1$, we perform lateral decomposition on the blocks $V^{(0)}_{0}$ and $Y^{(0)}_{0}$ of the block matrices $A_{0}^{(0)}$ and $D_{0}^{(0)}$ respectively, the only blocks of size other than $r_{0}$ resulting from the $0$-th lateral decomposition given in (\ref{eqn:lateral1}). In both cases, set $ m_{1}=m_{0}-r_{0}$ and $r_{1}=r_{0}$ so that $m_{1}-r_{1}=m_{0}-2r_{0}$. For $V^{(0)}_{0}$ this gives the decomposition
\begin{equation}\label{eqn:lateral2}
A_{0}^{(0)}=\left(\begin{array}{cc} U^{(0)}_{0} & 0 \\ 0 &
\left(\begin{array}{cc} U^{(1)}_{0} & 0 \\
0 & V^{(1)}_{0} \end{array}\right)
\left(\begin{array}{ccc} C_{0}^{(1)} & -S_{0}^{(1)} & 0 \\ S_{0}^{(1)} & C_{0}^{(1)}  & 0 \\
0 & 0 & I_{m_{0}-3r_{0}} \end{array}\right) \left(\begin{array}{cc} X^{(1)}_{0} &
0 \\ 0 & Y^{(1)}_{0}
\end{array}\right)
\end{array}\right)
\end{equation}
with $U^{(1)}_{0}, X^{(1)}_{0}, C_{0}^{(1)}$ and $S_{0}^{(1)}$ all of size
$r_{0}$, and $V^{(1)}_{0}$ and $Y^{(1)}_{0}$ of size $m_{1}-r_{1}$. All three matrices residing in the lower block diagonal of the matrix (\ref{eqn:lateral2}) are the same size. Therefore, by introducing identity matrices of size $r_{0}$ and factoring out at the matrix block level, $A_{0}^{(0)}$ will be updated to
\begin{equation}
A_{0}^{(0)}=A_{0}^{(1)}B_{0}^{(1)}D_{0}^{(1)}
\end{equation}
where
$$
A_{0}^{(1)}=\left(\begin{array}{ccc} U^{(0)}_{0} & 0 & 0 \\ 0 & U^{(1)}_{0} &  0 \\ 0 & 0 &
V^{(1)}_{0}
\end{array}\right),
B_{0}^{(1)}=\left(\begin{array}{cccc} I_{r_{0}} & 0 & 0 & 0\\ 0 & C_{0}^{(1)} & -S_{0}^{(1)}
& 0
\\ 0 & S_{0}^{(1)} & C_{0}^{(1)}  & 0 \\ 0 & 0 & 0 & I_{m_{0}-3r_{0}}
\end{array}\right),
$$
$$
D_{0}^{(1)}=\left(\begin{array}{ccc} I_{r_{0}} & 0 & 0 \\ 0 & X^{(1)}_{0}& 0 \\  0 & 0 &
Y^{(1)}_{0}
\end{array}\right)
$$
A similar lateral decomposition of the block $Y^{(0)}_{0}$ will update $D_{0}^{(0)}$ in (\ref{eqn:lateral1}) to
\begin{equation}
C_{0}^{(0)}=A_{1}^{(1)}B_{1}^{(1)}D_{1}^{(1)}
\end{equation}
where
$$
A_{1}^{(1)}=\left(\begin{array}{ccc} X^{(0)}_{0} & 0 & 0 \\ 0 & U^{(1)}_{1} &  0 \\ 0 & 0 &
V^{(1)}_{1} \end{array}\right),
B_{1}^{(1)}=\left(\begin{array}{cccc} I_{r_{0}} & 0 & 0 & 0\\ 0 & C_{1}^{(1)} & -S_{1}^{(1)}
& 0
\\ 0 & S_{1}^{(1)} & C_{1}^{(1)}  & 0 \\ 0 & 0 & 0 & I_{m_{0}-3r_{0}}
\end{array}\right),
$$
$$
D_{1}^{(1)}=\left(\begin{array}{ccc} I_{r_{0}} & 0 & 0 \\ 0 & X^{(1)}_{1} & 0 \\  0 & 0 &
Y^{(1)}_{1}
\end{array}\right)
$$
For iteration $j \neq 0$, perform lateral decomposition on the total
$2^{j}$ blocks $V^{(j-1)}_{k}$, $Y^{(j-1)}_{k}$, where $0 \leq k \leq 2^{(j-1)}-1$,
that occur in the global decomposition at the end of iteration $(j-1)$. For each
$V^{(j-1)}_{k}$, $Y^{(j-1)}_{k}$, set $r_{j}=r_{0}$, $m_{j}=m_{j-1}-r_{j-1}=m_{0}-jr_{0}$.
For each $V^{(j-1)}_{k}$, the lateral decomposition at level $j$ will give the following
\small{\begin{equation}\label{eqn:lateral3}
A_{k}^{(j-1)}=\left(\begin{array}{cc} \Delta^{(j-1)} & 0 \\ 0 &
\left(\begin{array}{cc} U^{(j)}_{k^{\prime}} & 0 \\
0 & V^{(j)}_{k^{\prime}} \end{array}\right)
\left(\begin{array}{ccc} C^{(j)}_{k^{\prime}} & -S^{(j)}_{k^{\prime}} & 0 \\ S^{(j)}_{k^{\prime}}& C^{(j)}_{k^{\prime}}
& 0 \\
0 & 0 & I_{m_{0}-(j+2)r_{0}} \end{array}\right) \left(\begin{array}{cc}
X^{(j)}_{k^{\prime}}& 0 \\ 0 & Y^{(j)}_{k^{\prime}}
\end{array}\right)
\end{array}\right)
\end{equation}}
{\normalsize where the $\Delta^{(j-1)}$ is the block diagonal matrix of size of $j\cdot
r_{0}$ arising from the lateral decomposition in the previous $j$ steps. The
blocks $U^{(j)}_{k^{\prime}}$, $X^{(j)}_{k^{\prime}}$, $C^{(j)}_{k^{\prime}}$ and
$S^{(j)}_{k^{\prime}}$ are all of size $r_{0}$, $0 \leq k^{\prime} \leq 2^{j}-1$.
The blocks $V^{(j)}_{k^{\prime}}$ and $Y^{(j)}_{k^{\prime}}$ are of size
$m_{j}-r_{j}$. The three matrices residing in the lower block diagonal of the matrix (\ref{eqn:lateral3}) are all of same size. Therefore, by introducing identity matrices of size $j\cdot r_{0}$ and factoring out at the block level, $A_{k}^{(j-1)}$ will be updated to
$$
A_{k}^{(j-1)}=A_{k^{\prime}}^{(j)}B_{k^{\prime}}^{(j)}D_{k^{\prime}}^{(j)}
$$
where
$$
A_{k}^{(j-1)}=\left(\begin{array}{ccc} \Delta^{(j-1)} & 0 & 0 \\ 0 & U^{(j)}_{k^{\prime}} &  0 \\
0 & 0 & V^{(j)}_{k^{\prime}}
\end{array}\right),
B_{k^{\prime}}^{(j)}=\left(\begin{array}{cccc} I_{j\cdot r_{0}} & 0 & 0 & 0\\ 0 &
C^{(j)}_{k^{\prime}} &
 -S^{(j)}_{k^{\prime}} & 0
\\ 0 &  S^{(j)}_{k^{\prime}} & C^{(j)}_{k^{\prime}}  & 0 \\ 0 & 0 & 0 & I_{m_{0}-(j+2)r_{0}}
\end{array}\right)
$$
$$
D_{k^{\prime}}^{(j)}=\left(\begin{array}{ccc} I_{j\cdot r_{0}} & 0 & 0 \\ 0 & X^{(j)}_{k^{\prime}} & 0 \\
0 & 0 & Y^{(j)}_{k^{\prime}}
\end{array}\right)
$$
For the next iteration, set $k=k^{\prime}$ and iterate. Upon completion of the lateral decomposition, repeat steps 1 and 2 for the synthesis of the circuit for the remaining $(n-1)$ qudits, with the restriction that each gate in the remaining circuit be decomposed with respect to the same control qudit identified in step 1.} 
\end{proof}

\begin{figure}
\centerline{\includegraphics[scale=0.19]{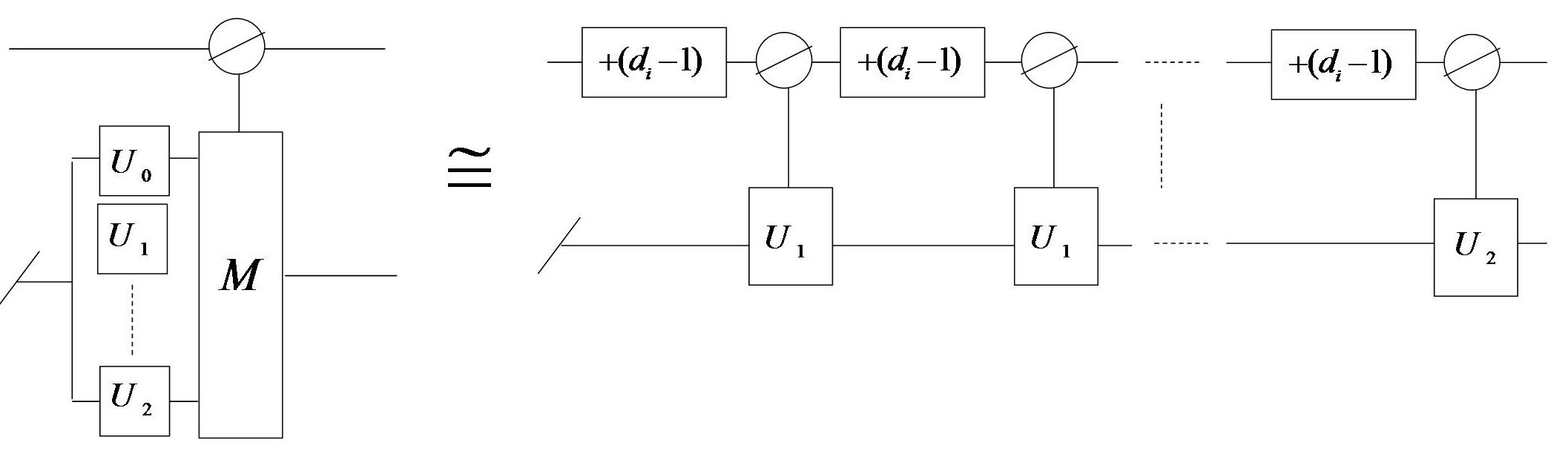}}
\caption{\small{An $n$ qudit hybrid quantum multiplexer, here realized in terms of Muthukrishnan-Stroud ($d$-valued controlled) gates. The top qudit has dimension $d_{i}$ and controls the remaining $(n-1)$ qudits of possibly distinct dimensions which are represented here by the symbol (/). The control $\oslash$ turns on for input value $\ket{d_{i}-1}$ mod $d_{i}$ of the controlling signal coming from the top qudit.The gates $+(d_{i}-1)$ shift the values of control qudit by $(d_{i}-1)$ mod $d_{i}$.}}\label{fig:hybrid QMUX}
\end{figure}

\begin{figure}
\centerline{\includegraphics[scale=0.23]{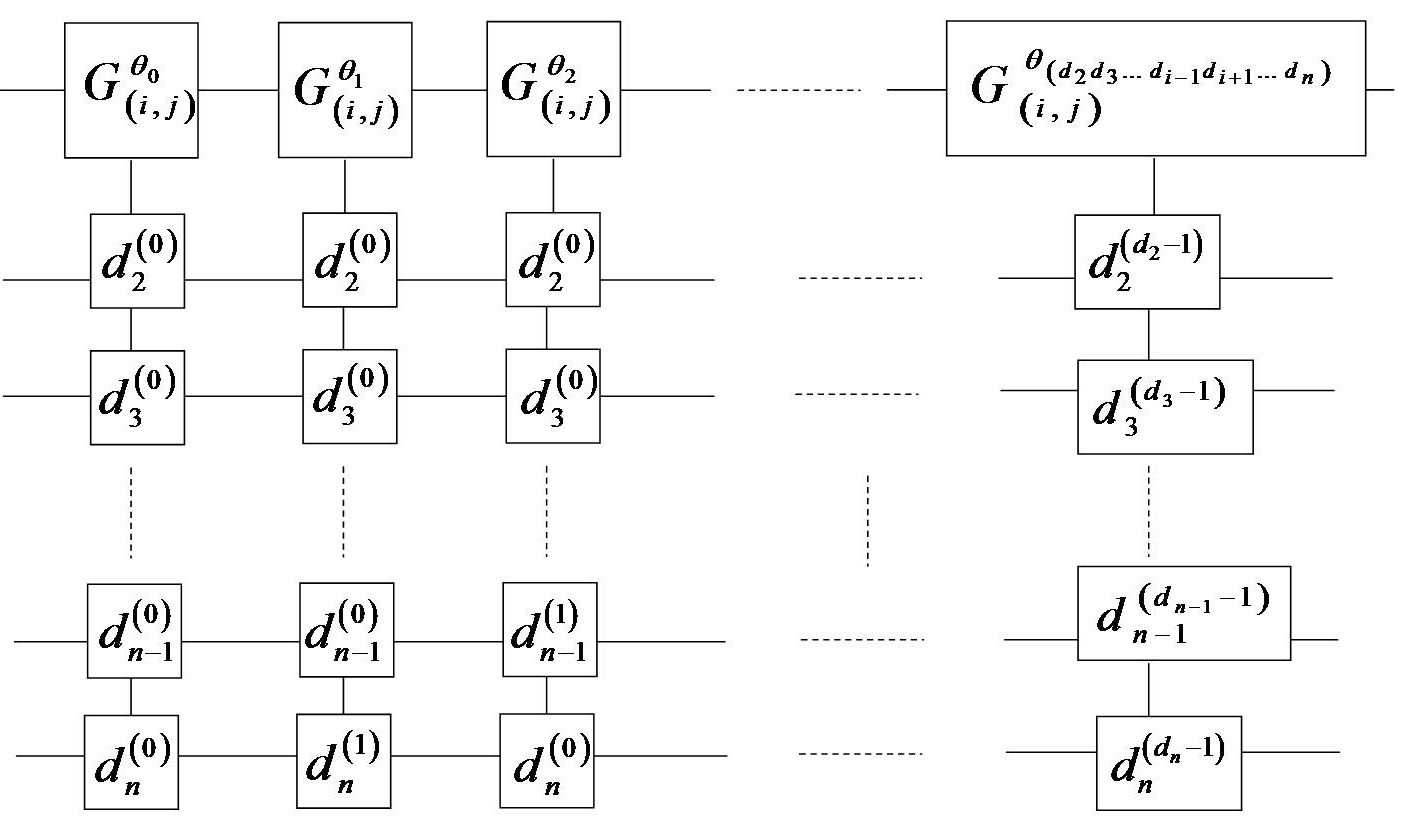}}
\caption{\small{A hybrid uniformly $(n-1)$-controlled Givens rotation. The lower $(n-1)$ qudits of dimensions $d_{2},d_{3}, \dots , d_{i-1},d_{i+1},\dots , d_{n}$, respectively, are the control qudits, and the top is the target qudit of dimension $d_{i}$. The control gate $d_{l}^{(k)}$ turns on whenever the control qudit of dimension $d_{l}$ takes on the value $k$ (mod) $d_{l}$.}}\label{fig:hybrid (n-1)-unifo control Givens}
\end{figure}

Since the basis for $\mathcal{H}$ was reordered in the beginning so that the control qudit was of the highest order, the block diagonal matrices with all blocks of size $r_{0} \times r_{0}$ are interpreted as quantum multiplexers and the cosine-sine matrices are interpreted as uniformly controlled Givens rotations. In Figures \ref{fig:hybrid QMUX} and \ref{fig:hybrid (n-1)-unifo control Givens}, we present the circuit diagrams of a hybrid quantum multiplexer and a uniformly controlled Givens rotation, respectively. A uniformly controlled Givens rotation matrix on $n$ qudits can be realized as the composition of various $(n-1)$-fold controlled Givens rotation matrices, $G_{(i,j)}^{\theta_{k}}$, acting on the top most qudit of the circuit with the angle of rotation depending on the basis state configuration, in their respective dimensions, of the lower $(n-1)$ qudits.

\subsection{$d$-valued Quantum Logic Circuits}\label{sect:d-valued}
Given the hybrid $n$ qudit synthesis, the case of $d$-valued synthesis becomes a special case of the former since by setting all $d_{i}=d$, the state space $\mathcal{H}=\mathcal{H}_{d_{1}} \otimes \mathcal{H}_{d_{2}} \otimes \dots \otimes \mathcal{H}_{d_{n}}$ reduces to the state space $H_{d}^{\otimes n}$. Unitary operators acting on the states in $H_{d}^{\otimes n}$ are unitary matrices of size $d^{n} \times d^{n}$. We give the following result for $d$-valued synthesis.

\begin{corollary}\label{coro1}{\rm
A $d$-valued $n$ qudit quantum logic gate can be synthesized in terms of quantum multiplexers and uniformly controlled Givens rotations.}
\end{corollary}
\noindent \textbf{Proof}: Since all the qudits are of the same dimension, there is no need to choose a control qudit. In the proof of Theorem \ref{thm1}, set $d_{i}=d$ for all $i$. Then $M=d_{1}d_{2}\dots d_{n}=d^{n}$. For iteration $j=0$ of the lateral decomposition, set $m_{0}=d^{n}$, $r_{0}=d^{n-1}$, so that $m_{0}-r_{0}=d^{n-1}(d-1)$. For $0 \leq j \leq (d-2)$, set $r_{j}=r_{0}=d^{n-1}$, and $m_{j}=m_{j-1}-r_{j-1}=d^{n-1}(d-(j+1))$.

For the $d$-valued case, we note that there are a total of $d^{n-1}(2^{d-1}-1)$ one qudit Givens rotations in the circuit at the $(n-1)$ level, each arising from the $\sum_{i=0}^{(d-2)}2^{i}=2^{d-1}-1$ uniformly controlled Givens rotations in the CS decomposition of an $n$ qudit gate. Moreover, in each uniformly controlled Givens rotation, there are $(n-1)d^{n-1}$ control symbols of which $(n-1)d^{n-2}$ correspond to control by the highest value of $d-1$. The latter controls do not require shift gates around them to increase the value of the signal qudit to $d-1$. Hence, there are $(n-1)d^{n-1}-(n-1)d^{n-2}=(n-1)(d^{n-1}-d^{n-2})$ control symbols that correspond to control by values other than $d-1$ and therefore need two shift gates (fig. 11) around them. This gives the total number of one qudit shift gates in each uniformly controlled rotation to be $2(n-1)(d^{n-1}-d^{n-2})$, whereby the total number of one qudit shifts and Givens rotations in the circuit at the $(n-1)$ level is $2(n-1)(d^{n-1}-d^{n-2})(2^{d-1}-1)+d^{n-1}(2^{d-1}-1)=(2^{d-1}-1)\left[2(n-1)(d^{n-1}-d^{n-2})+d^{n-1}\right]$.

There are $2^{d-1}$ quantum multiplexers in the decomposition, each consisting of a total of $2d$ shift and controlled gates. Hence, there are a total of  $d\cdot 2^d$ one qudit and controlled gates in the $(n-1)$ level circuit. This gives a total, worst case, one qudit and controlled gate count in the circuit at level $(n-1)$ to be $(2^{d-1}-1)\left[2(n-1)(d^{n-1}-d^{n-2})+d^{n-1}\right]+ d\cdot 2^d$.

\chapter{A Quaternionic Co-ordinatization of Binary Quantum Computation}\label{Quaterncoord}

A quaternionic coordinatization of the players' quantum strategies in certain quantized games by Landsburg in \cite{Landsburg} gives him a computational framework for classifying potential Nash equilibria in these games. This idea led Ahmed, Bleiler and Khan \cite{Ahmed} to construct a parallel coordinatization using octonions for another class of quantized games, giving the authors a computational framework for classifying potential Nash equilibria in these games. Motivated by these result, this chapter proposes a quaternionic coordinatization of binary quantum computation by putting quaternionic coordinates on the Lie group $SU(2)$ of quantum logic gates acting on one qubit and on the projective complex state space $\mathbb{C}P^1$ of one qubit, with the eventual goal of providing an enhanced computational capability for circuit analysis. 

In general, one qubit quantum logic gates are unitary matrices with determinant $1$ or $-1$. However, a $2 \times 2$ unitary matrix is equivalent to a special unitary matrix up to a factor of $\overline{i}$. That is, if 
$$
U=\left(\begin{array}{cc}
a & b \\ c & d
\end{array}\right)
$$
is a unitary matrix with determinant $ad-cb=-1$, then 
$$
U=\overline{i}U'=\overline{i}\left(\begin{array}{cc}
ia & ib\\ ic & id
\end{array}\right)
$$
where $U'$ has determinant $(ia)(id)-(ic)(ib)=-ad+cb=-1(ad-cb)=(-1)(-1)=1$ and is therefore special unitary. The factor $\overline{i}$ is regarded a unitary phase in any resulting calculations. For the remainder of this chapter, all instances of a unitary matrix $U$ with determinant $-1$ will be replaced with its equivalent special unitary matrix $U' \in SU(2)$. 

Now by identifying both $SU(2)$, the set of one qubit quantum logic gates, and $\mathbb{C}P^1$, the state space of a qubit, with unit quaternions $Sp(1)$, we develop a quaternionic co-ordinatization of binary quantum computation. In this chapter, we will use the notation $\mathds{1}$ and $\mathds{J}$ for the unit quaternions $1$ and $j$ respectively to emphasize their roles as control signals in the context of quantum computing.  
%

\section{Identifying $SU(2)$ with $Sp(1)$}
The Lie group $Sp(1)$ of unit quaternions can be considered as
$$
Sp(1)=\left\{u=u_0\mathds{1}+u_1\mathds{J}:\left|u\right|^2=\left|u_0\right|^2+\left|u_1\right|^2=(u'_0)^2+(u'_1)^2+(u''_0)^2+(u''_1)^2=1\right\}.
$$
The Lie group $SU(2)$ of $2 \times 2$ special unitary matrices is
$$
SU(2)=\left\{\left(\begin{array}{cc}
\alpha & -\overline{\beta} \\ \beta & \overline{\alpha} 
\end{array}\right):\alpha, \beta \in \mathbb{C} \hspace{.05in} {\rm and} \hspace{.05in} \left|\alpha\right|^2+\left|\beta\right|^2=1\right\}
$$
The special unitary requirement suggests a strong connection between $SU(2)$ and $Sp(1)$. Indeed, we can set up a one to one correspondence between $SU(2)$ and $Sp(1)$ as follows.
Consider $\mathbb{H}$ as $\mathbb{C}^2$ under the identification
$$
\alpha\mathds{1}+\beta \mathds{J} \longleftrightarrow \left(\begin{array}{c}
\alpha \\ \beta 
\end{array}\right)
$$
and let $y_1\mathds{1}+y_2\mathds{J} \in \mathbb{H}$ and $z_1\mathds{1}+z_2\mathds{J} \in Sp(1)$. Recall that $\alpha\mathds{J}=\mathds{J}\overline{\alpha}$ for all $\alpha \in \mathbb{C}$ and form the product
\begin{align*}
(y_1\mathds{1}+y_2\mathds{J})(z_1\mathds{1}+z_2\mathds{J})&=y_1z_1\mathds{1}+y_2\mathds{J}z_1+y_1z_2\mathds{J}+y_2\mathds{J}z_2\mathds{J}\\
&=(y_1z_1-y_2\overline{z}_2)\mathds{1}+(y_2\overline{z}_1+y_1z_2)\mathds{J}.
\end{align*}
Write this result as an element of $\mathbb{C}^2$ via the identification as
$$
 \left(\begin{array}{c}
y_1z_1-y_2\overline{z}_2 \\ y_2\overline{z}_1+y_1z_2
\end{array}\right).
$$
But 
$$
\left(\begin{array}{c}
y_1z_1-y_2\overline{z}_2 \\ y_2\overline{z}_1+y_1z_2
\end{array}\right) = \left(\begin{array}{cc}
z_1 & -\overline{z}_2 \\ z_2 & \overline{z}_1 
\end{array}\right)\left(\begin{array}{c}
y_1 \\ y_2 
\end{array}\right) 
$$
so the result of the quaternionic product, as an element of $\mathbb{C}^2$, is in the image of the special unitary transformation
$$
\left(\begin{array}{cc}
z_1 & -\overline{z}_2 \\ z_2 & \overline{z}_1 
\end{array}\right)
$$
acting on $\mathbb{C}^2$, establishing the following identification of $SU(2)$ and $Sp(1)$
$$
\left(\begin{array}{cc}
z_1 & -\overline{z}_2 \\ z_2 & \overline{z}_1 
\end{array}\right) \longleftrightarrow z_1\mathds{1} + z_2 \mathds{J}.
$$
In other words, {\it right} multiplication by a unit quaternion in $\mathbb{H}$ corresponds to the action {\it on the left} of the corresponding special unitary matrix on $\mathbb{C}^2$. 

In fact, it is possible to make the quaternionic product compatible with the left action of a $SU(2)$ element on $\mathbb{C}^2$. That is, the left action of a linear transformation on $\mathbb{C}^2$ can be made to correspond to multiplication on the {\it left} by a unit quaternion in $\mathbb{H}$ by writing quaternions with scalars {\it on the right}. For then, we get
\begin{align*}
(\mathds{1}z_1+\mathds{J}z_2)(\mathds{1}y_1+\mathds{J}y_2)&=\mathds{1}z_1y_1+\mathds{J}z_2y_1+z_1\mathds{J}y_2+\mathds{J}z_2\mathds{J}y_2\\
&=\mathds{1}(z_1y_1-\overline{z}_2y_2)+\mathds{J}(z_2y_1+z_1\overline{y}_2)
\end{align*}
which corresponds to
$$
\left(\begin{array}{c}
z_1y_1-\overline{z}_2y_2 \\ z_2y_1+z_1\overline{y}_2
\end{array}\right)=\left(\begin{array}{cc}
z_1 & -\overline{z}_2 \\ z_2 & \overline{z}_1 
\end{array}\right)\left(\begin{array}{c}
y_1 \\ y_2 
\end{array}\right) 
$$
with the identification of $SU(2)$ and $Sp(1)$ given by
$$
\left(\begin{array}{cc}
z_1 & -\overline{z}_2 \\ z_2 & \overline{z}_1 
\end{array}\right) \longleftrightarrow \mathds{1}z_1 + \mathds{J}z_2.
$$

Either of these two identifications of $SU(2)$ with $Sp(1)$ introduces quaternionic coordinates on $SU(2)$. We choose the latter due its salient property of keeping the quaternionic product compatible with the left action of $SU(2)$ on $\mathbb{C}^2$. In other words, we consider the quaternions as a right complex vector space.

It is an easy check that this identification preserves multiplication in $SU(2)$. If
$$
\left(\begin{array}{cc}
\alpha & -\overline{\beta} \\ \beta & \overline{\alpha} 
\end{array}\right) , \left(\begin{array}{cc}
\delta & -\overline{\gamma} \\ \gamma & \overline{\delta} 
\end{array}\right) \in SU(2),
$$
then their product in $SU(2)$ results in
$$
\left(\begin{array}{cc}
\alpha & -\overline{\beta} \\ \beta & \overline{\alpha} 
\end{array}\right) \left(\begin{array}{cc}
\delta & -\overline{\gamma} \\ \gamma & \overline{\delta} 
\end{array}\right) = 
\left(\begin{array}{cc}
\alpha\delta -\overline{\beta}\gamma & -\alpha\overline{\gamma} - \overline{\beta\delta} \\ \beta\delta + \overline{\alpha}\gamma & -\beta\overline{\gamma} + \overline{\alpha\delta}  
\end{array}\right)
$$
which is identified with the unit quaternion
\begin{equation}\label{quat}
\mathds{1}(\alpha\delta -\overline{\beta}\gamma)+\mathds{J}(\beta\delta + \overline{\alpha}\gamma),
\end{equation}
while identifying the $SU(2)$ elements with unit quaternions \emph{first} results in the quaternionic product  
$$
(\mathds{1}\alpha + \mathds{J}\beta)(\mathds{1}\delta + \mathds{J}\gamma)
                   =\mathds{1}(\alpha \delta - \overline{\beta}\gamma) + \mathds{J}(\beta\delta + \overline{\alpha} \gamma)
$$
the result of which is exactly the quaternion in (\ref{quat}). In fact, this identification sets up a Lie group isomorphism between $Sp(1)$ and $SU(2)$.

\section{Identifying $Sp(1)$ with $\mathbb{C}P^1$}

Observe that the Bloch sphere
$$
\mathbb{C}P^1 \equiv
\left(\mathbb{C}^2 - \left\{0\right\}\right)/\mathbb{C}^{*} \cong S^3/U(1)
$$
where $\mathbb{C}^{*}=\mathbb{R}^{+} \times U(1)$ and 
$$
\left(\begin{array}{c}
x \\ y
\end{array}\right)\equiv \left(\begin{array}{c}
x \lambda \\  y \lambda
\end{array}\right)=\left(\begin{array}{c}
x \\  y 
\end{array}\right)\lambda
$$
for $x,y \in \mathbb{C}$, both not equal to zero, and the scalar $\lambda \in U(1)$ and is called {\it phase}. Note that we scalar multiply elements of $\mathbb{C}P^1$ on the \emph{right} rather than the left, a convention that is necessary for differentiating between scalar multiplication and the action of $SU(2)$ on $\mathbb{C}P^1$ under the identifications. 

The Hopf map $H:S^3 \rightarrow \mathbb{C}P^1$ is defined here as 
$$
H:\left(\begin{array}{c}
x \\ y
\end{array}\right) \longmapsto yx^{-1}
$$ 
with $0^{-1}$ considered to be the number $\frac{1}{0}$. On the Bloch sphere, the pure states are represented by $\frac{0}{1}$ and $\frac{1}{0}$ corresponding to the vectors 
$$
\left(\begin{array}{c}
1 \\ 0
\end{array}\right), \left(\begin{array}{c}
0 \\ 1
\end{array}\right)
$$
respectively. In general, $\frac{\beta}{\alpha}=\beta \alpha^{-1}$ corresponds to the vector
$$
\left(\begin{array}{c}
\alpha  \\ \beta
\end{array}\right)
$$
and up to unitary phase
$$
\left(\begin{array}{c}
\alpha \lambda  \\ \beta \lambda
\end{array}\right) \equiv 
\left(\begin{array}{c}
\alpha  \\ \beta
\end{array}\right)
$$
We identify this element of the Bloch sphere with a unit quaternion representing its orbit in $S^3$. That is,
\begin{equation}\label{identification}
\left(\begin{array}{c}
x \\ y
\end{array}\right) \longmapsto \mathds{1}x+\mathds{J}y
\end{equation}
where
\begin{equation}\label{basis identification}
\left(\begin{array}{c}
1 \\ 0
\end{array}\right) \longmapsto \mathds{1}, \left(\begin{array}{c}
i \\ 0
\end{array}\right) \longmapsto  \mathds{I}, \left(\begin{array}{c}
0 \\ 1
\end{array}\right) \longmapsto  \mathds{\mathds{J}}, \left(\begin{array}{c}
0 \\ -i
\end{array}\right) \longmapsto  \mathds{K}
\end{equation}
is the identification of the basis of $\mathbb{C}^2$ (hence $\mathbb{C}P^1$) with the basis of $\mathbb{H}$ as complex vector spaces. The identifications in equations (\ref{identification}) and (\ref{basis identification}) induce a product between elements of $SU(2)$ and elements of $\mathbb{C}P^1$ via quaternionic multiplication that is consistent with the left action of an appropriate $SU(2)$ element on the elements of $\mathbb{C}P^1$. That is, for 
$$
A=\left(\begin{array}{c}
\alpha \\ \beta
\end{array}\right), \Delta=\left(\begin{array}{c}
\delta \\ \gamma
\end{array}\right) \in \mathbb{C}P^1, \left(\begin{array}{cc}
\alpha & -\overline{\beta} \\ \beta & \overline{\alpha} 
\end{array}\right) \in SU(2)
$$
quaternionic multiplication gives the product $\star$ between $A$ and $\Delta$ as follows.
\begin{align*}
A \star \Delta = 
\left(\begin{array}{c}
\alpha \\ \beta
\end{array}\right) \star \left(\begin{array}{c}
\delta \\ \gamma
\end{array}\right) &= ( \mathds{1}\alpha +  \mathds{\mathds{J}}\beta)( \mathds{1}\delta + \mathds{\mathds{J}}\gamma) \\
                   &=  \mathds{1}(\alpha \delta - \overline{\beta}\gamma) + \mathds{\mathds{J}}(\beta\delta + \overline{\alpha} \gamma)\\
                   &=\left(\begin{array}{c}
\alpha \delta - \overline{\beta}\gamma\\ \beta\delta + \overline{\alpha} \gamma
\end{array}\right)=\left(\begin{array}{cc}
\alpha & -\overline{\beta} \\ \beta & \overline{\alpha} 
\end{array}\right)\left(\begin{array}{c}
\delta \\ \gamma
\end{array}\right)
\end{align*}

\subsection{Action of $U(1)$ on $\mathbb{C}P^1$}

Note that the unit complex numbers $U(1)$ can be embedded into $SU(2)$ via
$$
\alpha \hookrightarrow \left(\begin{array}{cc}
\alpha & 0 \\ 0 & \overline{\alpha}
\end{array}\right)
$$
and in this form act on $\mathbb{C}P^1$ as linear transformation instead of scalar multiplication. Our identifications respect this fact, as the following example shows. 
\begin{align*}
 \left(\begin{array}{cc}
\alpha & 0 \\ 0 & \overline{\alpha}
\end{array}\right) \left(\begin{array}{c}
\delta \\ \gamma
\end{array}\right) =\left(\begin{array}{c}
\alpha\delta \\ \overline{\alpha}\gamma 
\end{array}\right) \longmapsto  \mathds{1}(\alpha\delta)+ \mathds{\mathds{J}}(\overline{\alpha}\gamma)&=\alpha \mathds{1}\delta+\alpha \mathds{\mathds{J}}\gamma \\
&=\alpha( \mathds{1}\delta+ \mathds{\mathds{J}}\gamma).
\end{align*}
Note that even though in the expression $\alpha( \mathds{1}\delta+\alpha \mathds{\mathds{J}}\gamma)$ the complex number $\alpha$ appears on the left, it does not represent scalar multiplication because of our convention that scalars multiply on the right. In fact, it's occurrence on the left of the quaternion $ \mathds{1}\delta+\alpha \mathds{\mathds{J}}\gamma$ tells us that it represents the action of $U(1)$ as a linear transformation under the embedding in $SU(2)$.

%


\chapter{Future Directions}

The proper quantization protocols developed in chapter \ref{Proper} for history dependent Parrondo games using certain quantum multiplexers lend a game theoretic perspective to the study of quantum logic circuits via quantum multiplexers. Indeed, the notion of the Parrondo effect is now attached to quantum circuits, and it is natural to raise the following question: 1) can a genuine ``quantum Parrondo effect'' be characterized in quantum circuits through this game theoretic perspective?

Moreover, to date there is no agreement in the literature on exactly what a quantum Markov process is. One difficulty lies in coming up with an appropriate definition of the ``quantum'' stable state. Our quantizations of history dependent Parrondo games are essentially specific quantized Markov processes involving specific elements of the Lie group $SU(2)$ and with stable states chosen game-theoretically. A more general set up is possible in which arbitrary elements of $SU(2)$ are utilized. In such a set up, is it possible to use quantum game theory to come up with a natural choice for the stable state? Moreover, is it possible to characterize a quantized version of the Parrondo effect in this general set up, and if so, what does it mean for quantum computation? 

To be more precise, the work in Chapter \ref{Proper} embeds classical history dependent Parrondo games into quantum multiplexers via embeddings of type 1 and 2. The resulting quantum multiplexers, when made to act upon a particular evaluative initial state, reproduce the payoff functions of the classical Parrondo games. Call such quantum multiplexers {\it mundane}. In other words, mundane quantum multiplexers reside in the image of the embeddings of type 1 or 2. However, the set of quantum multiplexers is much larger than the image of embeddings of either type; that is, there are quantum multiplexers that are outside such an image. Call such quantum multiplexers {\it exotic}. 

Clearly, the answer to question 1) above is in the affirmative for mundane quantum multiplexers based on the results of chapter \ref{Proper}. By taking quantum superpositions of the mundane quantum multiplexers associated with classical Parrondo games, the payoff function of the classical game can be reproduced by choosing a particular evaluative initial state such that the game is winning, even when the individual quantum games were losing with respect to appropriate evaluative initial states. For exotic quantum multiplexers, the answer is not clear cut since it is not known what an evaluative initial state for such a multiplexer should be. Therefore, a future study toward answering question 1) requires efforts into identifying such an appropriate initial state for exotic quantum multiplexers. In the context of quantum logic synthesis, how might an arbitrary quantum logic gate be synthesized via decomposition in a game theoretically meaningful way? Tha it, first assign a fixed number of qubits in the circuit to each player. Then, for an arbitrary quantum logic gate $U$, how might $U$ be decomposed into sets of one qubit gates, one for each player, and an initial state choosen, such that a given game theoretic outcome might be realized? 

%
%
%
%
%
%
%


\singlespace
\bibliography{refs} 
\bibliographystyle{plain} 

\doublespace
\chapterfont{\centering  \normalsize  }
\chapter*{Appendix A} \vspace{-.28in}
\begin{center}
{\bf QUATERNIONS}
\end{center} \vspace{.25in}
\addcontentsline{toc}{chapter}{Appendix A \hspace{.05in} Quaternions}
\chaptermark{Appendix}
\sectionmark{Appendix}
\markboth{Appendix A. Quaternions}{Appendix A. Quaternions}

\noindent Complex numbers are extension of real numbers. This fact motivates us to view quaternions as extension of the complex numbers, with the exception that the recipe for constructing the conjugate of a complex number needs modification when one tries to follow it to construct the conjugate of a quaternion. This modification is such that the quaternionic product is necessarily non-commutative and satisfies $zj=j\overline{z}$ for any complex number $z$ and the quaternion $j$. \vspace{.2in}

\noindent {\bf A.1 \quad Complex Numbers} \vspace{.2in}

\noindent The set of {\it complex numbers} is
$$
\mathbb{C}=\left\{a_0+a_1x : a_0,a_1 \in \mathbb{R} \hspace{.05in} {\rm and} \hspace{.05in} x^2= -1\right\}.
$$
Since complex numbers are just first degree polynomials, one defines binary operations of addition and multiplication on $\mathbb{C}$ via polynomial addition and multiplication respectively.
\begin{align*}
{\rm Addition}&: \hspace{.1in} (a_0+a_1x)+(b_0+ b_1x)=(a_0+b_0)+ (a_1+b_1)x\\
{\rm Multiplication}&: \hspace{.1in}  (a_0+a_1x)(b_0+ b_1x)=a_0b_0 + a_1b_0x+a_0b_1x+a_1b_1x^2
\end{align*}
The constraint $x^2=-1$ provides multiplicative closure to $\mathbb{C}$, yielding 
$$
(a_0+a_1x)(b_0+ b_1x)=(a_0b_0-a_1b_1)+(a_0b_1+a_1b_0)x
$$
The equation $x^2=-1$ has exactly two solutions, $x=\sqrt{-1}$ and $x=-\sqrt{-1}$. Setting $x=i=\sqrt{-1}$ leads to the conventional notation for the complex numbers
$$
\mathbb{C}=\left\{a_0+a_1i : a_0,a_1 \in \mathbb{R} \hspace{.05in} {\rm and} \hspace{.05in}i^2= -1\right\}.
$$
The solutions $i$ and $-i$ are called {\it imaginary numbers}. This terminology gives rise to the notion of the {\it real} part $a_0$ and the {\it imaginary} part $a_1$ of the complex number $a_0+a_1i$. Note that since $-i$ is also a solution to the equation $x^2=-1$, there are complex numbers in $\mathbb{C}$ of the form 
$$a_0+a_1(-i)=a_0-a_1i.
$$ 
The latter is called the {\it conjugate} of the complex number $a_0+a_1i$, and one checks that 
$$
(a_0+a_1i)(a_0-a_1i)=a_0^2+a_1^2 \in \mathbb{R}
$$
Clearly, the conjugate of $a_0-a_1i$ is the complex number $a_0+a_1i$; that is, double conjugation gives back the original complex number. The quantity 
$$
\left|a_0+a_1i\right|=\sqrt{a_0^2+a_1^2}
$$
defines the {\it length} of the complex number $a_0+a_1i$ (and of $a_0-a_1i$). It is an easy exercise to show that $\mathbb{C}$ in fact forms a field. \vspace{.2in}

\noindent {\bf A.2 \quad Quaternions} \vspace{.2in}

\noindent  The set of {\it quaternions} is
$$
\mathbb{H}=\left\{p_0+p_1y : p_0,p_1 \in \mathbb{C} \hspace{.05in} {\rm and} \hspace{.05in} y^2= -1\right\}.
$$
Again, addition and multiplication in $\mathbb{H}$ is defined as polynomial addition and multiplication, giving
\begin{align*}
{\rm Addition}&: \hspace{.1in} (p_0+p_1y)+(q_0+ q_1y)=(p_0+q_0)+ (p_1+q_1)y\\
{\rm Multiplication}&: \hspace{.1in}  (p_0+p_1y)(q_0+ q_1y)=p_0q_0 + (p_1q_0+p_0q_1)y+p_1q_1y^2
\end{align*}
Is $\mathbb{H}$ closed under multiplication? The answer is yes once we note that $p_0, p_1, q_0, q_1$ are all complex numbers and that this requires the use of both the constraints $y^2= -1$ and $x^2= -1$ in simplifying the quaternionic product. Let
$$
p_0=p'_0+p'_1i, \quad p_1=p''_0+p''_1i, \quad q_0=q'_0+q'_1i, \quad q_1=q''_0+q''_1i
$$
be complex numbers. Simplifying the quaternionic product now results in the expression
\begin{align*}
&(p_0+p_1y)(q_0+ q_1y)\\
&=(p'_0q'_0-p'_1q'_1)+(p'_0q'_1+p'_1q'_0)i\\
&+\left[(p''_0q'_0-p''_1q'_1)+(p''_0q'_1+p''_1q'_0)i+(p'_0q''_0-p'_1q''_1)+(p'_0q''_1+p'_1q''_0)i\right]y\\
&+\left[(p''_0q''_0-p''_1q''_1)+(p''_0q''_1+p''_1q''_0)i\right]y^2\\
&=(p'_0q'_0-p'_1q'_1-p''_0q''_0-p''_1q''_1)+(p'_0q'_1+p'_1q'_0-p''_0q''_1-p''_1q''_0)i\\
&+(p''_0q'_0-p''_1q'_1+p'_0q''_0-p'_1q''_1)y+(p''_0q'_1+p''_1q'_0+p'_0q''_1+p'_1q''_0)iy \\
&=(p'_0q'_0-p'_1q'_1-p''_0q''_0-p''_1q''_1)+(p'_0q'_1+p'_1q'_0-p''_0q''_1-p''_1q''_0)i\\
&+\left[(p''_0q'_0-p''_1q'_1+p'_0q''_0-p'_1q''_1)+(p''_0q'_1+p''_1q'_0+p'_0q''_1+p'_1q''_0)i\right]y \\
&=z_0+z_1y
\end{align*}
for complex numbers 
$$z_0=(p'_0q'_0-p'_1q'_1-p''_0q''_0-p''_1q''_1)+(p'_0q'_1+p'_1q'_0-p''_0q''_1-p''_1q''_0)i
$$ 
and 
$$
z_1= (p''_0q'_0-p''_1q'_1+p'_0q''_0-p'_1q''_1)+(p''_0q'_1+p''_1q'_0+p'_0q''_1+p'_1q''_0)i.
$$

It is important to note here that even though the variable $y$ is a square root of $-1$, it is {\it not} equal to $\pm i$. For if it were equal to $\pm i$, then the set $\mathbb{H}$ would equal the set $\mathbb{C}$! By analogy with the complex numbers, the variable $y$ might appropriately be called an {\it imaginary complex number}. It is commonly known as a {\it hypercomplex number}. Following convention, we replace $y$ with $j$ and write quaternions as $p_0+p_1j$. 

We next develop the notion of a conjugate of a quaternion; that is, for a given quaternion $p$, find a quaternion $q$ such that $pq \in \mathbb{R}$. Following the recipe that led to the definition of the complex conjugate naively we set $p_0 +p_1(-j)=p_0-p_1j$ as the {\it quaternionic conjugate} of the quaternion $p_0+p_1j$. This gives 
\begin{equation*}\tag{A.1}\label{quat conjugate}
(p_0+p_1j)(p_0-p_1j)=p_0^2+p_1p_0j-p_0p_1j+p_1^2. 
\end{equation*}
Multiplication of a complex number by its conjugate results in a real number that is the sum of the squares of two real numbers, namely the real and imaginary parts of the complex number. Since our definition of the quaternionic conjugate is motivated by the complex conjugate, we expect the right hand side of equation (\ref{quat conjugate}) to equal to the real number that results from the squares of the complex numbers $p_0$ and $p_1$. However, the fact that in general the square of a complex number is another complex number puts a kink in our plans. But all is not lost. Instead of insisting on the squares of the complex numbers $p_0$ and $p_1$ in our definition of the quaternionic conjugate, we are perfectly happy to work with the {\it squares of the lengths} of the complex numbers $p_0$ and $p_1$, which are both real numbers. This flexibility forces us to modify the proposed quaternionic conjugate to the quaternion $(\overline{p_0}-\overline{p_1}j)$ which gives
\begin{equation*}\tag{A.2}\label{moddified quat conjugate}
(p_0+p_1j)(\overline{p_0}-\overline{p_1}j)=
\left|p_0\right|^2+p_1\overline{p_0}j-p_0\overline{p_1}j+\left|p_1\right|^2
\end{equation*}
To eliminate the quaternionic part from the right hand side of equation (\ref{moddified quat conjugate}) we are forced to set 
$$
p_1\overline{p_0}=p_0\overline{p_1}=\overline{p_1}p_0
$$ 
which means that $p_1\overline{p_0}$ is in fact a real number, sacrificing the generality of our argument. 

At this stage, one wonders whether the recipe for the complex conjugate that has been followed thus far with a slight modification to develop the quaternionic conjugate needs to be changed drastically. Indeed, if we leave out the major ingredient of commutativity from the recipe and assume that for a complex number $z$, 
\begin{equation*}\tag{A.3}\label{non-comm condition}
zj=j\overline{z},
\end{equation*}
then equation (\ref{quat conjugate}) must be re-written as 
\begin{align*}\label{quat conjugate non-comm}
(p_0+p_1j)(p_0-p_1j)&=p_0^2+p_1jp_0-p_0p_1j+p_1jp_1j\\
&=p_0^2+p_1\overline{p_0}j-p_0p_1j+p_1\overline{p_1}jj\\
&=p_0^2+p_1\overline{p_0}j-p_0p_1j+\left|p_1\right|^2
\end{align*}
The occurrence of $\left|p_1\right|^2$ in the preceding equation is glaring, and suggests that we modify the proposed quaternionic conjugate yet again to be 
$\overline{p_0}-p_1j$ which upon multiplication with $p_0+p_1j$ and after using the non-commutativity condition $zj=j\overline{z}$ leads to
\begin{align*}
(p_0+p_1j)(\overline{p_0}-p_1j)&=\left|p_0\right|^2+p_1j\overline{p_0}-p_0p_1j+\left|p_1\right|^2\\
&=\left|p_0\right|^2+p_1p_0j-p_0p_1j+\left|p_1\right|^2\\
&=\left|p_0\right|^2+\left|p_1\right|^2\\
&=(p'_0)^2+(p'_1)^2+(p''_0)^2+(p''_1)^2
\end{align*}
The quaternionic conjugate defined this way behaves much like the complex conjugate. For example, the quaternionic conjugate of $\overline{p_0}-p_1j$ is $p_0+p_1j$. Moreover, as with the complex conjugate, the product of a quaternion with its conjugate is expressible as the sum of squares of four real numbers. We use the latter to define the {\it lenght} of a quaternion as
$$
\left|p_0+p_1j\right|=\sqrt{\left|p_0\right|^2+\left|p_1\right|^2}=\sqrt{(p'_0)^2+(p'_1)^2+(p''_0)^2+(p''_1)^2}.
$$

Rewriting $p_0+p_1j$ as 
\begin{equation*}\tag{A.4}\label{quat real}
p_0+p_1j=(p'_0+p'_1i)+(p''_0+p''_1i)j=p'_0+p'_1i+p''_0j+p''_1ij
\end{equation*}
introduces the term $ij$ which the non-commutativity condition of equation (\ref{non-comm condition}) shows to be a square root of $-1$. For convenience, set $k=ij$. Then one computes 
$$
k^2=(ij)^2=(ij)(ij)=(ij)(j(-i))=i(-1)(-i)=i^2=-1
$$
Complex number arithmetic together with equation (\ref{non-comm condition}) establish the following identities as well.
$$
ik=i(ij)=i^2j=-j
$$
$$
jk=j(ij)=(-i)j^2=i
$$
The last two identities and the identity $ij=k$ establish the {\it right-hand rule} for quaternionic multiplication which is conveniently represented in the picture below. 
This rule is summed up in {\it Hamilton's Relation} $i^2=j^2=k^2=ijk=-1$.  

\begin{figure}[htp]
\centering
\includegraphics[scale=0.5]{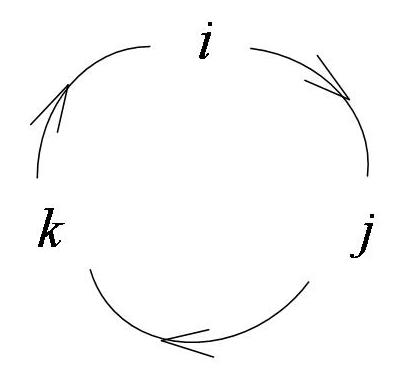}
\end{figure}

One can verify that the quaternions satisfy all the axioms of a field except commutativity, and therefore form a division ring. Our definition of the quaternions in fact shows that the quaternions form a two dimensional algebra over the complex numbers with basis $\left\{1, j\right\}$. Equation (\ref{quat real}) shows that the quaternions form a four dimensional algebra over the reals with basis $\left\{1, i,j,k\right\}$.

\doublespace
\chapterfont{\centering  \normalsize  }
\chapter*{Appendix B}\vspace{-.28in}
\begin{center}
{\bf COSINE SINE DECOMPOSITION OF UNITARY MATRICES}
\end{center} \vspace{.25in}
\addcontentsline{toc}{chapter}{Appendix B \hspace{.05in} Cosine Sine Decomposition of Unitary Matrices}
\chaptermark{Appendix}
\sectionmark{Appendix}
\markboth{Appendix B. Cosine Sine Decomposition of Unitary Matrices}{Appendix B. Cosine Sine Decomposition of Unitary Matrices}

As we shall see, the cosine sine decomposition (CSD) is essentially the well known singular value decomposition (SVD) of a unitary matrix implemented at the block matrix level. The reader is cautioned that for a given matrix, the CSD is {\it not} unique. The material presented in this appendix is not new. The discussion of the SVD is based on lecture notes of Professor Bin Jiang at Portland State University and the CSD discussion is based on the account given in \cite{Stewart:90} on pages 37-40. \vspace{.2in}

\noindent {\bf B.1 \quad Singular Value Decomposition} \vspace{.2in}

\noindent Begin with the vector and matrix $2$-norms, described below. \vspace{.15in} 

\noindent {\bf Definition B.1.}\label{vect 2 norm}
The $2$-norm of a vector $x \in \mathbb{C}^n$ is the function $\left\| \hspace{0.05in}\right\|_2: \mathbb{C}^n \rightarrow \mathbb{R}$ defined by
$$
\left\|x\right\|_2=\left(x^{\dag}x\right)^{\frac{1}{2}}=\left( \sum_{i=1}^n|x_i|^2\right)^{\frac{1}{2}}
$$

Here, $x^{\dag}=(x_1^*,x_2^*, \dots, x_n^*)^T$ and $|x_i|^2=x_ix_i^*$ for $x_i \in \mathbb{C}$.\vspace{.15in} 

\noindent {\bf Definition B.2.}\label{mat 2 norm}
The $2$-norm of a matrix $A \in \mathbb{C}^{m \times n}$ is the function $\left\| \hspace{0.05in}\right\|_2: \mathbb{C}^{m \times n} \rightarrow \mathbb{R}$ defined by
$$
\left\|A \right\|_2={\rm max}_{\left\|x\right\| \neq 0}\frac{\left\|Ax\right\|_2}{\left\|x\right\|_2}={\rm max}_{\left\|x \right\|_{2}=1}\left\|Ax \right\|_2
$$

Since we will not refer to any other norms that can be defined on vectors and matrices, from now on we will use the $\left\| \hspace{0.05in} \right\|$ instead of the more explicit $\left\| \hspace{0.05in} \right\|_2$ to simplify notation. Also, for $A \in \mathbb{C}^{m \times n }$, denote by $A^{\dag}$ the conjugate transpose of $A$. Recall that a matrix $A$ is {\it unitary} if $AA^{\dag}=A^{\dag}A=I$. Equivalently, the action of a unitary matrix preserves vector norm.\vspace{.15in} 

\noindent {\bf Lemma B.3.}\label{lem1}
Vector and matrix $2$-norms are invariant under unitary transformations.

\begin{proof} Let $U \in \mathbb{C}^{n \times n}$ be a unitary transformation, and $x \in \mathbb{C}^n$. Then 
$$
\left\|Ux\right\|=\left((Ux)^{\dag}(Ux) \right)^{\frac{1}{2}}=\left(x^{\dag}U^{\dag}Ux \right)^{\frac{1}{2}}=\left(x^{\dag}x \right)^{\frac{1}{2}}=\left\|x\right\|
$$
Now let $A \in \mathbb{C}^{m \times n}$. Then
$$
\left\|AU\right\|={\rm max}_{\left\|x \right\|=1}\left\|AUx \right\|={\rm max}_{\left\|Ux \right\|=1}\left\|AUx \right\|={\rm max}_{\left\|y \right\|=1}\left\|Ay\right\|=\left\|A \right\|
$$
If $A \in \mathbb{C}^{n \times n}$. Then
$$
\left\|UA\right\|={\rm max}_{\left\|x \right\|=1}\left\|(UA)x \right\|={\rm max}_{\left\|x \right\|=1}\left\|U(Ax) \right\|={\rm max}_{\left\|x \right\|=1}\left\|Ax \right\|=\left\|A \right\|
$$
\end{proof}

We are now ready to prove the existence of a singular value decomposition.\vspace{.15in} 

\noindent {\bf Proposition B.4.}\label{SVD} If $A \in \mathbb{C}^{m \times n}$, then there exists unitary matrices $U \in \mathbb{C}^{m \times m}$ and $V \in \mathbb{C}^{n \times n}$, and a matrix $\Sigma={\rm diag}(\sigma_1, \sigma_2, \dots, \sigma_p, 0, \dots, 0) \in \mathbb{R}^{m \times n}$, $p={\rm min}(m,n)$, such that 
$$
A=U \Sigma V^{\dag}.
$$

The $\sigma_i$ are called {\it singular values} of $A$ and are typically ordered so that
$$
\sigma_1 \geq \sigma_2 \geq \dots \sigma_p \geq 0.
$$ 
\begin{proof} The proof will be inductive. Let $\sigma=\left\|A\right\|$. Since 
$$
\left\|A\right\|={\rm max}_{\left\|x\right\|=1}\left\|Ax\right\|,
$$ 
there exists a unit norm $x \in \mathbb{C}^n$ such that $\sigma = \left\|Ax\right\|$; therefore, $Ax=\sigma y$ for some $y \in \mathbb{C}^m$ with $\left\|y\right\|=1$.

If 
$$V_1=\left(v_1 \quad v_2 \quad \dots \quad v_r\right) \in \mathbb{C}^{m \times r}, \quad r<m
$$
has orthonormal columns $v_i$, then applying Gram-Schimdt process we can always find 
$$
V_2=\left(v_{r+1} \quad v_{r+2} \quad \dots \quad v_{m}\right) \in \mathbb{C}^{m \times (m-r)}
$$
so that $(V_1, V_2)$ is unitary and ${\rm rank}(V_1)^{\perp}={\rm rank}(V_2)$. From this fact we conclude that there exist $V_1' \in \mathbb{C}^{n \times (n-1)}$ and $V_1' \in \mathbb{C}^{m \times (m-1)}$ such that $V_1=(x \quad V_1') \in \mathbb{C}^{n \times n}$ and $U_1=(y \quad U_1') \in \mathbb{C}^{m \times (m-1)}$ are unitary. Hence,
\begin{align*}
U_1^{\dag}AV_1=\left(\begin{array}{c}
 y^T  \\ U_1^T \end{array}\right)A(x \quad V_1')&=\left(\begin{array}{cc}
y^TAx & y^TAV_1' \\ (U_1')^TAx & (U_1')^TAV_1'
\end{array}\right)\\
&=\left(\begin{array}{cc}
y^T \sigma y & y^TAV_1' \\ (U_1')^T \sigma y & (U_1')^TAV_1'
\end{array}\right)\\
&=\left(\begin{array}{cc}
\sigma & w^T \\
0 & B
\end{array}\right)\equiv A_1
\end{align*}
where $w^T \in \mathbb{R}^{(n-1)}$.

In fact $w=0$. For by lemma {\bf B.3.}, $\left\|A_1\right\|=\left\|A\right\|=\sigma$ and 
\begin{align*}
\left\|A_1\right\|&={\rm max}_{\left\|x\right\| \neq 0}\frac{\left\|A_1x\right\|}{\left\|x\right\|} \\ 
& \geq \frac{\left\|A_1 \left(\begin{array}{c}
 \sigma \\ w \end{array}\right)\right\|}{\left\|\left(\begin{array}{c}
 \sigma \\ w \end{array}\right)\right\|} \\
 &= \frac{\left\| \left(\begin{array}{c}
\sigma^2 + w^Tw \\ Bw \end{array}\right)\right\|}{\sqrt{\sigma^2+w^Tw}}\\
&\geq \frac{\sqrt{\left(\sigma^2+w^Tw\right)^2}}{\sqrt{\sigma^2+w^Tw}}\\
&=\sqrt{\sigma^2+w^Tw}
\end{align*}
Therefore, $\sigma \geq \sqrt{\sigma^2+w^Tw}$ and hence $w^Tw=0$ which implies that $w=0$. 

We now have that 
\begin{equation*}\tag{B.1}\label{SVD proof}
U_1^{\dag}AV_1=\left(\begin{array}{cc}
\sigma & 0 \\
0 & B
\end{array}\right)
\end{equation*}

Now applying the same method to $B$ and the resulting blocks $B'$ inductively, we have
$$
U_p^{\dag} \dots U_2^{\dag}U_1^{\dag}AV_1V_2 \dots V_p={\rm diag}(\sigma_1, \sigma_2, \dots \sigma_p, 0, \dots, 0)
$$
Let $U=U_1U_2 \dots U_p$ and $VV_1 V_2 \dots V_p$. Then both $U$ and $V$ are unitary and 
$$
A=U \Sigma V^{\dag}.
$$
\end{proof}

\noindent {\bf B.2 \quad Cosine Sine Decomposition} \vspace{.2in}

\noindent {\bf Proposition B.5.}
Let the unitary matrix $\textit{W}\in \textbf{C}^{n\times n}$ be partitioned in $2 \times 2$ block form as
$$
W=\bordermatrix {  &l      & n-r    \cr
                 l &W_{11} & W_{12} \cr
               n-l &W_{21} & W_{22} \cr}
$$
with $2l\leq n$. Then there exist unitary
matrices $U={\rm diag}(U_{11}, U_{22})$ and $V={\rm diag}(V_{11},V_{22})$ with $U_{11}, V_{11} \in \mathbb{C}^{l \times l}$ such that
\begin{equation*}\tag{B.2}\label{CSD}
U^{\dag}WV  = \bordermatrix {  &l      &l     &n-2l   \cr
                 l &C & -S  &0 \cr
                 l &S & C &0   \cr
              n-2l &0      & 0     &I }  
\end{equation*}
where 
$$
C = {\rm diag}(\cos \theta_{1}, \cos\theta_{2}, \dots,\cos\theta_{l})
$$
$$
S = {\rm diag}(\sin \theta_{1}, \sin \theta_{2},\dots,\sin \theta_{l})
$$ 
such that $\sin^{2}\theta_{i}+\cos^{2}\theta_{i}=1$ for some $\theta_{i}$, $1 \leq i \leq l$.

\begin{proof} Let 
$$
U_{11}^{\dag}W_{11}V_{11}=C
$$
be a singular value decomposition of the block $W_{11}$ of $W$ and suppose that 
$$
C={\rm diag}(C_1,I_{l-k})
$$
where the diagonal elements of $C_1$ satisfy 
$$
0 \leq c_1 \leq c_2 \leq \dots c_k <1.
$$
Note that since $W$ is unitary, the singular values cannot be greater than 1.  Clearly, the columns of the matrix 
$$
\left(\begin{array}{c}
W_{11} \\ W_{21}
\end{array}\right)V_{11}
$$ 
are orthonormal. Therefore, 
$$
I=\left[\left(\begin{array}{c}
W_{11} \\ W_{21}
\end{array}\right)V_{11}\right]^{\dag}\left[\left(\begin{array}{c}
W_{11} \\ W_{21}
\end{array}\right)V_{11}\right]=C^2+\left(W_{21}V_{11} \right)^{\dag}\left( W_{21}V_{11} \right);
$$
that is, 
$$
(W_{21}V_{11})^{\dag}(W_{21}V_{11})={\rm diag}(I-C_1^2,0_{l-k})
$$
The columns of $W_{21}V_{11}$ are orthogonal with the last $(l-k)$ of them being 0. Thus, there exists a unitary matrix $\widehat{U}_{22} \in \mathbb{C}^{(n-l) \times (n-l)}$ such that
$$
\widehat{U}_{22}^{\dag}W_{21}V_{11}=\left(\begin{array}{c}
S \\ 0\end{array}\right)
$$  
where 
\begin{equation*}\tag{B.3}\label{S condition}
S={\rm diag}(s_1, s_2, \dots, s_k, 0, \dots, 0)={\rm diag}(S',0)
\end{equation*}
with $S'$ consisting of $k$ rows and the all $0$'s block consisting of $(r-k)$ rows. Since 
$$
{\rm diag}(U_{11}, \widehat{U}_{22})^{\dag}\left(\begin{array}{c}
W_{11} \\ W_{21}
\end{array}\right)V_{11}=\left(\begin{array}{c}
C\\ S \\0
\end{array}\right)
$$
has orthogonal columns, it follows that for $1 \leq i \leq l$
\begin{equation*}\tag{B.4}\label{cs condition}
c_i^2+s_i^2=1.
\end{equation*}
In particular, $S'$ is non-singular.

Similarly, we may determine a unitary matrix $V_{22} \in \mathbb{C}^{(n-l) \times (n-l)}$ such that
$$
U_{11}^{\dag}W_{12}V_{22}=(T,0)
$$
where $T={\rm diag}(t_1, t_2, \dots, t_l)$ with $t_i \leq 0$. Since 
$$
U_{11}^{\dag}(W_{11} \quad W_{12}){\rm diag}(V_{11}, V_{22})=(C \quad T \quad 0)
$$
has orthogonal rows, it must be that $c_i^2+t_i^2=1$, and it follows from (\ref{S condition}) and (\ref{cs condition})that $T=-S$.

Now set $\widehat{U}={\rm diag}(U_{11}, \widehat{U}_{22})$ and $V={\rm diag}(V_{11}, V_{22})$. Then it follows from the preceding discussion that 
$$
X=\widehat{U}^{\dag}WV
$$
can be partitioned as 
\begin{equation*}\tag{B.5}\label{X matrix}
X=\bordermatrix {    &k      & l-k    &k        &l-k         &n-2l  \cr
                 k   &C_1     & 0     &-S_1      &0           &0     \cr
               l-k   &0      & I      &0        &0           &0      \cr
               k     &S_1     &0      &X_{33}   &X_{34}      &X_{35}  \cr
               l-k   &0      &0       &X_{43}   &X_{44}      &X_{45}   \cr
               n-2l  &0      &0       &X_{53}   &X_{54}      &X_{55}   \cr}
\end{equation*}
Since $X$ is unitary and $\Sigma_1$ has positive diagonal elements, we have $X_{33}=C_1$. Moreover, $X_{34}$, $X_{35}$, $X_{43}$, and $X_{53}$ are zero. Therefore, the partition of $X$ in (\ref{X matrix}) can now be updated to
%
\begin{equation*}\tag{B.6}\label{X matrix update}
X=\bordermatrix {    &k      & l-k    &k        &l-k         &n-2l  \cr
                 k   &C'     & 0      &-S'      &0           &0     \cr
               l-k   &0      & I      &0        &0           &0      \cr
               k     &S'     &0       &C'       &0           &0  \cr
               l-k   &0      &0       &0   &X_{44}      &X_{45}   \cr
               n-2l  &0      &0       &0   &X_{54}      &X_{55}   \cr}
\end{equation*}
and the the matrix 
$$
U_{33}=\left(\begin{array}{cc}
X_{44} & X_{45} \\
X_{54} & X_{55} \end{array}\right) \in \mathbb{C}^{(n-l-k) \times (n-l-k)}
$$
is unitary. 

Now we have
\begin{align*}
{\rm diag}(I^{(l+k)}, U_{33}^{\dag})X&=\left(\begin{array}{ccccc}
C_1 & 0 & -S_1 & 0 & 0 \\
0 & I & 0 & 0 & 0\\
S_1 & 0 & C_1 & 0 & 0\\
0 & 0 & 0 & I & 0\\
0 & 0 & 0 & 0 & I\end{array}\right)\\
&= \bordermatrix {  &l      &l     &n-2l   \cr
                 l &C & -S  &0 \cr
                 l &S & C &0   \cr
              n-2l &0      & 0     &I }  
\end{align*}
Note that 
$$
{\rm diag}(I^{(l+k)}, U^{\dag}_{33})X={\rm diag}(I^{(l+k)}, U^{\dag}_{33})U^{\dag}WV.
$$
Hence, if we set
\begin{align*}
U&=\widehat{U}{\rm diag}(I^{(l+k)}, U_{33})\\
&={\rm diag}(U_{11},\widehat{U}_{22}){\rm diag}(I^{(l)}, {\rm diag}(I^{(k)}, U_{33}))\\
&={\rm diag}(U_{11}, \widehat{U}_{22}\cdot{\rm diag}(I^{(k)}, U_{33}))\\
&={\rm diag}(U_{11}, U_{22})
\end{align*}
Set 
$$
U_2={\rm diag}(I_k,\widehat{U}_3)\widehat{U}_2
$$
and
$$
U={\rm diag}(U_1,U_2)
$$
Then
$$
U^{\dag}WV={\rm diag}(I_{r+k},\widehat{U}_3)X,
$$
then $U^{\dag}WV$ has the form (\ref{eqn:CSD1}), where $U$ and $V$ are block diagonal unitary matrices. 
\end{proof}

\doublespace
\chapterfont{\centering  \normalsize  }
\chapter*{Appendix C}\vspace{-.28in}
\begin{center}
{\bf LIST OF NOTATIONS AND NOMENCLATURE}
\end{center} \vspace{.25in}
\addcontentsline{toc}{chapter}{Appendix C \hspace{.05in} List of Notations and Nomenclature}
\chaptermark{Appendix}
\sectionmark{Appendix}
\markboth{Appendix C. List of Notations and Nomenclature}{Appendix C. List of Notations and Nomenclature}

\begin{itemize}
\item The state space of one qubit is the two dimensional complex projective Hilbert space $\mathbb{C}P^1$. As is the convention in quantum mechanics, an element $\psi$ of the state space is denoted in Dirac notation by $\ket{\psi}$ and is called a ``ket'' vector.    
\item $\ket{0}=(1,0)^T$ and $\ket{1}=(0,1)^T$ are elements of the orthonormal computational basis of $\mathbb{C}P^1$. We point out that every ket is a column vector, however, as is the case here, it is sometimes written as the transpose of the appropriate row vector for notational convinience.  
\item $\ket{\psi}=\psi_0\ket{0}+\psi_1\ket{1}=(\psi_0, \psi_1)^T$ is a quantum superposition of the elements of the computational basis, with $\left|\psi_0\right|^2+\left|\psi_1 \right|^2=1$. In the language of linear algebra, $\ket{\psi}$ is a linear combination of the elements of the computational basis.
\item The dual of $\ket{\psi}$ is the ``bra'' vector $\bra{\psi}=(\overline{\psi_0} \quad \overline{\psi_1})$, where $\overline{\psi_i}$ is the complex conjugate of the complex number $\psi_i$. Note that a bra vector is a row vector. 
\item For $\ket{\psi}=(\psi_0,\psi_1)^T$ and $\ket{\phi}=(\phi_0,\phi_1)^T$ in $\mathbb{C}P^1$, their inner product is given by $(\ket{\psi},\ket{\phi})=(\overline{\psi_0} \quad \overline{\psi_1})(\phi_0, \phi_1)^T$ and is denoted in the bra-ket notation by $\bra{\psi}\ket{\phi}$ or just $\left\langle \psi | \phi \right\rangle$.
\item The outer product of $\ket{\psi}$ and $\ket{\phi}$ is denoted by $\ket{\psi}\bra{\phi}$ and is used to construct measurment operators.
\item If $M$ is a matrix, then $M^{\dag}$ is the conjugate transpose of $M$. If $M$ is unitary, then $M^\dag=M^{-1}$. 
\item The trace ${\rm trace}(A)$ of a square matrix $A$ is the sum of its diagonal elements. 
\end{itemize}   

\end{document}